\documentclass[fleqn,usenatbib]{mnras}

\usepackage{newtxtext,newtxmath}
\usepackage[T1]{fontenc}

\usepackage{xcolor}

\DeclareRobustCommand{\VAN}[3]{#2}
\let\VANthebibliography\thebibliography
\def\thebibliography{\DeclareRobustCommand{\VAN}[3]{##3}\VANthebibliography}

\usepackage{graphicx}	% Including figure files
\usepackage{amsmath}	% Advanced maths commands

\newcommand{\teff}{$T_{\rm eff}$} 
\newcommand{\logg}{$\log g$} 
\newcommand{\kms}{km s$^{-1}$}

\title[Magnesium Isotope Ratios in M 22]{The complex stellar system M 22: constraining the chemical enrichment from AGB stars using magnesium isotope ratios}

\author[M. McKenzie et al.]{M. McKenzie,$^{1,2}$\thanks{E-mail: madeleine.mckenzie@anu.edu.au}
D. Yong,$^{1,2}$
A. I. Karakas,$^{3,2}$
E. Wang,$^{1,2}$
S. Monty,$^{4}$
A. F. Marino,$^{5,6}$
A. P. Milone,$^{7,5}$ \newauthor
T. Nordlander,$^{1,2}$
A. Mura-Guzmán,$^{1,2}$
S. Martell$^{8,2}$
M. Carlos$^{9}$\\
$^{1}$Research School of Astronomy \& Astrophysics, Australian National University, Canberra, ACT 2611, Australia, \\
$^{2}$ARC Centre of Excellence for Astrophysics in Three Dimensions (ASTRO-3D), Canberra 2611, Australia,\\
$^{3}$School of Physics \& Astronomy, Monash University, Clayton VIC 3800, Australia, \\
$^{4}$Institute of Astronomy, University of Cambridge, Madingley Rd, Cambridge, CB3 0HA, UK, \\
$^{5}$Istituto Nazionale di Astrofisica — Osservatorio Astronomico di Padova, Vicolo dell’Osservatorio 5, Padova, IT-35122,\\
$^{6}$Istituto Nazionale di Astrofisica — Osservatorio Astrofisico di Arcetri, Largo Enrico Fermi, 5, Firenze, IT-50125,\\
$^{7}$Dipartimento di Fisica e Astronomia “Galileo Galilei,” Università di Padova, Vicolo dell’Osservatorio 3, Padova, IT-35122,\\ 
$^{8}$School of Physics, University of New South Wales, Sydney, NSW 2052, Australia,\\
$^{9}$Department of Physics and Astronomy, Uppsala University, Box 516, SE-751 20 Uppsala, Sweden \\
}

\date{Accepted 2023 September 27. Received 2023 September 27; in original form 2023 August 22}

\pubyear{2023}

\begin{document}
\label{firstpage}
\pagerange{\pageref{firstpage}--\pageref{lastpage}}
\maketitle

% Abstract of the paper
\begin{abstract}

The complex star cluster M 22 (NGC 6656) provides a unique opportunity for studying the slow neutron capture ($s$-)process nucleosynthesis at low metallicity due to its two stellar groups with distinct iron-peak and neutron capture element abundances. Previous studies attribute these abundance differences to pollution from $3-6 \ \rm{M}_{\odot}$ asymptotic giant branch (AGB) stars which produce significant quantities of the neutron-rich Mg isotopes $^{25}$Mg and $^{26}$Mg. We report the first-ever measurements of Mg isotopic abundance ratios at $\rm{[Fe/H]} \ \sim -2$ in a globular cluster-like system using very high-resolution and signal-to-noise spectra (R = 110 000, S/N = 300 per pixel at 514 nm) from the VLT/UVES spectrograph for six stars; three in each $s$-process group. Despite the presence of star-to-star variations in $^{24}$Mg, $^{25}$Mg, and $^{26}$Mg, we find no correlation with heavy element abundances, implying that the nucleosynthetic source of $s$-process enrichment must not influence Mg isotope ratios. Instead, a key result of this work is that we identify correlations between $^{26}$Mg/$^{24}$Mg and some light elements. Using a custom suite of AGB nucleosynthesis yields tailored to the metallicity of M 22, we find that low mass ($\sim 1 \rm{-} 3 \  \rm{M}_{\odot}$) AGB stars are capable of reproducing the observed $s$-process abundances of M 22 and that the absence of any difference in Mg isotope ratios between the two $s$-process groups precludes AGBs with masses above $\sim3 \ \rm{M}_{\odot}$. This places tighter constraints on possible formation scenarios and suggests an age difference of at least $\sim280 \rm{-} 480 \ \rm{Myr}$ between the two populations which is independent of isochrone fitting. 

\end{abstract}

% Select between one and six entries from the list of approved keywords.
% Don't make up new ones.

%%%%%%%%%%%%%%%%%%%%%%%%%%%%%%%%%%%%%%%%%%%%%%%%%%

%%%%%%%%%%%%%%%%% BODY OF PAPER %%%%%%%%%%%%%%%%%%

\section{Introduction}

Galactic globular clusters (GCs) are among the oldest objects for which reliable ages can be obtained \citep[e.g.,][]{VandenBerg+2002, Salaris_Weiss2002, MarinFranch+2009, VandenBerg+2013, VandenBerg+2016} and thus provide a remarkable opportunity to study stellar nucleosynthesis, star formation and galactic assembly in the early Universe. For many decades, GCs have served as ideal laboratories for testing the predictions of stellar evolution theory, as they represent the closest approximation to simple stellar populations, i.e., single age, helium abundance, metallicity, and initial mass function \citep{Renzini_Buzzoni1986}. However, this vast over-simplification neglects the inescapable reality that GCs are composed of multiple stellar populations \citep[recent reviews on this topic include][] {Gratton+2012, Bastian_Lardo2018, Gratton+2019, Milone_Marino_2022}. Chemical abundance measurements and high-precision photometry of GCs have revealed several intriguing results that continue to challenge our knowledge of stellar nucleosynthesis.

Most clusters \citep[known as Type I GCs;][]{Milone+2017MSP} exhibit two main populations and are, to first order, homogeneous in elements heavier than iron.
Each population is characterised by star-to-star abundance variations in light elements, most notably O and Na (but also He, Li, C, N, F, Mg and/or Al) which have been detected in every well-studied cluster \citep[e.g.,][] {Kraft_1994, Carretta+2009GIRAFFE, Carretta+2009UVES, Piotto+2015, Milone+2017MSP, Meszaros+2020}.
In addition to these light element abundance variations, approximately $17\%$ of clusters exhibit a dispersion in iron peak elements and sometimes even in slow ($s$-) and rapid ($r$-) neutron capture process elements \citep[Type II GCs;][]{Milone+2017MSP, Milone_Marino_2022}. [Fe/H] abundances can range from as little as 0.1 dex \citep[e.g.,][]{Marino+2021, Monty+2023} to almost 2 dex \citep[e.g.,][]{Johnson_Pilachowski2010, Johnson+2020, Nitschai+2023}. These objects must have experienced a complex star formation history and have been hypothesised to be possible dwarf galaxy remnants \citep{DaCosta_2016}. The main examples of this phenomenon include $\omega$ Centauri \citep{Norris_DaCosta1995} and M 54 \citep{Carretta+2010_m54}. Although, a large portion of this population has no obvious dwarf progenitor (e.g., NGC 5286; \citealt{Marino+2015_5286}, Terzan 5; \citealt{Ferraro+2009, McKenzie_Bekki2018}, and M 2; \citealt{Yong+2014_m2}) suggesting that there must be multiple formation channels for creating chemically anomalous clusters.

One Type II cluster with evidence of heavy element abundance variations is M 22 \citep{Peterson_1980, Pilachowski+1982, Brown+1990, Lehnert+1991, Brown_Wallerstein1992, DaCosta+2009, Marino+2009, Lee+2009, Marino+2011, Roederer+2011, Alves-Brito+2012, Joo_Lee2013, Gratton+2014, Lim+2015, Lee_2016}. In our previous work on this cluster \citep[][hereafter Paper I]{Mckenzie2022}, high precision abundance measurements with uncertainties as low as $\sim 0.01 \ \rm{dex}$ allowed us to unambiguously demonstrate the presence of abundance spreads in $\alpha$, iron peak, and also $s$-process elements. These data confirmed that every element heavier than silicon exhibits a star-to-star abundance variation that can be used to divide the stars into two groups.

Based on the observed range in Paper I, iron abundance variations within the cluster must be $\geq 0.24 \ \rm dex$, while differences in the $s$-process element Yttrium were as large as $0.65 \ \rm{} dex$. Additionally, for every pair of elements, there are abundance correlations of high statistical significance. This striking bimodality between $s$-process elements gave rise to the nomenclature of $s$-process rich and $s$-process poor stars, which are equivalent to the iron-rich and iron-poor populations. Using photometric techniques, \citet{Lee_2020} determined that the $s$-process/iron-poor population slightly dominates the cluster, constituting 63\% of the population. Furthermore, \citet{Milone+2017MSP} found consistent results using Hubble Space Telescope data, with $40.3\pm 2.1 \%$ of the population belonging to the $s$-process rich population

The source of these anomalous abundances is currently unknown. In Paper I, we describe three main scenarios; (i) M 22 could be a nuclear remnant of an accreted galaxy \citep[e.g.,][]{DaCosta+2009}. (ii) Two globular clusters could have merged in a dwarf host, thus generating the bimodal pattern seen for some elements \citep[e.g.,][]{Lee_2020}. (iii) M 22 may have been born from clumpy substructure during the Milky Way's infancy, thus this cluster may be one of the original building blocks of our Galaxy. However, none of these scenarios can provide a comprehensive picture of M 22's formation. Hence further constraints are necessary to identify and preclude possible mechanisms that were instrumental in generating such anomalous abundances.

Almost a decade ago, two independent teams \citep{Straniero+2014, Shingles+2014} examined M 22 and reached the same conclusion: the $s$-process abundance differences between the two stellar groups can be attributed to pollution from asymptotic giant branch (AGB) stars with masses in the range $3 \rm{-} 6 \rm{M}_{\odot}$. AGB ejecta is commonly suggested as the source of CNO enhancement in Type I GCs \citep[i.e., the AGB scenario; e.g.,][]{Cottrell_DaCosta_1981, DErcole+2008, DAntona+2016, DErcole+2016}. In addition, AGB stars may also be the source of $s$-process element enhancement. Near the metallicity of M 22, AGB models in this mass range are predicted to produce significant quantities of the heavy magnesium isotopes \citep[e.g.,][]{Karakas2010, Ventura+2018}.
Mg has three stable isotopes, $^{24}$Mg, $^{25}$ Mg and $^{26}$Mg, which have a terrestrial benchmark ratio of $^{24}$Mg : $^{25}$ Mg : $^{26}$Mg = 78.99 : 10.00 : 11.01 \citep{DeBievre_Barnes1985}.  The stellar yields from AGB stars, and the Mg isotopes in particular, are dependent upon the choice of input physics \citep[e.g.,][]{Ventura+2011}. Models suggest that while the Mg isotope ratios predicted do not always vary monotonically with mass, generally the amounts of $^{25}$Mg and $^{26}$Mg in the ejecta increase with increasing stellar mass \citep{Karakas_Lattanzio2003, Ventura+2009, Fishlock+2014, Doherty+2014}.

The goal of this study is to measure the isotopic ratios of Mg in the two $s$-process groups in M 22 to place independent constraints upon nucleosynthesis in AGB stars at low metallicity. This will (i) test the hypothesis that $3 \rm{-} 6 \rm{M}_{\odot}$ AGB stars are responsible for the $s$-process abundance differences and (ii) place a more precise limit upon the mass range of the AGB stars. While the metallicity spread in M 22 requires a contribution from supernovae, our study focuses on the crucial role of AGBs and represents an essential step towards a complete understanding of the chemical evolution of this stellar system, and perhaps other similar objects. Mg isotopic ratios have yielded fascinating results in other clusters such as NGC 6752 \citep{Yong+2003}, $\omega$ Centauri \citep{DaCosta+2013}, 47 Tucanae \citep{Thygesen+2016}, M13 and M71 \citep{Shetrone1996, Yong+2006, Melendez_Cohen2009} as well as open clusters \citep{Yong+2004} and dwarf halo stars \citep{Yong+2003_coolstars, Melendez_Cohen2007, Carlos+2018}. Stars within M 22 are more metal-poor than any globular cluster for which Mg isotope ratios have been previously measured, therefore these data will provide important new observational insights into light element abundance variations at the isotopic level at low metallicity.

This work calculates Mg isotope ratios between the two $s$-process groups to isolate the mass of the AGB stars responsible for this difference. We then use these results to infer an age difference between the populations which is independent of isochrone fitting but is dependent on the choice of AGB model adopted for the analysis.

In Section \ref{sec:Obs_analysis} we discuss our sample of stars and our approach for calculating isotopic abundances, in Section \ref{sec:results} we illustrate the relationship between $s$-process enhancement and Mg isotopes, and discuss these results in Section \ref{sec:discussion}. Our conclusions are presented in Section \ref{sec:conclusion}.

\section{Observations and Analysis}
\label{sec:Obs_analysis}

\subsection{Target selection and observations}
\label{sec:targets} 

\begin{figure*}
    \includegraphics[width=\textwidth]{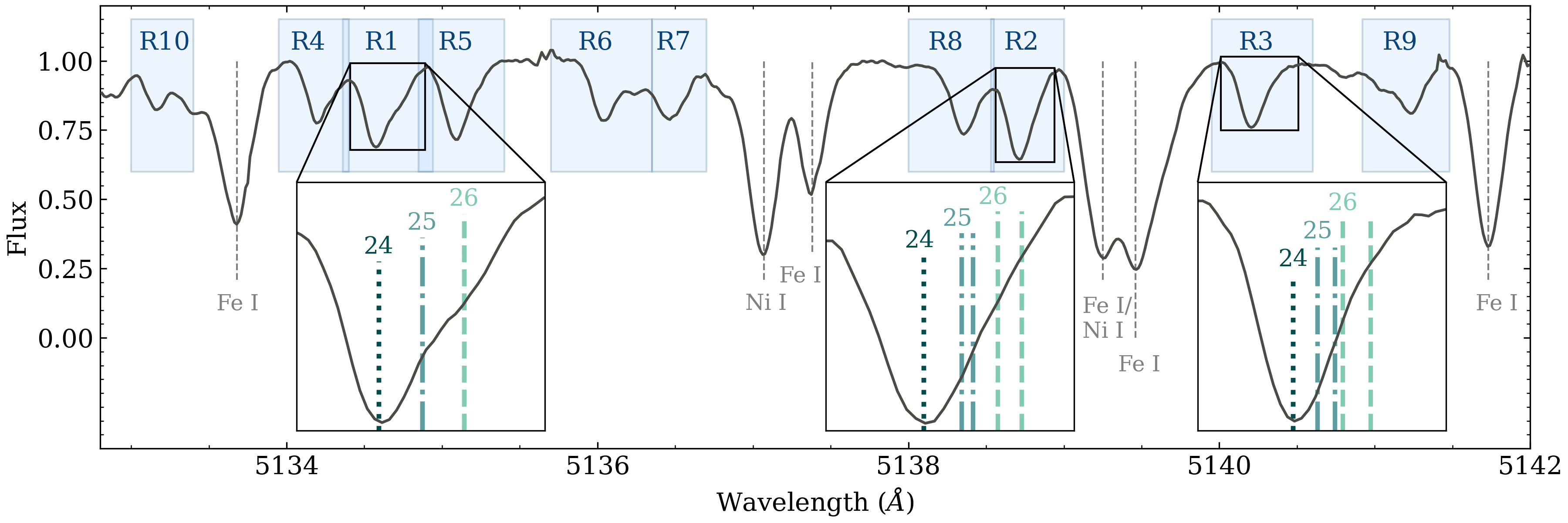}
    \caption{The MgH line region between 5133 \AA{} to 5142 \AA{} for star C with each region highlighted in blue. R1, R2 and R3 (the three regions used in previous studies) are given in the inset plots. The location of the $^{24}$Mg, $^{25}$Mg and $^{26}$Mg isotopes from the G\&L line list are shown in the inset plots in dark, medium and light green respectively.  Prominent lines in the spectra have been labelled in grey. We do not see the C$_2$ line feature in our spectra at 5135.5 \AA{} as discussed in \protect\cite{McWilliam_Lambert1988}.}
    \label{fig:line_region}
\end{figure*}

Stellar parameters were derived using a line-by-line differential analysis \citep[see, e.g.,][for an overview of this approach]{Nissen_Gustafsson2018}. Paper I provides a comprehensive discussion of our target stars, and we briefly describe them here for completeness. Our program stars were sourced from \citet{Marino+2011} and each star was visually examined to identify those with detectable MgH molecular lines. These lines were visible for six stars sitting at the tip of the red giant branch (RGB), three in each $s$-process group. Measuring Mg isotopic abundances requires both high resolution and signal-to-noise ratio (S/N) spectra, thus these stars were re-observed with UVES \citep{Dekker+2000} on the ESO VLT UT2 telescope. The observations were taken using image slicer \#3 and the 0\farcs3 slit. The spectra for each star have a resolution of $R = 110,000$ and S/N $\geqslant 300$ per pixel near the MgH $5140$ \AA\ lines. We used the 580 nm setting which provided wavelength coverage from approximately $4800$ \AA\ to $6800$ \AA\ with a small gap near $5800$ \AA\ due to the space between the two CCDs in the UVES camera. No detected neighbours were present within the entrance aperture (1.5×2.0 arcsec) of the image slicer which minimizes contamination. Exposure times for each star ranged from 1.5 to 2.1h. The spectra were reduced using the ESO pipeline and initial values for the radial velocities were estimated using IRAF.

\subsection{Stellar parameters and chemical abundances}
\label{sec:stellar_params}
Paper I provides a comprehensive description of our differential abundance analysis and abundance dispersions between the two $s$-process populations. We also discuss our determination of stellar parameters and compare them to previous values in the literature. In this work, O, Na, Mg, Al, Si and Ca abundances from \cite{Marino+2011} are used in conjunction with our differential abundances calculated in Paper I. Table \ref{tab:stelar_params} provides the stellar parameters and a subset of our differential abundances. We only list elements that were well measured by several lines and represent different nucleosynthetic channels. The errors in these abundances are included and discussed at length in Paper I. 
We also use the stellar spectra of Arcturus and NGC6752-mg9 from \cite{Yong+2003} to validate our method for determining Mg isotopes. For our Arcturus model atmosphere, we use the stellar parameters; \teff$ = 4300 \ \rm{K}$, \logg$ = 1.5 \ \rm{cm \ s^{-2}}$, $\rm{[Fe/H]} = -0.5$ and $\xi = 2$ \ \kms\ 
 . For NGC6752-mg9 \citep{Yong+2013}, we use the parameters; \teff$ = 4288 \ \rm{K}$, \logg$ = 0.91 \ \rm{cm \ s^{-2}}$, $\rm{[Fe/H]} = -1.66$, $\xi = 1.72$ \ \kms\ .

\begin{table*}
\centering
    \caption{Stellar parameters and a subset of differential chemical abundances from Paper I. These stellar parameters and chemical abundances are determined with respect to our reference star NGC6752-mg9. Our differential notation $\Delta^{\rm{X}}$ is analogous to square bracket notation [X/H], but with respect to NGC6752-mg9. We also include whether a star is a member of the \textit{s}-process rich group. We refer the reader to Paper I for the quantification and description of the errors.}
    \label{tab:stelar_params}
\begin{tabular}{lccccccccccc}
\hline
ID &
  \teff \ (K) &
  \logg \ (cm s$^{-2}$)&
  {[}Fe/H{]} &
  \textit{s}-process rich? &
  $\Delta^{\rm{FeI}}$ &
  $\Delta^{\rm{SiI}}$ &
  $\Delta^{\rm{CaI}}$ &
  $\Delta^{\rm{YII}}$ &
  $\Delta^{\rm{LaII}}$ &
  $\Delta^{\rm{NdII}}$ &
  $\Delta^{\rm{EuII}}$ \\ \hline
C      & 3912 & 0.105 & -1.696 & \checkmark & -0.033 & 0.118  & 0.044  & 0.251  & 0.092  & 0.294  & -0.100 \\
III-3  & 4041 & 0.250 & -1.778 & \checkmark & -0.106 & 0.077  & -0.121 & 0.257  & 0.060  & 0.137  & -0.080 \\
III-14 & 4038 & 0.120 & -1.87  & $\times$                  & -0.183 & -0.145 & -0.262 & -0.326 & -0.347 & -0.131 & -0.150 \\
III-15 & 4136 & 0.450 & -1.825 & \checkmark & -0.147 & -0.144 & -0.190 & -0.232 & -0.255 & -0.116 & -0.070 \\
III-52 & 4100 & 0.510 & -1.707 & $\times$                  & -0.049 & 0.037  & 0.042  & 0.211  & 0.150  & 0.195  & -0.080 \\
IV-102 & 4043 & 0.100 & -1.973 & $\times$                  & -0.268 & -0.223 & -0.343 & -0.420 & -0.505 & -0.308 & -0.300 \\ \hline
\end{tabular}
\end{table*}

\subsection{MgH line selection and line list}
\label{sec:MgH_and_ll}
Traditionally, three molecular MgH features are used for the derivation of magnesium isotopes which we label as R1, R2 and R3\footnote{We note that R is short for "region" and should not be confused with the R branch when dealing with rovibrational spectra.} (see Table \ref{tab:Mg_lines} and Fig. \ref{fig:line_region}). The lines appear asymmetric due to the trailing red wing as a result of the neutron-rich isotopologues of $^{25}$MgH and $^{26}$MgH. Each of these Mg features suffers from blends with both atomic lines and molecular lines of C$_2$, CN and CH. However, \citet{McWilliam_Lambert1988} determined that these three features are the least impacted by blends compared to other MgH transitions in the vicinity.

To determine the Mg isotopic ratios, we use the line list created in \cite{Gay_Lambert2000} by calculating the wavelength shift of molecular isotopologues. The effectiveness of this line list has been demonstrated in several previous studies (e.g., \citealt{Yong+2003}; \citealt{DaCosta+2013}), however, we note that this line list has been tuned to fit the \textit{dwarf} star Gmb 1830 analysed in \cite{Gay_Lambert2000}. As the stars in our study are all on the tip of the red giant branch, we test this line list (which we call the `G\&L' line list) against an additional new line list from the program \textsc{linemake}\footnote{\url{https://github.com/vmplacco/linemake}} \citep{Placco+2021} (our `linemake' line list). We select the option to include hydride molecules, CN and C$_{2}$ for this new list. The MgH transitions from this list originate from \cite{Hinkle+2013}, and have been previously used in studies such as \cite{Thygesen+2016}. We apply the MOOG \citep{Sneden_1973} routine \texttt{weedout} to remove very weak lines from this line list. A model atmosphere based on the Arcturus stellar parameters was used and we set the minimum line/continuum opacity to be 0.01. This reduces the number of lines in the original line list from linemake from 8617 to 2051 lines over the region from 5100 to 5150 \AA{}. When generating our line list, we tested the option to include another molecule found in this region, TiO. However, all TiO lines were too weak and were removed by this \texttt{weedout} step. When cropped to the same wavelength region (from 5130 to 5142 \AA{}), this new linemake line list contains twice the number of lines in the G\&L line list.

As well as these three conventional MgH regions, we explore additional lines that would hopefully provide a larger sample size of MgH regions and thus Mg isotopic fractions. As our stars have lower Fe abundances compared to stars used in previous studies, the influence of blended features will be reduced, yielding a larger sample of MgH features to measure. Furthermore, our linemake line list spans a larger wavelength range, potentially including even more lines that can be used to build a more statistically significant sample of measurements within each of our stars. These additional lines are described in Table \ref{tab:Mg_lines}.

\begin{table}
\centering
    \caption{Locations of the MgH line regions used in the present work to derive Mg isotope ratios. `Standard line' refers to MgH lines that are most commonly used in the literature and were introduced in \protect\cite{McWilliam_Lambert1988}. Citations refer to specific studies that have previously used this line, and `New line' denotes lines that have been used only in this work. Wavelengths are all given in \AA.}
    \label{tab:Mg_lines}
\begin{tabular}{lccl} \hline
R   & $\lambda$ $^{24}$MgH & Identification     & Comments     \\ \hline 
1 & 5134.6                     & 0-0Q$_1$(23) \& 0-0R$_2$(11) & Standard line \\
2 & 5138.7                     & 0-0Q$_1$(22) \& 1-1Q$_2$(14) & Standard line \\
3 & 5140.2                     & 0-0R$_1$(10) \& 1-1R$_2$(14) & Standard line \\ 
4 & 5134.2                     & 0-0Q$_2$(23)                  & \cite{Melendez_Cohen2009}\\
5 & 5135.1                     & 0-0R$_1$(11)                  & \cite{Thygesen+2016}\\
6 & 5136.1                     & 1-1Q$_2$(15) \& 1-1R$_2$(5)  & New line      \\
7 & 5136.4                     & 1-1Q$_1$(15)                  & New line      \\ 
8 & 5138.4                     & 0-0Q$_2$(22)                  & New line      \\
9 & 5141.0                     & 1-1R$_1$(4) \& 1-1Q$_2$(13)  & New line      \\
10& 5133.2                     & 1-1Q$_2$(16)                  & New line      \\ \hline
\end{tabular}
\end{table}

\subsection{RATIO}
\label{sec:RATIO}

To calculate the isotopic ratios, we developed, tested and applied our own wrapper for \textsc{moog} \citep{Sneden+1997} which we call \textsc{ratio}\footnote{\url{https://github.com/madeleine-mckenzie/RAtIO}} (Rapid, AuTomatic Isotope Optimisation). We use the \textsc{moog17scat}\footnote{\url{https://github.com/alexji/moog17scat}} implementation of \textsc{moogsilent} from Alex Ji which includes a proper treatment of scattering from \cite{Sobeck+2011}\footnote{We use this implementation of \textsc{moog} over more recent versions as this is the only version which successfully installed and ran on both a laptop and the Mt Stromlo AVATAR compute clusters.}. We also use the \textsc{moog} wrapper pyMOOGi \citep{Adamow_2017} to estimate initial values for testing purposes and visualisation. We used one-dimensional plane-parallel local thermodynamic equilibrium (LTE) model atmospheres from the grid of \cite{Castelli_Kurucz2003}. The six parameters which we optimise for include the total amount of Mg (log$\epsilon$(Mg)), broadening from macroturbulence (which we call the broadening, or S, for simplicity), the isotopic ratios $^{25}$Mg/$^{24}$Mg and $^{26}$Mg/$^{24}$Mg, the placement of the continuum and the radial velocity correction (see Table \ref{tab:MCMC_params}). As we are using spectra with $\rm{R} = 110,000$ at 5140 \AA{}, we model the instrumental profile using a Gaussian function where the FWHM $= 5140/110,000 = 0.047 $ \AA{}. We also include a macroturbulence parameter which for our giant stars is around $\sim$7-8 \kms.

\subsubsection{Implementation}

\begin{table}
    \centering
    \caption{Parameters used by \textsc{ratio} in the MCMC fitting routine. $\mathcal{U}$(\textit{a,b}) represents a uniform prior between values \textit{a} and \textit{b}. $\mathcal{N}$($\mu$,$\sigma$) represents a Gaussian prior with mean $\mu$ and standard deviation $\sigma$. }
    \label{tab:MCMC_params}
    \begin{tabular}{llll}
    \hline
    Parameters                 & Abbreviation & Priors               & Units      \\
    \hline
    log$\epsilon$(Mg)          & Mg           & $\mathcal{U}$(-1,1)  & -          \\
    Macroturbulent broadening  & S            & $\mathcal{U}$(0,10)  & kms$^{-1}$ \\
    $^{25}$Mg/$^{24}$Mg        & $\frac{25}{24}$ (or 25/24)& $\mathcal{U}$(0,2)   & -          \\
    $^{26}$Mg/$^{24}$Mg        & $\frac{26}{24}$ (or 26/24)& $\mathcal{U}$(0,2)   & -          \\
    Continuum correction       & C            & $\mathcal{N}$(0,0.2) & -          \\
    Radial velocity correction & Rv           & $\mathcal{N}$(0,0.5) & kms$^{-1}$ \\
    \hline
    \end{tabular}
\end{table}

\begin{table*}
\centering
    \caption{The isotopic ratios from synthetic spectra generated by MOOG and analysed using our isotopic analysis code \textsc{ratio}. The top four rows are the results using the G\&L line list and the bottom four rows are using our linemake line list which uses updated MgH transitions (see Sec. \ref{sec:MgH_and_ll}). Both line lists yield very similar results. The  `weighted mean' and `Posteriors' columns refer to different analysis techniques. Both again give similar results, however, the `Posteriors' approach gives larger (and more realistic) errors. `\# of lines' refers to the number of lines that were used to obtain each ratio.}
    \label{tab:synth_summary}
\begin{tabular}{lccccc}
\hline
 & True values & Weighted mean & \# of lines & Posteriors & \# of lines \\ \hline
&\multicolumn{5}{c}{\textbf{G \& L}} \\ \hline
Synth C 1 & 68 : 20 : 12 & 67 ($\pm0.4$) : 21 ($\pm0.2$) : 12  ($\pm0.2$) & 8 & 67 ($\pm3.7$) : 21 ($\pm2.6$) : 12 ($\pm1.1$) & 6 \\
Synth C 2 & 80 : 10 : 10 & 79 ($\pm0.4$) : 11 ($\pm0.2$) : 10 ($\pm0.2$) & 8 & 79 ($\pm3.5$) : 11 ($\pm2.6$) : 10 ($\pm0.9$) & 8 \\
Synth C 3 & 94 : 2 : 4 & 95 ($\pm1.4$) : 2 ($\pm0.6$) : 3 ($\pm0.8$) & 7 & 94 ($\pm3.3$) : 3 ($\pm2.4$) : 3 ($\pm1.0$) & 8 \\
Synth IV-102 & 71 : 20 : 9 & 70 ($\pm1.1$) : 23 ($\pm0.2$) : 7 ($\pm0.9$) & 8 & 70 ($\pm7.5$) : 22 ($\pm4.8$) : 8($\pm2.7$) & 8 \\ \hline
&\multicolumn{5}{c}{\textbf{linemake}} \\
\hline
Synth C 1 & 68 : 20 : 12 & 66 ($\pm0.6$) : 22 ($\pm0.4$) : 12 ($\pm0.2$) & 10 & 66 ($\pm3.3$) : 22 ($\pm2.2$) : 12 ($\pm1.0$) & 10 \\
Synth C 2 & 80 : 10 : 10 & 79 ($\pm0.6$) : 11 ($\pm0.4$) : 10($\pm0.2$) & 10 & 79 ($\pm3.2$) : 11 ($\pm2.4$) : 10 ($\pm0.9$) & 10 \\
Synth C 3 & 94 : 2 : 4 & 95 ($\pm1.0$) : 2 ($\pm0.6$) : 3 ($\pm0.4$) & 10 & 93 ($\pm2.8$) : 4 ($\pm2.0$) : 3 ($\pm1.0$) & 10 \\
\hline
\end{tabular}
\end{table*}

Previous studies determining isotopic ratios either fit the line profile by eye or implemented a grid-based search over the total Mg abundance, $^{25}$Mg/$^{24}$Mg and $^{26}$Mg/$^{24}$Mg based on some initial guess. We improve upon these previous approaches and present extensive testing of a range of approaches and methodologies in Appendix \ref{app:alternative_methods}. The method adopted for this study utilises the Python package \texttt{emcee} \citep{Foreman-Mackey+2013} which implements the Goodman \& Weare's Affine Invariant Markov Chain Monte Carlo (MCMC) Ensemble sampler. The model generates the synthetic spectra from \textsc{moogsilent} and our log-likelihood function is based on the reduced $\chi^2$ value. We summarise our parameters and the priors we use for each in Table \ref{tab:MCMC_params}. The total Mg value, broadening and isotopic ratios all use uniform priors whereas the continuum and radial velocity correction use Gaussian priors as we assume we have successfully pre-processed the spectra to be approximately the optimum value.

We use 600 walkers for 1800 steps to generate our posterior distribution. We then use the python package \texttt{chainconsumer} \citep{Hinton2016} to analyse the output from \texttt{emcee} and generate summary statistics from marginalised posterior distributions. We confirm that our chains pass the Gelman Rubin diagnostic criteria. In many cases, we find that the resulting distributions are not Gaussian and thus we use a maximum posterior point rather than the mean value for our parameters. We use the `Max Shortest' statistic (i.e., see panel two of fig. 6 from \citealt{Andrae2010}) for our 68.3\% confidence intervals as this gave consistent results in the case of bimodal distributions. This method consistently gives good fits to the spectra (based on a `by eye' approach), however in some cases, it has a tendency to underestimate the depth of the line which we suspect is due to NLTE effects (e.g. \citealt{Mashonkina+2007}). The computational time and accuracy of the fits to the isotopic features make \texttt{emcee} the preferred choice in comparison to the other methods discussed in Appendix \ref{app:alternative_methods} for fitting Mg isotope ratios. 

Some of our MgH fitting regions contain blends with other elements, the most common being \ion{Fe}{I} and $\rm{C}_2$. The linemake line list includes updated laboratory values from \cite{Ram+2014}, however, none of our stars showed significant $\rm{C}_2$ absorption at 5135.6 \AA{}. To ensure that these blends do not have a significant contribution to our isotopic ratios, we test to see if including the total Fe and C abundances as additional parameters to the \texttt{emcee} fitting routine changes the isotopic ratio. The resulting posterior distributions for both parameters are predominantly uniform and do not significantly change the final isotopic abundances within their error margins. Therefore, Mg is the only element that we include in our fitting routine.

Stars with lower \teff\ or \logg\ generally have larger macroturbulence \citep{Carney+2008}, making RGB tip stars some of the most challenging to analyse for their isotopic ratios. 
Previous studies have noted the difficulty in analysing this type of star (compared to dwarf stars; e.g., \citealt{Yong+2004}) and the degeneracies of this problem, but the extent of quantifying the isotopic errors has been based on the $1\sigma$ confidence limit as defined in \cite{Bevington_Robinson1992}; $\Delta\chi^2 = \chi^2 - \chi^2_{\rm{min}} = 1$ while holding all other parameters constant. 
However, through testing our previous methods discussed in Appendix \ref{app:alternative_methods}, it was clear that the $1\sigma$ confidence limit was an underestimate of the true distribution of the system. For example, in \citet{Yong+2003}, they state that `the formal statistical errors are dwarfed by the systematic uncertainties. We conservatively estimate errors in $^{24}$Mg:$^{25}$Mg:$^{26}$Mg = (100 - b - c):b:c as b $\pm$ 5 and c $\pm$ 5'. Using MCMC, we can, for the first time, place \textit{realistic} errors on not only the isotopic ratios but all other optimised parameters.

\subsubsection{Methodology}

% Big corner plot figure
\begin{figure*}
    \includegraphics[width=0.9\textwidth]{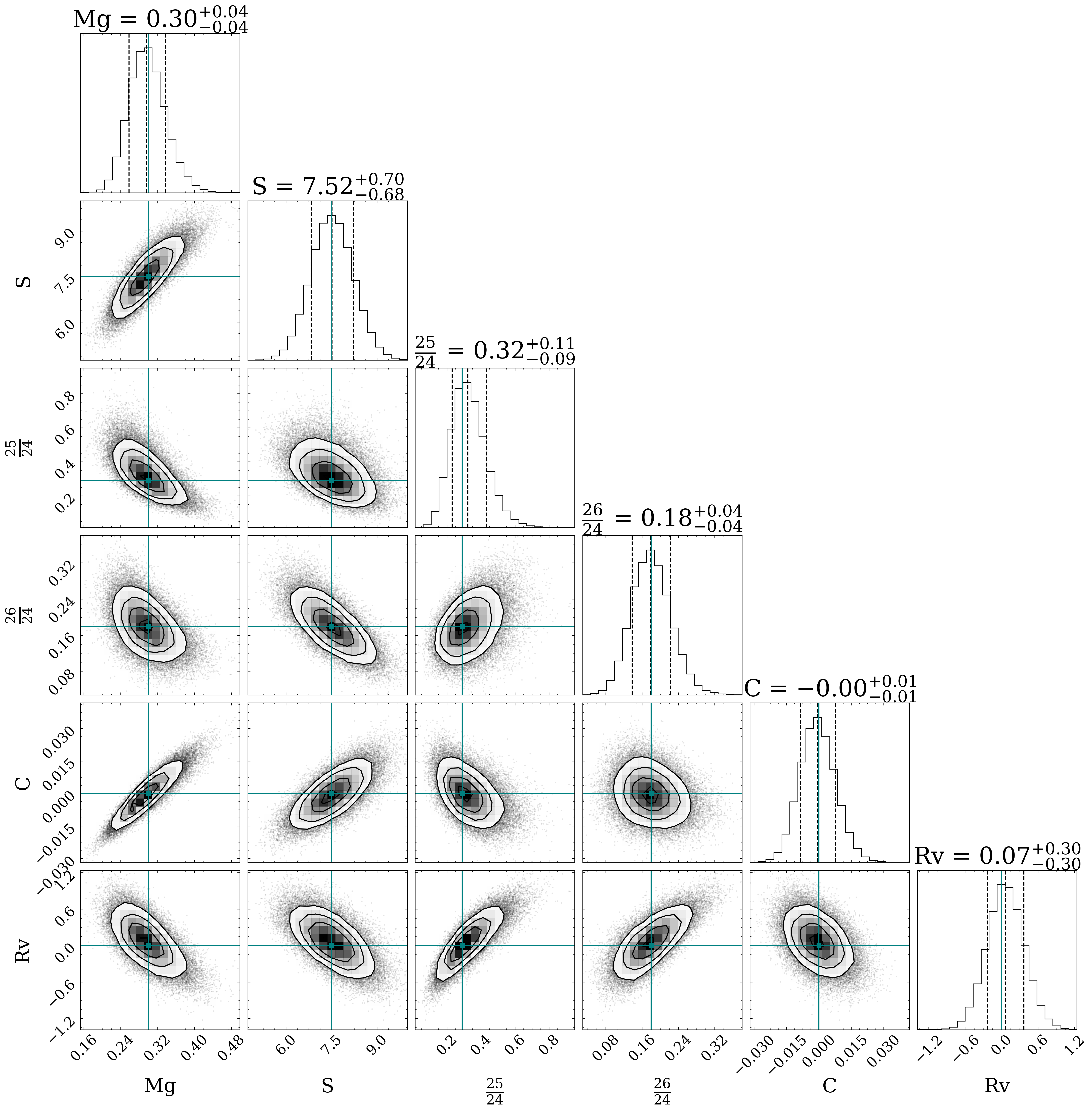}
    \caption{A corner plot made using the python package \texttt{corner} for the synthetic spectra `Synth C 1' (see Table 
    \ref{tab:synth_summary}) with an isotopic ratio 68:20:12 analysed using the linemake line list. The true values used to generate the synthetic spectra for each of the parameters are shown in green and the dashed lines are the 16th, 50th and 84th percentiles. As described in Table \ref{tab:MCMC_params}, Mg represents the total Mg abundance, S refers to the macroturbulent broadening, the two Mg isotope ratios $^{25}$Mg/$^{24}$Mg and $^{26}$Mg/$^{24}$Mg are simplified to $\frac{25}{24}$ and $\frac{26}{24}$, C represents the correction to the continuum and Rv is the radial velocity correction.}
    \label{fig:corner}
\end{figure*}

To assess the performance of the MCMC implementation of \textsc{ratio}, we compare our results to a set of synthetic spectra generated by \textsc{moog} with similar absorption features to our target stars, the stellar benchmark star Arcturus and our reference star from Paper I; NGC6752-mg9. In contrast to previous methods for calculating isotopic ratios, we have access to more than triple the number of previously used lines. Therefore we have the luxury of including only well-fitted lines for our final ratio. Additionally, our Bayesian approach means that, for the first time, we have access to the posterior distributions for all our parameters.

We choose to accept or reject the synthetic model of a line based on a set of criteria. Firstly, the best model returned by the maximum posterior point must be a good fit to the data (as judged by eye). Secondly, the posterior distribution for all parameters must resemble a Gaussian function with a single optimum value. Finally, we remove any isotopic ratios that are clear outliers compared to the rest of the regions due to unknown blends. We avoid using the final isotopic ratio as a reference for whether we keep or remove a line to help alleviate any unconscious biases (for example, trying to find consistent results between regions). 

\subsubsection{Synthetic spectra}
In Table \ref{tab:synth_summary} we introduce two methods for finding the overall ratios in each of our stars; our `weighted mean' and `posteriors' methods. Previous approaches in the literature weight the mean value of the isotopes based on the $\chi^2_{min} +1$ error and do not report any formal uncertainties onto the final isotopic ratio. For our `weighted mean' method, we use a similar approach to the literature of averaging the best value from each region, but instead of weighting by the $\chi^2_{min} +1$ error, we use our 68.3\% confidence interval (i.e.,
isotopic measurements with a large confidence interval will have less of an effect on the final isotopic ratio compared to lines with smaller confidence intervals). We give the final error on our ratios to be the standard error of the mean. For our `posteriors' method, we take the product of the kernel density estimates of the sampled posteriors normalised by the area. For the error, we use the maximum posterior point of the resulting distribution as our final ratio, and the larger of the distance between the median, and the $16^{\rm{th}}$ and $84^{\rm{th}}$ percentiles. We take the larger of the two values to better represent the systematics involved in our measurements.

To quantify the differences and select a preferred approach for reporting the errors, we generate a set of spectra using the model atmosphere based on our most metal-rich star, C (at $\rm{[Fe/H]} = -1.7$), and our most metal-poor star, IV-102 (at $\rm{[Fe/H] =} -2.0$). Despite C being our most metal-rich star, we remind the reader that there has only been \textit{one} more metal-poor globular cluster (or metal complex/nuclear star cluster) star that has ever been successfully analysed with isotopic analysis ($\omega$ Cen ROA 94 with $\rm{[Fe/H]} = -1.78$). IV-102 represents the most difficult case for isotopic analysis of an RGB star documented in the literature.

Each synthetic spectrum was generated using the two different line lists and sampled to mimic the resolution of our M 22 targets. We then analyse these spectra using the methods described above to test how well \textsc{ratio} can recover the original isotopic ratio used to create the data. The isotope ratios used to generate Synth C 1 were based on one of the most $^{24}$Mg poor stars in the literature from \cite{Yong+2004}. Synth C 2 was based on the isotopic ratio of Arcturus (see below), and Synth C 3 represents an almost purely $^{24}$Mg star. Synth IV-102 is our most challenging case with a MgH line depth of only 0.1 relative to the continuum, making it very difficult to disentangle the relative effects of the total Mg, broadening, and isotopic ratio. Each synthesised star is analysed with the same line list they were made with, so theoretically should be able to perfectly reproduce the values they were created with. Deviations from the true value represent the systematic uncertainty associated with our analysis. We present the weighted average mean of the measured lines in Table \ref{tab:synth_summary} and find that the final isotopic ratio across all synthetic spectra differs by at most 2\% for the Synth C spectra. In the more challenging case of Synth IV-102, \textsc{ratio} still performs remarkably well with a difference of at most 3\%, however, the difficulty of these measurements is reflected by the larger errors, especially for the case of the Posteriors method.

Fig. \ref{fig:corner} gives the corner plot made using the Python package \texttt{corner} \citep{corner} for Synth C 1 using the linemake line list. The code almost perfectly recovers the original values used to synthesise the spectra of Mg = 0.3, S = 7.52, 25/24 = 0.29, 26/24 = 0.18, C = 0, Rv = 0. This corner plot illustrates the degeneracies between the parameters, with every parameter correlating with each other, and the challenge of measuring isotopic ratios via asymmetries of intrinsically weak spectral lines.

\subsubsection{Arcturus}
The synthesised spectra represent the easiest isotopic analysis test case. Running \textsc{ratio} on real spectra that contain both known and unknown blends adds additional layers of complexity. Arcturus spectra from the NOAO Arcturus Spectral Atlas \citep{Hinkle+2000} have been measured by \cite{Hinkle+2013} to have an isotopic ratio of $80:10:10$ with probable uncertainties of $\sim \pm2$ in the minor isotope percentages. \cite{Thygesen+2016} also gives two estimates of $82.0:9.8:8.2$ and $83.6:9.2:7.2$. For the G\&L line list using lines R1, 3, 4, 5 and 8, we recover a final ratio of $83(\pm 5.0):9(\pm4.4):8(\pm0.6)$ for the weighted mean method, and $81(\pm 2.6):10(\pm1.9):9(\pm0.8)$ for the posterior method. For the linemake line list and only excluding lines R2 and R7, the weighted mean method gave $83 (\pm 3.7):7 (\pm1.8):10 (\pm1.9)$ and the posterior method gave $82 (\pm 2.2):8 (\pm1.4):10 (\pm0.8)$. The differences in errors between measurement approaches arise as a result of the weighted mean method being more sensitive to outlier results.

\subsubsection{NGC6752-mg9}
Finally, we analyse the spectra of NGC6751-mg9, our reference star in Paper I, which was measured by 
\cite{Yong+2003} to have an isotopic ratio of $72:17:10$. For our analysis, the G\&L line list using the weighted mean method is identical to that used by \cite{Yong+2003}, differing only by the method used to identify the optimum ratio and the lines used. We obtain the ratio of $74(\pm3.0):12(\pm2.7):14(\pm0.4)$ using the lines which passed our diagnostic criteria; R1, 4 and 8. We find similar results using the posterior method; $72(\pm7.3):13(\pm4.6):15(\pm2.6)$. Using the linemake line list, we find larger discrepancies between the previously published result, with the weighted mean method returning a value of $79(\pm5.7):7(\pm3.1):14(\pm2.6)$ and the posteriors method giving $76(\pm8.1):10(\pm5.0):14(\pm3.1)$. Given the quoted errors these values agree with previous results in the literature.

\subsection{AGB models}
\label{sec:AGB_models}
The thermally-pulsing phase of low and intermediate-mass stars is a rich source of stellar nucleosynthesis (reviews include \citealt{Iben_Renzini1983}; \citealt{Meyer1994}; \citealt{Busso+1999}; \citealt{Karakas_Lattanzio_2014}). Elements heavier than iron can be synthesised via the $s$-process in the He-burning shell and mixed to the surface via repeated third dredge-up events. Observations and models suggest that low-mass stars ($M \approx 1-3 M_{\odot})$ are the dominant source of $s$-process elements such as Y, Ba, La and Pb in galaxies but their long lifetimes preclude their contribution to GCs, which likely formed on timescales considerably shorter than the $\sim $~Gyr timescale required for even a $2 \ \rm{M}_{\odot}$ AGB low-metallicity star. Intermediate-mass stars over about 3~$M_{\odot}$ are also predicted to synthesise $s$-process elements on shorter timescales, where a 3.5$M_{\odot}$ star with [Fe/H] $\approx -2.2$ will evolve to become a white dwarf in under 200~Myr \citep[e.g.,][]{Karakas2010}. 

However with increasing stellar mass, the amount of $s$-process enrichment at the surface of AGB models is predicted to decrease, owing to an increasingly massive envelope, leading to large amounts of dilution, and also a much smaller He-intershell region \citep[e.g.,][]{Straniero+2014, Karakas_Lugaro2016}. Furthermore, there is discussion in the literature about the amount of third dredge-up experienced by the most massive AGB stars (including super-AGB stars) with some models predicting little to none \citep{Karakas2010, Ventura+2011, Straniero+2014, Marigo2022}. Little to no mixing equates to no production of heavy elements. 

Intermediate-mass AGB stars also experience hot bottom burning (HBB), which is when the base of the convective envelope becomes hot enough to sustain proton-capture nucleosynthesis. The peak temperature at the base of the envelope increases with increasing mass, up to 100 million K (100 MK) or higher in AGB stars near the carbon-burning threshold \citep{Ventura+2011, Doherty+2014}. AGB models near the minimum mass threshold for HBB, which is mass dependent but around $3.5 \ \rm{M}_{\odot}$ for $\rm{[Fe/H] =} -1.8$, likely experience enough third dredge-up to make $s$-process elements. However, the most massive AGB stars with extreme HBB are not theorised to be efficient producers of heavy elements. 

The neutron-rich Mg isotopes, $^{25}$Mg and $^{26}$Mg, can be produced during convective He-shell burning in intermediate-mass AGB stars by $\alpha$ capture onto $^{22}$Ne in almost equal abundances by the reactions $^{22}\rm{Ne}(\alpha, n)^{25}\rm{Mg}$ and $^{22}\rm{Ne}(\alpha, \gamma)^{26}\rm{Mg}$. Both reactions are highly dependent on temperature during thermal pulses, requiring $T > 300$~MK, which is only found in the most massive AGB models. Similar to $s$-process elements, third dredge-up is required to mix the $^{25}$Mg and $^{26}$Mg to the surface. However the Mg isotopes can also be altered in the envelopes of intermediate-mass AGB stars by HBB, if temperatures are high enough for activation of the Mg-Al reactions \citep[see discussions in][]{Karakas_Lattanzio2003, Ventura+2011, Ventura+2018}. Here the dominant isotope $^{24}$Mg can be destroyed to produce the neutron-rich isotopes, $^{25}$Mg and $^{26}$Mg, along with Al (e.g. \citealt{Karakas_Lattanzio2003}; \citealt{Prantzos_Charbonnel_Iliadis2007}; \citealt{Karakas2010}; \citealt{Ventura+2011}; \citealt{Ventura+2018}).

The AGB models we use here span an initial mass range from $0.9 \rm{-} 6.5 \  \rm{M}_{\odot}$ and have an initial $Z = 2.2 \times 10^{-4}$ ([Fe/H] = $-1.82$). The input physics used in the calculations and the methodology is the same as described in \citet{Karakas+2018},  with the exception that we calculated a scaled-solar and an $\alpha$-enhanced set of models for each mass. The $\alpha$-enhanced models include radiative and low-temperature opacity tables that match the initial composition, chosen to reflect the composition of Galactic halo dwarf stars, with $\rm[\alpha/Fe] \approx 0.30$ \citep[abundances taken from][]{Reggiani+2017}.
The models do not include mixing by rotation nor any non-standard mixing processes.

For models below about 3 M$_{\odot}$ we adopted the mass loss rate from \cite{Vassiliadis_Wood1993}, which is a good match for carbon stars. For models above about 3 M$_{\odot}$, we adopted \cite{Bloecker+1995} mass-loss, with $\eta_{\rm B} = 0.02$. At the transition mass we calculated models with both mass-loss prescriptions. The production of $s$-process elements in low-mass AGB stars requires a $^{13}$C pocket, in order for the $^{13}\rm{C}(\alpha, n)^{16}\rm{O}$ reaction to be efficiently activated \citep{Gallino+1998}. In our models we include a {\em partially mixed zone} (PMZ) where protons are ingested into the top of the He-intershell region after thermal pulses. This is done in models up to $4 \ \rm{M}_{\odot}$, with the mass of the PMZ decreasing as a function of increasing mass (see description in \citealt{Karakas_Lugaro2016}). We used:
\begin{itemize}
    \item M $\leq$ 2.5: PMZ = 2$\times10^{-3}\ \rm{M}_{\odot}$,
    \item M > 4.0: no PMZ; neutrons entirely from $^{22}\rm{Ne}(\alpha, \rm{n})^{25}\rm{Mg}$ reaction,
    \item 3.0 < M < 4.0: a transition region where we used a smaller pocket of PMZ = $1\times10^{-4}\ \rm{M}_{\odot}$,
    \item 2.5 < M < 3.0: a region where we used PMZ = $1\times10^{-3}$ M$_{\odot}$.
\end{itemize}

Note that the mass extent of the resultant $^{13}$C pocket is smaller than the size of the mass region over which we mix in protons (the PMZ), where a $^{14}$N-rich region forms closest to the top of the He-shell and inhibits $s$-process element production (owing to $^{14}$N being an efficient neutron poison). Also, the mass regimes are shifted somewhat with respect to solar metallicity, because metal-poor models have larger H-exhausted cores at a given stellar mass.

\section{Results}
\label{sec:results}

\subsection{Isotopic analysis}
\begin{table*}
\centering
    \caption{The isotopic ratios for our target stars using the linemake line list and using the posterior method to calculate the final ratios.}
    \label{tab:LP_tab}
\begin{tabular}{llllll} 
\hline
id     & $^{25}$Mg/$^{24}$Mg  & $^{26}$Mg/$^{24}$Mg  & Final ratios                                 & \# of lines & Included   \\
\hline
C      & 0.18  ($\pm 0.041$)  & 0.196  ($\pm0.019$)  & 73 ($\pm3.2$) : 13 ($\pm2.4$) : 14 ($\pm0.8$) & 6 & 1, 2, 3, 4, 7, 8 \\
III-3  & 0.11  ($\pm 0.079$)  & 0.077  ($\pm 0.042$) & 84 ($\pm7.5$) : 9 ($\pm5$) : 7 ($\pm2.8$)    & 4           & 1, 5, 7, 8 \\
III-14 & 0.125  ($\pm 0.075$) & 0.041  ($\pm 0.031$) & 86 ($\pm6.9$) : 11 ($\pm4.9$) : 3 ($\pm2.0$)  & 5 & 1, 3, 4, 5, 7    \\
III-15 & 0.058  ($\pm 0.075$) & 0.195  ($\pm0.048$)  & 80 ($\pm6.7$) : 5 ($\pm4.7$) : 15 ($\pm2.7$) & 4           & 1, 2, 4, 8 \\
III-52 & 0.155  ($\pm 0.053$) & 0.045  ($\pm0.02$)   & 83 ($\pm5.3$) : 13 ($\pm3.8$) : 4 ($\pm1.5$)  & 5 & 1, 3, 4, 5, 7    \\
IV-102 & 0.079  ($\pm 0.085$) & 0.021  ($\pm 0.035$) & 91 ($\pm8.1$) : 7 ($\pm5.7$) : 2 ($\pm2.4$)  & 4           & 1, 4, 5, 8 \\
\hline
\end{tabular}
\end{table*}

\begin{figure}
    \includegraphics[width=\columnwidth]{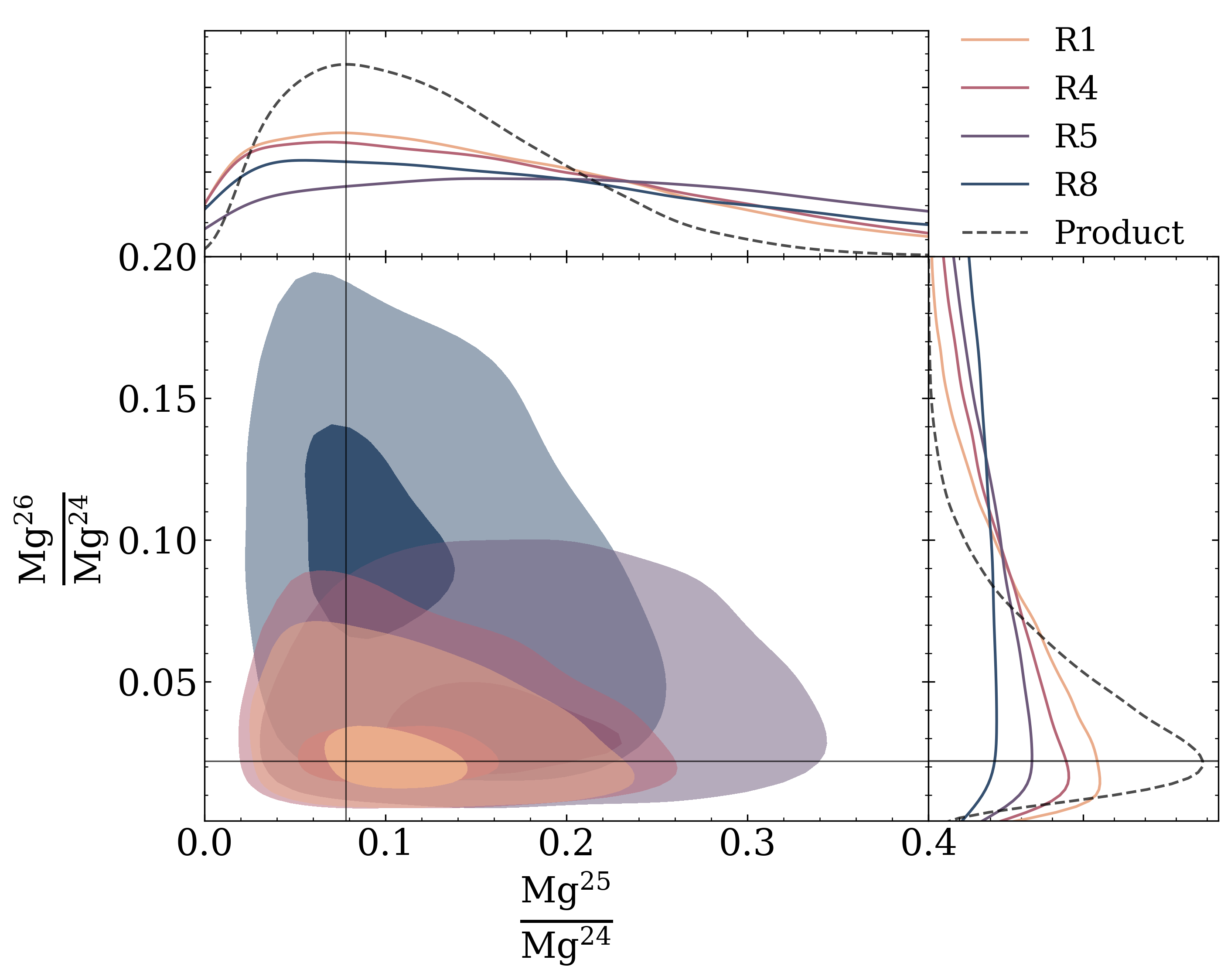}
    \caption{The marginalised posterior distributions used to calculate the final ratios for the star IV-102. Each distribution represents one run of \textsc{ratio} for one line in the star. This is analogous to the $\frac{26}{24}$ as a function of $\frac{25}{24}$ in Fig. \ref{fig:corner}. The shaded regions represent the 68\% and 95\% confidence intervals.}
    \label{fig:posteriors}
\end{figure}

The isotopic abundances listed in Table \ref{tab:LP_tab} and accompanying figures all use the linemake line list and the posteriors method for determining the final ratio.
In Fig. \ref{fig:posteriors}, we plot the marginalised posteriors for $^{25}$Mg/$^{24}$Mg and $^{26}$Mg/$^{24}$Mg for the star IV-102, using lines in R1, 4, 5, and 8 (see Table \ref{tab:LP_tab}). This highlights that the posteriors method determines the optimum isotopic ratio across multiple regions. The distributions from each MgH line are roughly in agreement. The narrower range of $^{26}$Mg/$^{24}$Mg compared to $^{25}$Mg/$^{24}$Mg demonstrates that $^{26}$Mg/$^{24}$Mg is far better constrained than $^{25}$Mg/$^{24}$Mg and results in smaller $^{26}$Mg errors.

In Table \ref{tab:LP_tab} we present our final isotopic ratios as calculated by the listed regions. 
Due to unknown blends and uncertainties in our line list, we reject regions R6, 9 and 10 for all our target stars as even the best-fit model is still a poor representation of the data. However, this new approach of employing between 4 and 6 lines for each of our metal-poor stars marks a substantial advancement over previous studies and represents a crucial step towards precisely determining isotopic ratios.

\begin{figure*}
    \includegraphics[width=\textwidth]{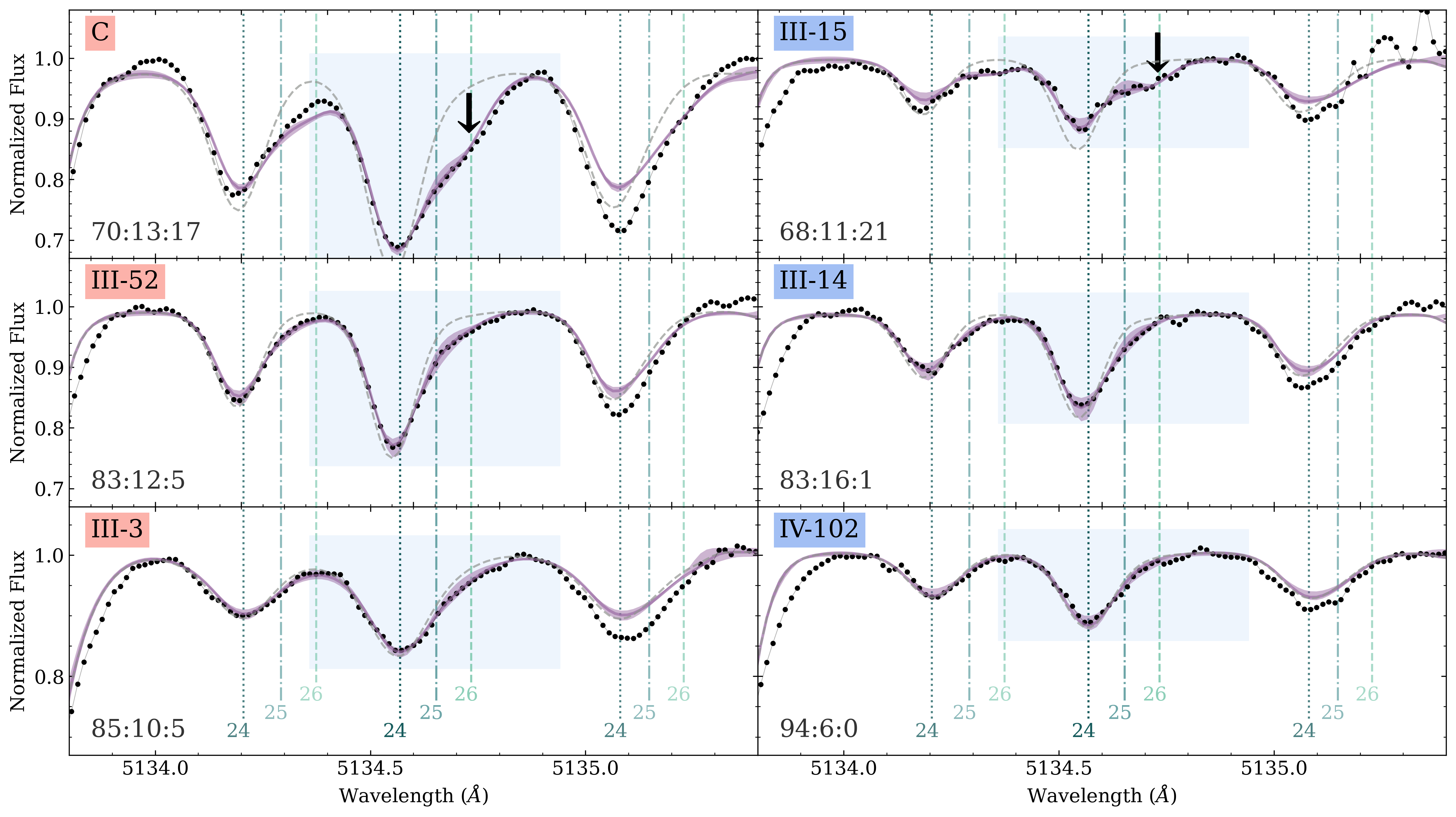}
    \caption{Fits for the 5134.6 \AA{}\ region, R1, for our target stars using the linemake line list. The $s$-process rich stars are to the left with red labels and the $s$-process poor stars are to the right with blue labels. The stars decrease in metallicity from top to bottom for each $s$-process group. The light blue rectangle is the section of spectra that we use for fitting the line. The shaded purple region is the 68\% confidence interval of the best fit as determined by \textsc{ratio}. The $s$-process abundance does not influence the relative isotopic ratios, however, the stars in the top panels, C and III-15, are Na rich and O poor which results in an enhancement in $^{26}$Mg, as visible in the spectra, illustrated by the black arrow. This is especially obvious when compared to the dashed grey line using the best-fit parameters but with a ratio of 99.8:0.1:0.1 (i.e., a purely $^{24}$Mg model). The ratios in the bottom left correspond to $^{24}$Mg: $^{25}$Mg: $^{26}$Mg as determined for this line.}
    \label{fig:r1_fits}
\end{figure*}

\begin{figure}
    \includegraphics[width=\columnwidth]{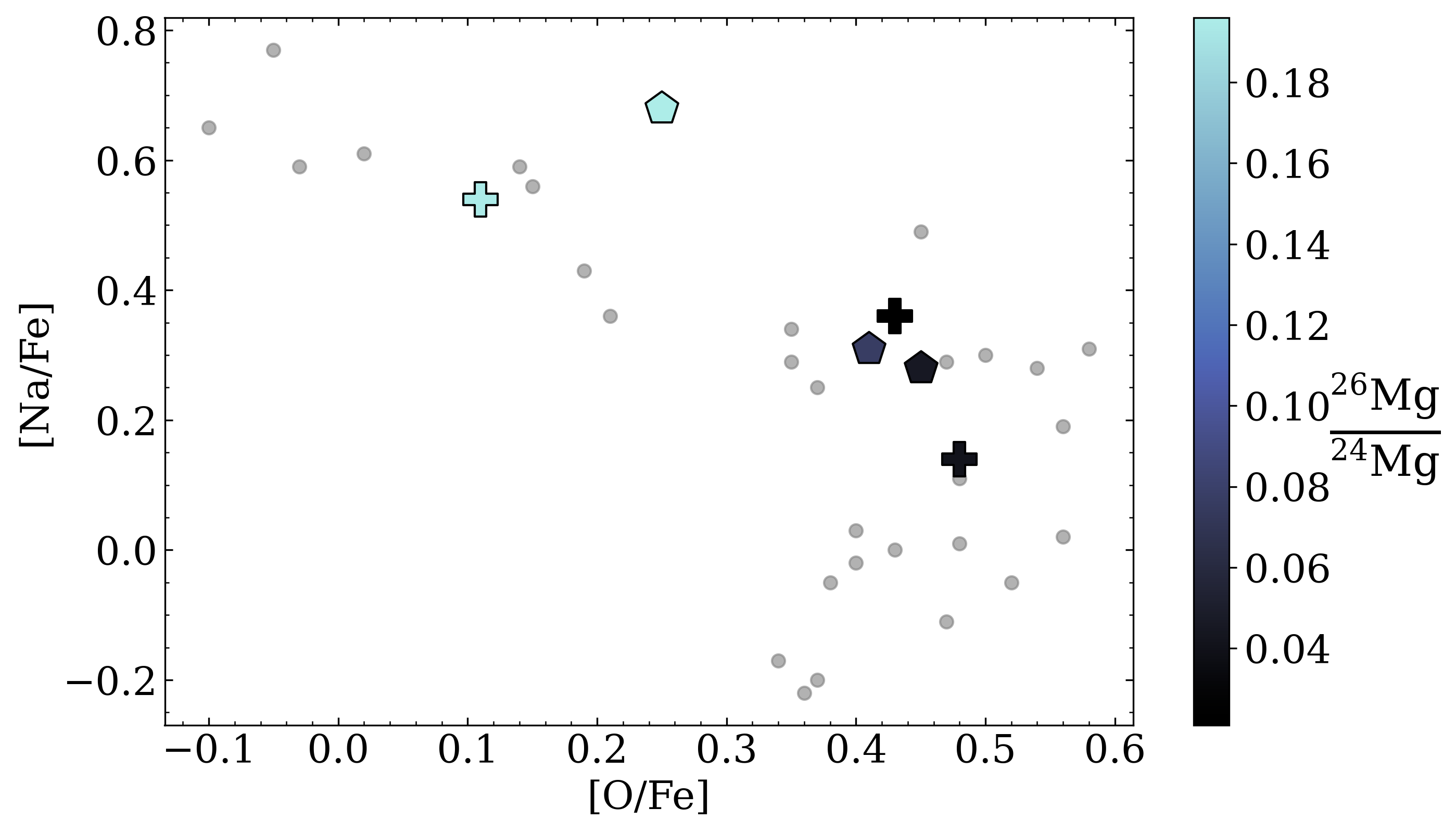}
    \caption{The Na-O anti-correlation using abundances from \protect\cite{Marino+2011} with the stars in this study coloured by their $^{26}$Mg/$^{24}$Mg isotopic ratios. $s$-process rich stars have pentagon markers whereas the $s$-process poor stars have plus markers. One of the key results from our study is that our Mg isotopic ratios correlate with the light element abundances rather than with the neutron capture elements.}
    \label{fig:NaO}
\end{figure}

The strength of the MgH line directly enhances our ability to accurately determine each isotope's contribution. R1 is the strongest and least blended MgH region and has been recognised in previous works as being the best representation of the total Mg isotopic ratio. Visually, the best-fit model in each of our stars is in excellent agreement with our data. Fig. \ref{fig:r1_fits} plots the spectra for each target star around R1 and the corresponding model determined by \textsc{ratio}. The isotopic ratio for R1 is given in the bottom left-hand corner of each panel. Although not exact, these values are all very similar to the final ratio given in Table \ref{tab:LP_tab}. The line to the left, R4, is also well fit by the solution to R1. However, it is understood in the literature that the line to the right, R5, does not match the depth of the observations given the best-fit parameters to R1 \citep[e.g.,][]{Yong+2003, DaCosta+2013}. The stars C and III-15, originating from the $s$-process rich and poor populations respectively, have a notable enhancement in $^{26}$Mg as illustrated by the black arrow and the final isotopic ratio in the bottom left corner in each panel. The reason for these anomalous ratios becomes clear in Fig. \ref{fig:NaO}, which gives the Na-O anti-correlation based on abundances from \cite{Marino+2011}. C and III-15 both belong to the Na-enhanced and O-poor populations, leading to the conclusion that Mg isotopic ratios correlate with light element abundances, rather than their $s$-process abundances.

\begin{figure}
    \includegraphics[width=\columnwidth]{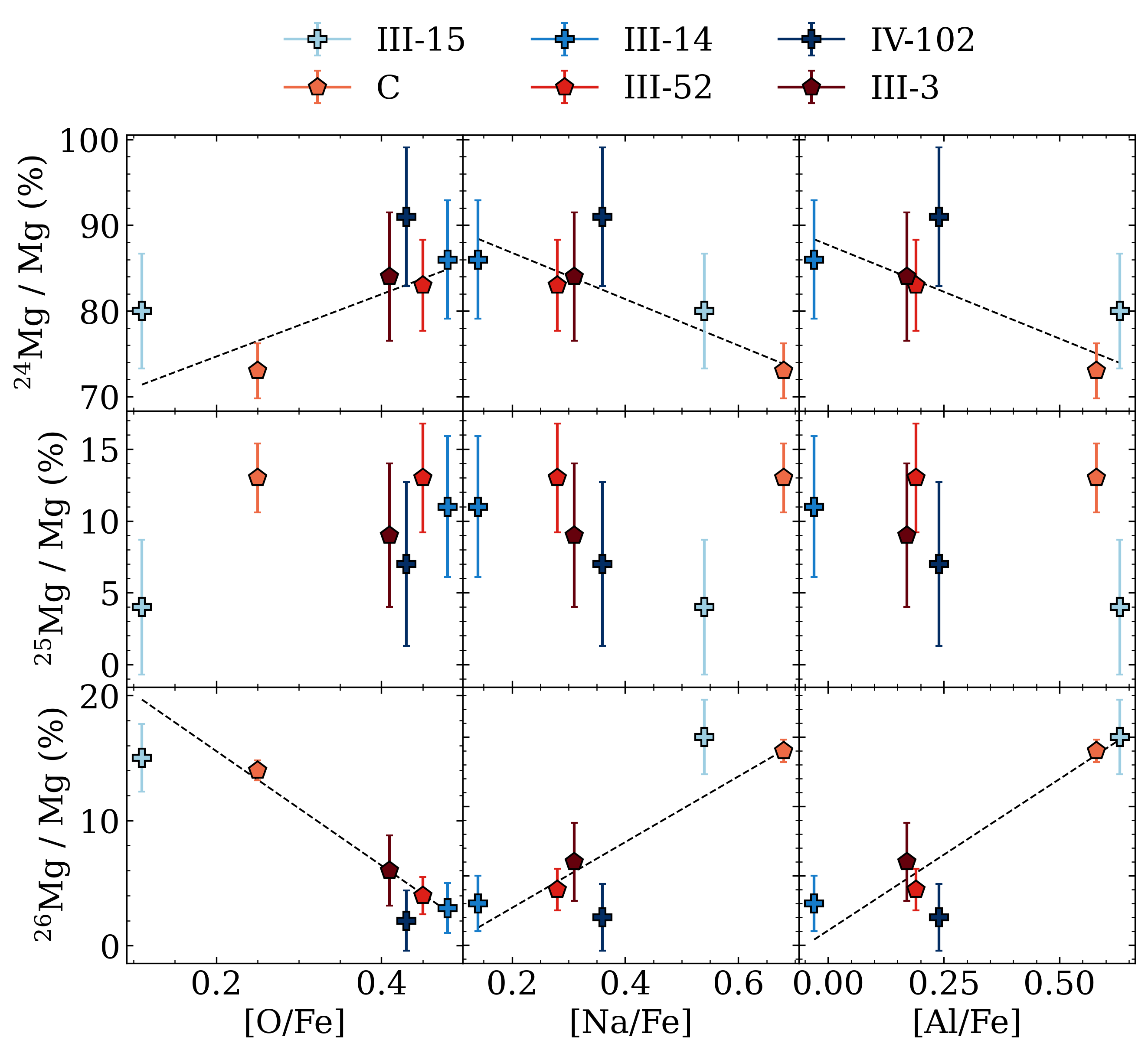}
    \caption{Mg isotope ratios as a function of the light elements O, Na and Al using \protect\cite{Marino+2011} abundances. A line of best fit (taking isotopic errors into account) is plotted for $^{24}$Mg and $^{26}$Mg.}
    \label{fig:iso_anti_light}
\end{figure}

This conclusion is further supported by Fig. \ref{fig:iso_anti_light}, which plots isotopic ratios against the light elements O, Na and Al from \cite{Marino+2011}. We use the same colour scheme as Paper I, with $s$-process rich and poor stars in red and blue respectively, and increasing iron abundance corresponding to an increase in the lightness of the colour. $^{24}$Mg positively correlates with O and negatively correlates with Na and Al. $^{25}$Mg shows no correlations with any of the light elements, but $^{26}$Mg shows the opposite behaviour to $^{24}$Mg; negatively correlating with O and positively correlating with Na and Al. Out of the four methods for calculating the final isotopic ratio, the linemake line list with the posteriors method plotted here gives the strongest correlations to the light element abundances based on the $\rm{R}^2$ coefficient. We do not observe significant correlations between the other light elements provided by \citet[][Si, Ca and Mg]{Marino+2011}, which we plot in Fig. \ref{fig:iso_anti_light_no_corr} in the Appendix. We plan to explore the origins of the relationships between these elements and different Mg isotopes in future works. $^{26}$Mg is constrained by the placement of the continuum and has a smaller dependence on the total Mg or the broadening (e.g., see the corner plot in Fig. \ref{fig:corner}) leading to smaller errors. It also has the strongest correlations with these light elements and thus must play an important role in nucleosynthesis.

\begin{figure}
    \includegraphics[width=\columnwidth]{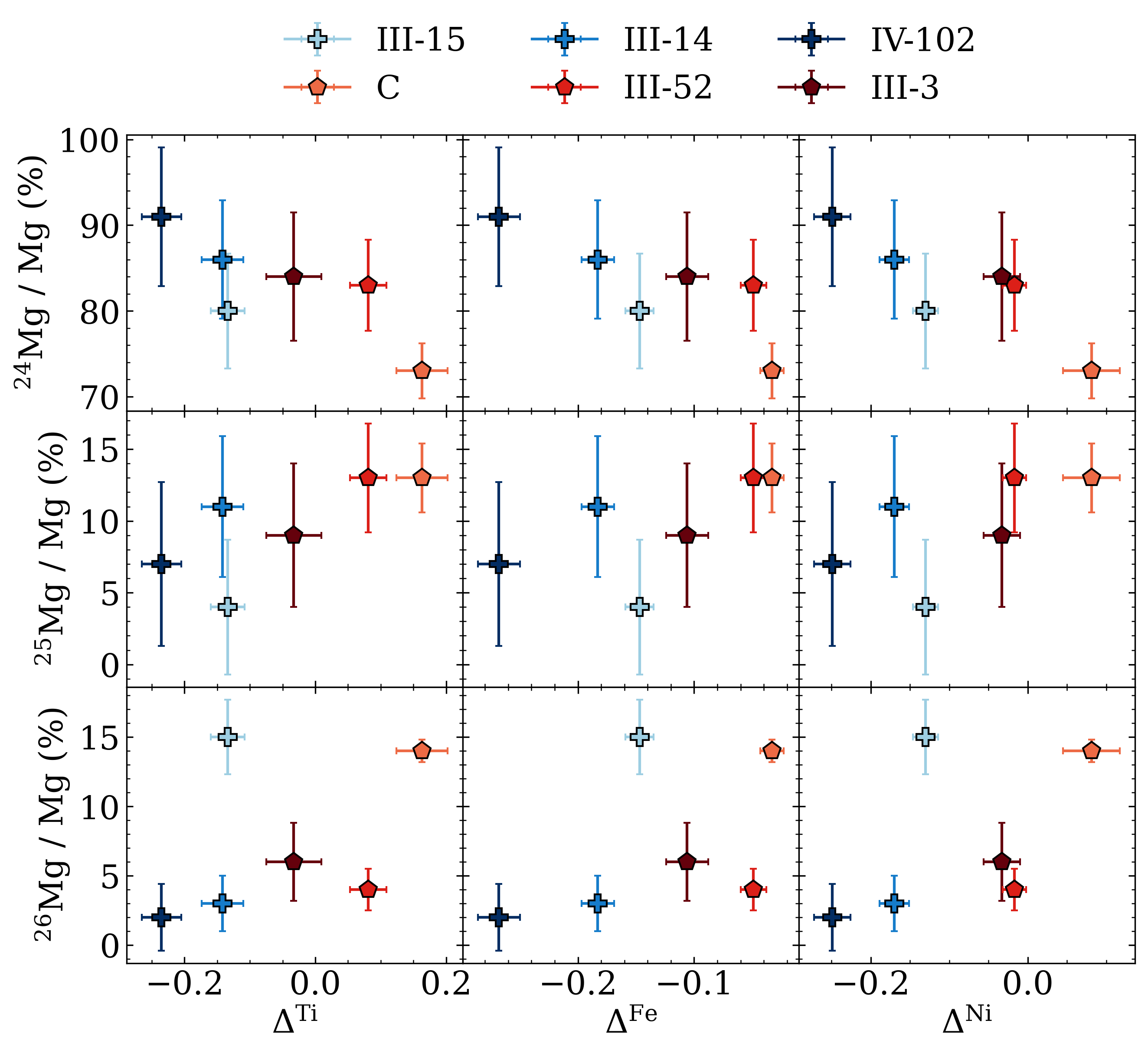}
    \caption{Mg isotope ratios as a function of iron peak elements Ti, Fe and Ni based on the differential abundances from Paper I. }
    \label{fig:iso_anti_ironPeak}
\end{figure}

Using our high-precision differential abundances from Paper I, we explore the relationship between iron peak and $s$-process elements. Ti, Fe, and Ni are chosen to represent the iron peak family as they are well measured over a large number of lines, have small errors ($\sim \ 0.02 \ \rm{dex}$), and are produced at different ratios depending on their source. Fig. \ref{fig:iso_anti_ironPeak} shows consistent trends: $^{24}$Mg decreases, $^{26}$Mg increases with iron peak abundance, and there are no significant trends in $^{25}$Mg. However, there are no strong trends within each of the $s$-process groups. The most Fe-rich star in each $s$-process group shows the highest $^{26}$Mg. Assuming that iron peak element abundance negatively correlates with age (i.e., the more iron-rich stars are younger), $^{26}$Mg production (but not $^{25}$Mg) is a time-dependent phenomenon. 

\begin{figure}
    \includegraphics[width=\columnwidth]{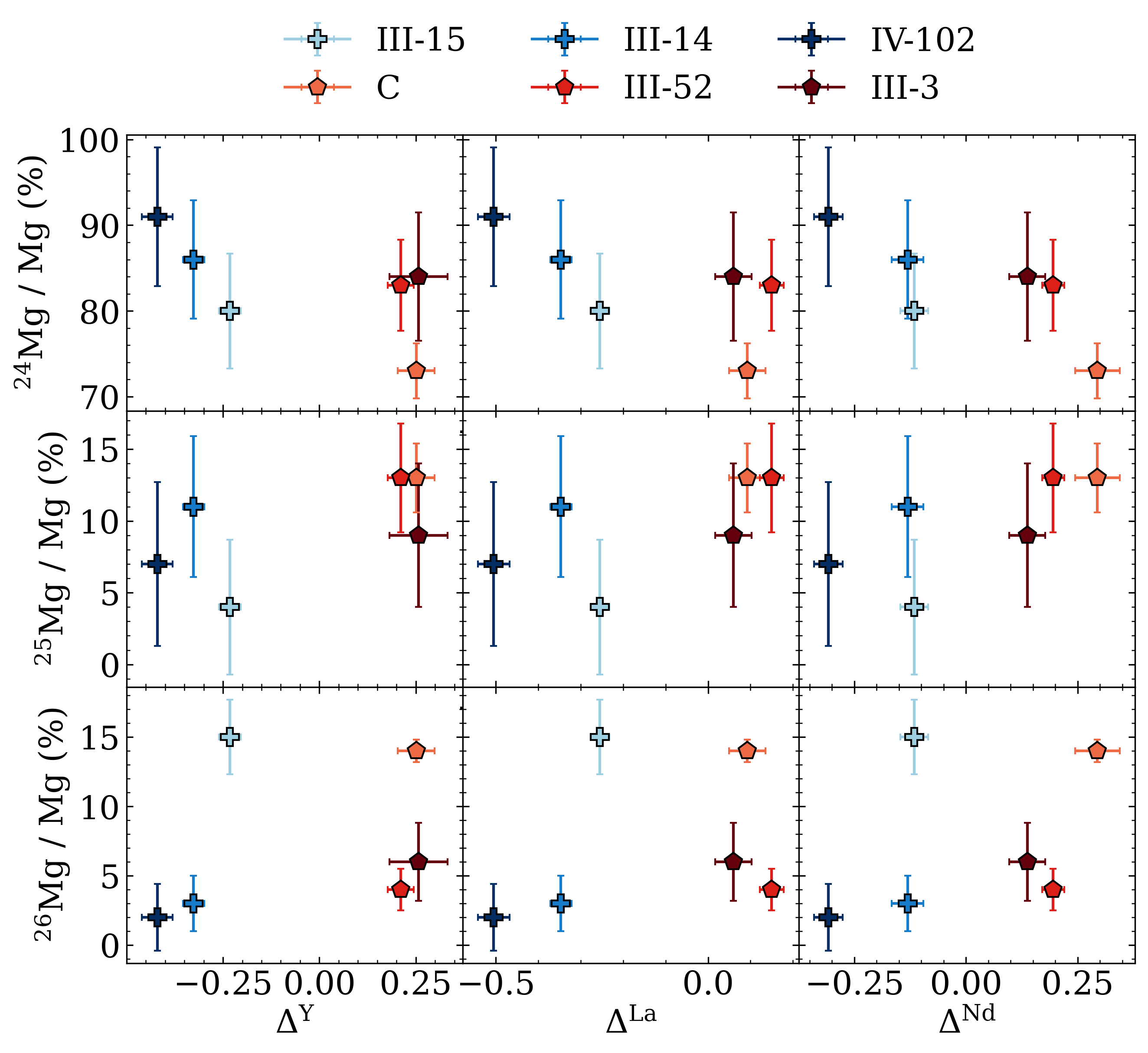}
    \caption{Mg isotope ratios as a function of $s$-process elements Y, La and Nd based on the differential abundances from Paper I. We see similar abundance patterns to Fig. \ref{fig:iso_anti_ironPeak}, however as expected, there is a greater separation between the two $s$-process groups.}
    \label{fig:iso_anti_heavy}
\end{figure}

Finally, the $s$-process elements in Fig. \ref{fig:iso_anti_heavy}, Y, La and Nd, exhibit a similar pattern to the iron-peak elements. On average, the $s$-process poor population is more enhanced in $^{24}$Mg, and depleted in $^{25}$Mg and $^{26}$Mg compared to the $s$-process rich population. We continue to see the pattern of $^{26}$Mg increasing with increasing $s$-process production within each population. There is no discernible difference in the isotopic ratios between the $1^{\rm{st}}$-peak $s$-process element Y, compared to the $2^{\rm{nd}}$ peak elements La and Nd. Furthermore, there is no significant trend for each of the s-process populations.

\begin{figure}
    \includegraphics[width=\columnwidth]{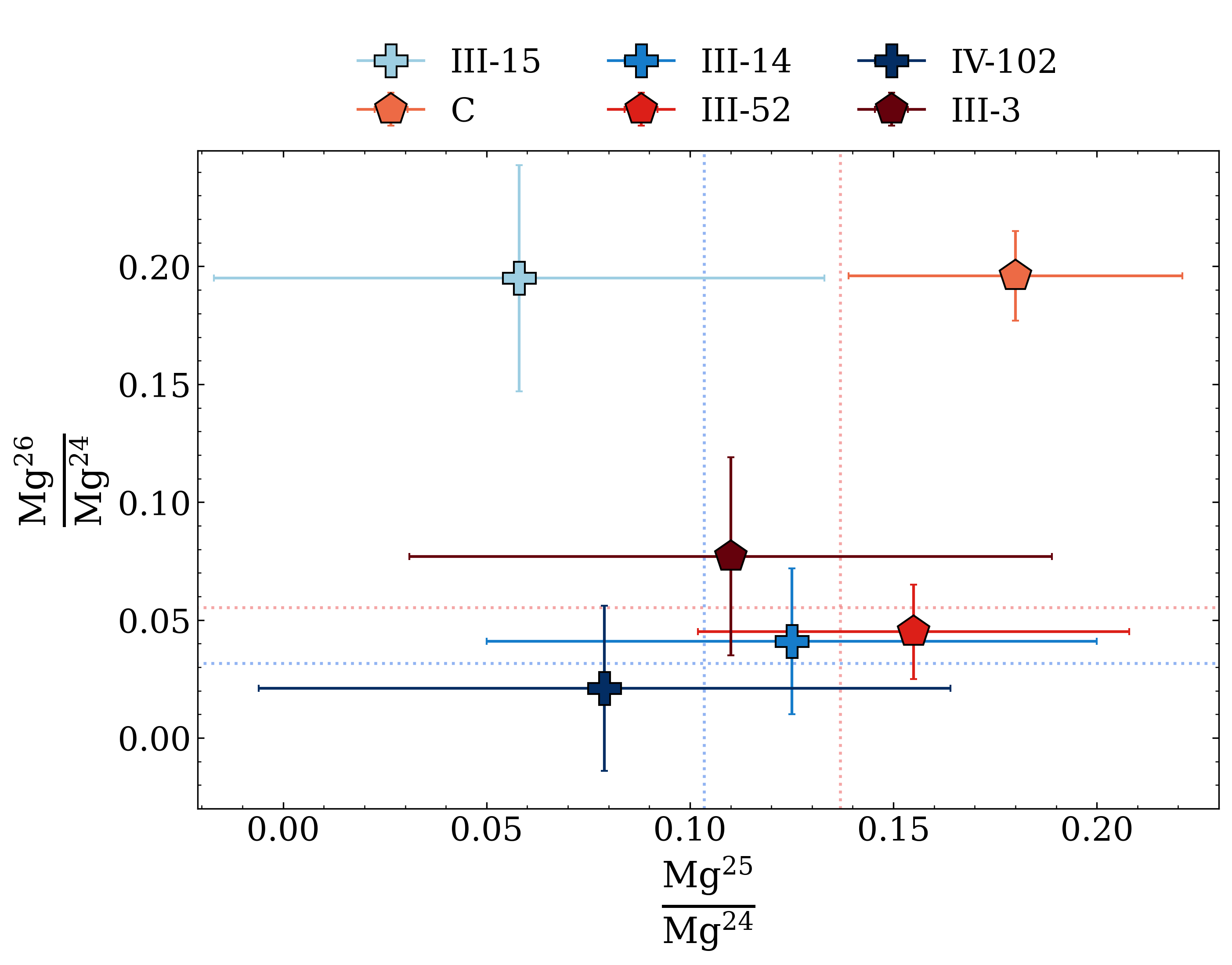}
    \caption{Isotopic ratios for $^{25}$Mg/$^{24}$Mg and $^{26}$Mg/$^{24}$Mg. The red and blue vertical dotted lines represent the weighted average value for the $s$-process rich and poor isotope ratios respectively. We do not include the Na-enhanced, O-depleted stars III-15 and C in our weighted average to exclude effects from light element abundance variations. The $s$-process rich population is slightly more enhanced in both $^{25}$Mg and $^{26}$Mg compared to the $s$-process poor population.}
    \label{fig:iso_ratio}
\end{figure}

In Fig. \ref{fig:iso_ratio}, we plot the isotopic ratios for $^{25}$Mg/$^{24}$Mg and $^{26}$Mg/$^{24}$Mg with their associated errors. The stars III-15 and C clearly stand out in $^{26}$Mg/$^{24}$Mg. To calculate the average difference between the populations, we remove these stars as there must be additional nucleosynthetic processes influencing their isotopic ratio. We find that the $s$-process rich population has more $^{25}$Mg and $^{26}$Mg compared to the $s$-process poor population; with $(^{25}\rm{Mg}/^{24}\rm{Mg})_{\rm{rich}} - (^{25}\rm{Mg}/^{24}\rm{Mg})_{\rm{poor}}  = 0.033$ and $(^{26}\rm{Mg}/^{24}\rm{Mg})_{\rm{rich}} - (^{26}\rm{Mg}/^{24}\rm{Mg})_{\rm{poor}} = 0.024$. These differences are on the same order of magnitude as our errors, therefore we assume that $s$-process production has a minimal impact on the resulting isotopic ratios (although a larger sample is needed to confirm this).

\subsection{Constraining the mass range of AGB stars}

\begin{figure*}
    \includegraphics[width=\textwidth]{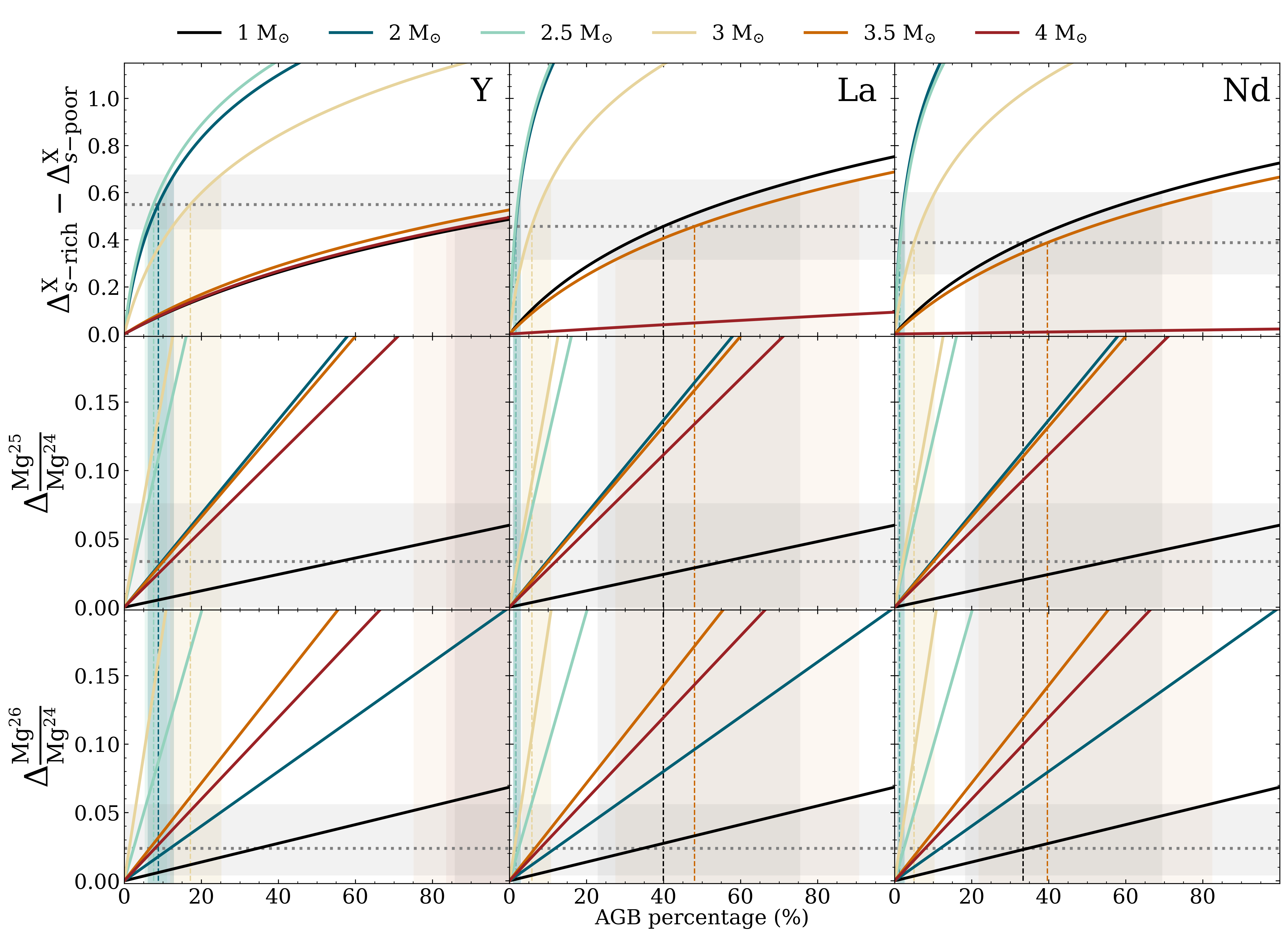}
    \caption{The production of $s$-process elements $\Delta^{\rm{Y}}$, $\Delta^{\rm{La}}$ and $\Delta^{\rm{Nd}}$ (top panels), $^{25}$Mg/$^{24}$Mg (middle panels) and $^{26}$Mg/$^{24}$Mg (bottom panels) as a function of the percentage of AGB ejecta. The average abundance in the wind of 1 M$_{\odot}$, 2 M$_{\odot}$, 3 M$_{\odot}$, 3.5 M$_{\odot}$, 4 M$_{\odot}$ AGB models are plotted in each panel. The dashed grey line represents the average difference between the $s$-process rich and poor populations for each quantity. The shaded grey region is the minimum and maximum range in $s$-process abundances given our target stars. Only low mass ($\leq$ 3.5 M$_{\odot}$) AGB models are capable of manufacturing enough La to reproduce the observed differences in M 22's stellar populations. The vertical lines correspond to the required dilution at which each AGB model produces the difference in the two $s$-process abundance groups. The shaded vertical lines of the same colour are the upper and lower limits. This can then be used to predict the approximate isotopic abundance in the wind expected for both $^{25}$Mg/$^{24}$Mg and $^{26}$Mg/$^{24}$Mg.
    }
    \label{fig:agb_y}
\end{figure*}

This subsection explores whether a certain mass range of AGB stars can replicate both $s$-process abundance differences and M 22's Mg isotope ratios. We suggest AGBs as the primary source of the $s$-process and neutron-rich Mg isotopes, however, we briefly discuss alternative scenarios in the following sections. 
We compare our results to custom AGB models at the metallicity of M 22; $\rm{[Fe/H]} = -1.8$. Fig. \ref{fig:agb_y} plots the percentage of AGB ejecta required to reproduce the difference in $s$-process elements $\Delta^{\rm{Y}}$, $\Delta^{\rm{La}}$ and $\Delta^{\rm{Nd}}$ respectively (i.e., see Fig. \ref{fig:iso_anti_heavy}). On the x-axis, we give the degree of dilution of AGB ejecta with pristine material (i.e., gas with the same abundances as the $s$-process poor population) as a percentage. A value of 0 represents the abundance equivalent to the $s$-process poor population while a value of 100 represents the value corresponding to pure AGB ejecta. The top panels in each figure give the enhancement in Y, La and Nd as a function of dilution for the AGB masses of 1, 2, 2.5, 3, 3.5 and 4 M$_{\odot}$. The 1, 2, 2.5 and 3 M$_{\odot}$ mass models use the \cite{Vassiliadis_Wood1993} mass loss rate, whereas the 3.5 and 4 M$_{\odot}$ models use the \cite{Bloecker+1995} rate. The grey dotted horizontal line denotes the measured $s$-process enhancement between the two populations based on the differential abundances in Paper I. The shaded grey region represents the minimum and maximum difference based on the range of the $s$-process abundance measurements. For clarity, we explicitly write out the y-axis label as $\Delta^{\rm{X}}_{s\rm{-rich}}-\Delta^{\rm{X}}_{s\rm{-poor}}$
for our three representative $s$-process elements given in the left hand corner of the top panels. The middle and bottom panels illustrate the production of $^{25}$Mg/$^{24}$Mg and $^{26}$Mg/$^{24}$Mg between the $s$-process populations respectively.  
The grey dotted horizontal lines illustrate the observed differences in the isotopic ratio between the two $s$-process populations as demonstrated by Fig. \ref{fig:iso_ratio}. Again, the shaded grey regions represent the minimum and maximum range in our measurements. Vertical dashed lines are plotted at the level of dilution required to reproduce the $s$-process difference and the corresponding shaded region of the same colour represents the possible values that can satisfy our observations. For La and Nd, AGB models above 3.5 M$_{\odot}$ can not reproduce the observed $s$-process abundance differences.

\begin{figure}
    \includegraphics[width=\columnwidth]{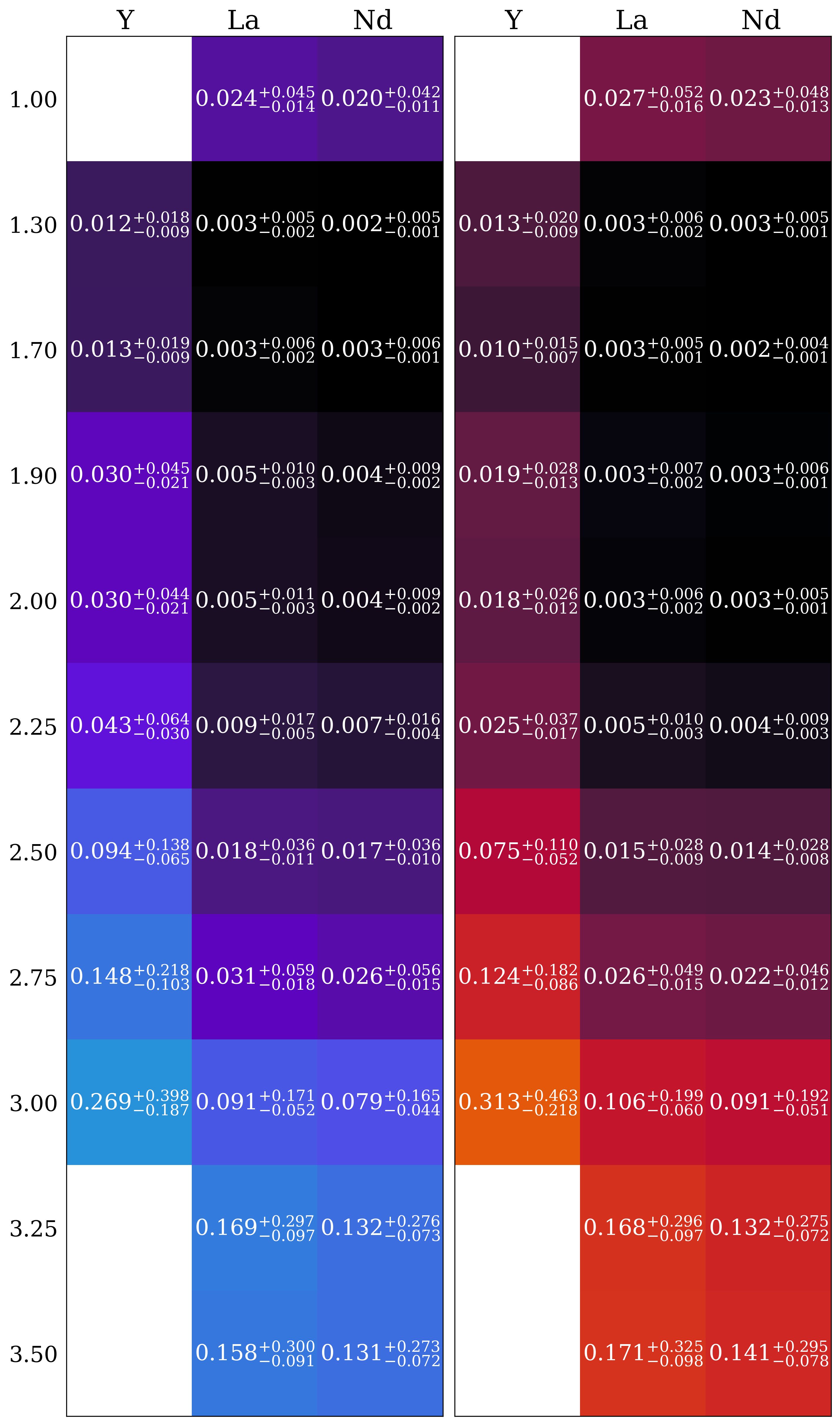}
    \caption{The predicted changes in $^{25}$Mg/$^{24}$Mg (left, blue) and $^{26}$Mg/$^{24}$Mg (right, orange) given the enrichment required to match $s$-process population differences. 
    Columns represent $s$-process elements Y, La, and Nd, and rows denote AGB model masses in $\rm{M}_{\odot}$. Colour intensity indicates the $^{25}$Mg and $^{26}$Mg enhancement level, with lighter colours marking larger differences between $s$-process poor and rich populations. Super and subscript values provide upper and lower limits based on $s$-process abundances. Empty entries signal models that can't generate enough $s$-process material to reproduce our observations. For comparison, the average difference in $^{25}$Mg/$^{24}$Mg is 0.033 and has a range of up to 0.076.With average differences of 0.033 (up to 0.076) and 0.024 (maximum 0.057) for $^{25}$Mg/$^{24}$Mg and $^{26}$Mg/$^{24}$Mg respectively, 2.75 $\rm{M}_{\odot}$ AGB stars best reproduce our isotopic observations.}
    \label{fig:agb_prediction}
\end{figure}

For each AGB model, we take the amount of dilution which reproduces observations to calculate the expected isotopic ratio. For instance, a 3 M$_{\odot}$ model producing $\sim$0.45 dex of La alters $^{25}\rm{Mg}/^{24}\rm{Mg}$ and $^{26}\rm{Mg}/^{24}\rm{Mg}$ by $\sim \ 0.15$, and $\sim \ 0.16$ respectively. However, Fig. \ref{fig:iso_ratio} demonstrates that a difference between $s$-process populations of $\sim$ 0.15 should be detectable given our errors. The isotopic ratios given the difference in $s$-process abundances are shown in Fig. \ref{fig:agb_prediction}. Each row represents the expected isotopic enrichment from each $s$-process element, and the different AGB mass models are given as rows. Rows are coloured by their predicted isotopic yields with blue and orange colours corresponding to higher $^{25}\rm{Mg}/^{24}\rm{Mg}$ and $^{26}\rm{Mg}/^{24}\rm{Mg}$ yields respectively. La and Nd give very similar isotope predictions which are expected as they are both $2^{\rm{nd}}$ peak $s$-process elements. One of the main conclusions of this paper is that both these elements favour models between $\sim1\rm{-}3 \ \rm{M}_{\odot}$, with the best match occurring at $\sim2.75 \ \rm{M}_{\odot}$.

\section{Discussion}
\label{sec:discussion}

\subsection{Correlations with light elements}
\cite{Marino+2011} demonstrated that M 22 stars have a similar Na-O anticorrelation and Na-Al correlations to the general population of `normal' GCs \citep[e.g.,][] {Carretta+2009} and that the Na-O anticorrelation is present in each $s$-group. This is reflected in our sample, with C and III-3 both belonging to the Na-enhanced populations. Employing the same technique used to determine the AGB mass range responsible for the two $s$-process populations but instead using Na, we can investigate the mass range of AGB stars that enriched this Na-enhanced population. Unlike the $s$ element groups where there is a more distinct cut-off between the populations, it is harder to define a threshold at which Na abundances become `enhanced'. However, if we assume a $\sim0.4 \rm{dex}$ increase in Na between C and III-3 and the other measured stars, AGB stars with masses between $3.5 \rm{-} 4.5 \ \rm{M}_{\odot}$ have the potential to fulfil the observed isotopic distribution. These stars have a total stellar lifetime of $\sim100 \rm{-} 200 \ \rm{Myr}$. However, we caution that these measurements have limited diagnostic power as they are based on our small sample size and a somewhat arbitrary Na difference. However, there is the potential to extend this to multiple elements using a larger sample size of stars in future works.

This is the first study to provide a comprehensive comparison between isotopic ratios and chemical abundances for a wide range of elements. The measured correlations suggest that the nucleosynthetic processes responsible for processing Mg isotopes must also act on [O/Fe] and [Na/Fe] and [Al/Fe] (Fig. \ref{fig:iso_anti_light}), but not on [Mg/Fe], [Si/Fe], and [Ca/Fe] (Fig. \ref{fig:iso_anti_light_no_corr}). Future works aim to expand on the number of clusters with high-quality abundance and isotopic measurements to investigate whether this is a universal feature amongst GCs and other complex star clusters. 

\subsection{A comparison to previous literature results}
M13 (\citealt{Shetrone1996}; \citealt{Yong+2006}), NGC6752 \citep{Yong+2003}, M71 (\citealt{Melendez+2009}; \citealt{Yong+2006}), M4 \citep{DaCosta+2013}, $\omega$ Centauri \citep{DaCosta+2013} and 47 Tucanae \citep{Thygesen+2016} are the only clusters with Mg isotopic measurements. Mg isotopes are most often compared with Al abundances, and it is clear that as Al abundance increases, $^{24}$Mg decreases, $^{26}$Mg increases and there is no significant change in $^{25}$Mg. Therefore, our results are consistent with the literature results for Al.

The APOGEE survey is one study that has demonstrated that the Mg-Al anti-correlation is metallicity dependent \citep{Meszaros+2020}. NGC6752 has a similar [Fe/H] to M 22 (i.e.,$\sim-1.66$; \citealt{Yong+2013}), making it an ideal reference cluster for this work. To compare, we use the isotopic values from \cite{Yong+2003} and differential abundances from \cite{Yong+2013} (which were analysed using the same reference star as Paper I; NGC6752-mg9). Na is the only light element with differential abundances that we see trends with isotopic ratio. Both $s$-process populations in M 22 share an overlapping $^{26}$Mg-Na correlation and $^{24}$Mg-Na anticorrelation with NGC6752. In NGC6752 however, Na-rich stars reach much higher $^{26}$Mg values than any of the M 22 stars. This could suggest that more massive AGB stars (in comparison to the AGB mass range given for M 22) were responsible for generating the isotopic trends in NGC6752, as this would allow for the production of the heavier isotopes, while not influencing the $s$-process abundance. A plot with the comparison between the isotopic ratios in M 22 and NGC6752 is included in Appendix \ref{app:light_elements}. Similarly to M 22, there are no trends in Si or La.

In the study by \cite{Thygesen+2016} on the cluster 47 Tuc which used 3D hydrodynamic models to derive their isotope ratios, they found that there was a significant enhancement in the ratio of $^{26}$Mg/$^{24}$Mg, but $^{25}$Mg/$^{24}$Mg essentially remained unchanged.

In their theoretical study of 3D line formation, \citet{Thygesen+2017} found only negligible effects on the isotopic ratios, smaller than 1\,\%, for a red giant model with $\rm[Fe/H] = -2$ and with temperature similar to that of our stars. 
We therefore do not investigate 3D NLTE effects further at this time.

\subsection{Alternative explanations}

\cite{Ventura+2018} provides an extensive review of Mg isotope ratios and their significance in understanding the enrichment processes within GCs. The authors discuss that Mg is one of the key elements traced by stars within GCs as it provides invaluable insights into the characteristics of the polluting source responsible for the formation of 2G stars. As Mg is unaffected by mixing episodes during both the RGB and progenitor AGB phases, its abundance reflects the conditions the processed gas was exposed to. Therefore this allows for the determination of the degree and extent of nucleosynthesis.

Supermassive stars (SM; e.g., \cite{Gieles+2018}) and the hot bottom burning envelopes of AGBs are the only stellar p-burning environments reaching high enough temperatures to process $^{24}$Mg (\citealt{Prantzos_Charbonnel_Iliadis2007}). AGB stars have been used by both \cite{Shingles+2014} and \cite{Straniero+2014} to explain the $s$-process enhancement in M 22, but the massive stars with rotation model \citep{Hirschi_2007} used in \cite{Shingles+2014} could not reproduce the $s$-process enhancement within the cluster. However, they note that there are still large uncertainties in $s$-process yields in these models and that other sets of yields may change their conclusions. To complicate matters further, \cite{Denissenkov+2015} argued against it due to the large expected $^{25}$Mg content predicted by \cite{Ventura+2011} and championed the SM star model. 

The winds of fast-rotating massive stars (FRMS; \citealt{Decressin+2007}) have also been proposed as a solution to the multiple stellar populations problem in GCs. FRMS can generate a large amount of $s$-process products at low metallicity, where the dominant neutron source is  $^{22}\rm{Ne}(\alpha, n)^{25}\rm{Mg}$ in the convective He-burning core and in the subsequent convective C-burning shell (\citealt{Pignatari+2008}; \citealt{Frischknecht+2012}; \citealt{Frischknecht+2016}). However, \cite{Ventura+2018} argues that only AGB and very massive stars can reproduce the Mg-Al anticorrelation. Scenarios that invoke binary star interactions (e.g., \citealt{deMink+2009}, \citealt{Bekki2023}) are cited as a potential solution for the multiple stellar population problem in conventional GCs, but do not consider any potential $s$-process element abundance differences.

One caveat of these scenarios is that they mostly refer to Type I GCs, rather than the metal-complex Type II clusters. The existence of a [Fe/H] spread in M 22 demands that some fraction of ejecta used to make the $s$-process rich population was also enriched by SNe. However, as SNe are not major contributors to the $s$-process or heavy Mg isotopes, we save this discussion for a more comprehensive chemical evolution analysis in the future.

\subsection{Implications for \textit{s}-process elements}

Our AGB mass range is marginally lower than the results from \cite{Shingles+2014}, who employed chemical evolution modelling to determine a lower limit on AGB masses of $2.75$-$3.25\ \rm{M}_{\odot}$, corresponding to a minimum enrichment timescale of $240$-$360 \ \rm{Myr}$. \cite{Straniero+2014} suggests an even higher limit on the AGB mass; $4\pm 0.5 \  \rm{M}_{\odot}$. By using isochrone fitting of observations of the sub-giant branch of M 22, \cite{marino+2012} found that the age difference between the two $s$-process populations is no more than $\sim 300 \ \rm{Myr}$. Additionally, employing Yonsei–Yale isochrones, \cite{Joo_Lee2013} discovered that the age difference is $300 \ \rm{Myr}  \ \pm 400 \ \rm{Myr}$.

The total stellar lifetime for our best fit AGB model, $\sim2.75 \  \rm{M}_{\odot}$, is $\sim 350 \ \rm{Myr}$. 
This age represents the lower limit on the age difference as it does not account for the gas cooling timescale for both the AGB ejecta and the pristine gas diluting this ejecta. Still, this difference is already larger than predictions from isochrone fitting. A larger AGB mass would better agree with the enrichment timescale set by isochrone fitting. However, $\sim3 \ \rm{M}_{\odot}$ with a stellar lifetime of $\sim 280 \ \rm{Myr}$ predicts a difference in $\rm{Mg}^{25}/\rm{Mg}^{24}$ and $\rm{Mg}^{26}/\rm{Mg}^{24}$ of $\sim0.15$ and $\sim0.17$ respectively averaged across the $s$-process elements. These values are roughly a factor of 5 larger than the measured differences and would be detectable given our errors. Future chemical evolution models should incorporate a more comprehensive approach to determining the optimum AGB mass, incorporating not only Mg isotopes but other light elements such as C, N and O to better constrain these measurements.

A significant amount of dilution is required for each model. For example, the contribution from AGB ejecta to reproduce the $s$-process abundance enhancement for our $2.75 \ \rm{M}_{\odot}$ model is roughly $5\%$. The accretion of pristine gas has been a mechanism suggested in GC formation scenarios \citep{DErcole+2016} and has been theoretically demonstrated to be a viable mechanism for creating additional stellar populations (\citealt{McKenzie_Bekki2021}; \citealt{Lacchin+2021}). Additionally, gas accretion can help to solve a variation of the "mass budget problem" (\citealt{Renzini_2008}; \citealt{Renzini+2015}) present in this cluster of how the $s$-process poor population can generate enough mass to form the $s$-process rich population (where \citealt{Milone+2017MSP} and \citealt{Lee_2020} estimate that the mass ratio between the $s$-process poor to rich populations to be $\approx60:40$).

Uncertainties stemming from constraining AGB mass using isotopic analysis are difficult to assess. While we have made significant efforts to minimise errors from the $s$-process element abundance measurements, errors in the isotopic measurements (e.g., errors originating from the model atmospheres, line lists and 1D radiative transfer) and AGB modelling (e.g., due to numerical treatments of convection, mass loss, and reaction rates and low-temperature opacities) possess considerable systematic uncertainties. However, we are still able to constrain previous measurements of AGB mass ranges to the low-mass end.

Follow-up studies will investigate the yields from different AGB models and test whether different scenarios, such as the FRMS scenario, can reproduce these observations.

\section{Conclusions}
\label{sec:conclusion}

% Potentially change to numbered conclusions
The formation of M 22 is still an unsolved mystery, however, chemical abundance analysis on the isotopic level offers a new, unexplored perspective on the problem. Given the considerable variations in the yields of high and low-mass AGB stars, we leverage these differences to determine the specific AGB mass range that served as progenitors for the $s$-process rich population. Furthermore, this allows us to estimate an age difference between the two populations in a way that is independent of isochrone fitting. Low-mass AGB stars manufacture a significant quantity of $s$-process elements compared to their high-mass counterparts while having a minimal impact on the production of heavy Mg isotopes. This study suggests that AGB stars with a mass ranging from $\sim1$-$3 \ \rm{M}_{\odot}$ best reproduce both the $s$-process and isotopic abundances.

This work introduces a new approach to calculating Mg isotope ratios, measuring more than three times the number of lines traditionally used, leveraging MCMC methods to identify the optimal fit for the lines, and subsequently combining these lines based on their probability distributions to determine the overall Mg isotope ratio of the star. This innovative approach not only delivers realistic uncertainties for the first time but also enables the isotopic measurement of the most metal-poor RGB star ever discussed in the literature. 

Interestingly, we identify correlations between the Mg isotopic ratios and the lighter elements O, Na, and Al, but not with Mg, Si, or Ca. Correlations in Al are consistent with previous studies of GCs, however, other light elements are not commonly discussed. There appear to be weaker correlations between each of the $s$-process groups for both the iron peak and $s$-process elements, however, there is no distinguishable difference between the two stellar populations. 

To explore potential sources for these abundance variations, we use custom-made AGB models at the metallicity of M 22. Our findings suggest that only low-mass AGB models ($\sim 2.75 \ \rm{M}_{\odot}$) are capable of reproducing both the observed $s$-process enhancement while not influencing the isotopic ratio. This places significant constraints on previous estimates of the AGB mass ranges for the cluster and implies a more considerable age difference between the two populations than first thought. This novel approach represents an important advancement in our understanding of the formation of anomalous star clusters on the isotopic level. Furthermore, the potential of Mg isotopic ratios warrants further exploration in different areas of stellar astrophysics, serving as a promising tool to expand our understanding of stellar nucleosynthesis.

\section*{Acknowledgements}

All the authors thank the referee, Chris Sneden, for his thoughtful and insightful comments which helped improve this work. MM thanks Professor Mark Krumholz for his helpful suggestions regarding the statistical analysis and Patrick Armstrong for recommending the package \texttt{chainconsumer} \citep{Hinton2016} for this analysis. Additionally, we thank Ian Roederer for his input into the initial idea for this project and his helpful comments. This research is supported by an Australian Government Research Training Program (RTP) Scholarship and by the Australian Research Council Centre of Excellence for All Sky Astrophysics in 3 Dimensions (ASTRO 3D), through project number CE170100013. AFM and APM acknowledge support from INAF Research GTO-Grant Normal RSN2-1.05.12.05.10 - Understanding the formation of globular clusters with their multiple stellar generations (ref. Anna F. Marino) of the ``Bando INAF per il Finanziamento della Ricerca Fondamentale 2022".

This work makes use of Jupyter notebooks \citep{Kluyver2016jupyter} as well as the Python packages \texttt{NumPy} \citep{harris2020array}, \texttt{Matplotlib} \citep{Hunter2007}, \texttt{Pandas} \citep{reback2020pandas}, and \texttt{emcee} \citep{Foreman-Mackey+2013}.

%%%%%%%%%%%%%%%%%%%%%%%%%%%%%%%%%%%%%%%%%%%%%%%%%%
\section*{Data Availability}
The spectra used in this study are available from the ESO website with program ID 095.D-0027(A). All other data, including tables containing the isotopic results from the alternative methods of determining the final ratio, are available upon reasonable request.

%%%%%%%%%%%%%%%%%%%% REFERENCES %%%%%%%%%%%%%%%%%%

% The best way to enter references is to use BibTeX:

\bibliographystyle{mnras}
\bibliography{refs} % if your bibtex file is called example.bib

@ARTICLE{Yong+2003,
       author = {{Yong}, D. and {Grundahl}, F. and {Lambert}, D.~L. and {Nissen}, P.~E. and {Shetrone}, M.~D.},
        title = "{Mg isotopic ratios in giant stars of the globular cluster NGC 6752}",
      journal = {\aap},
         year = 2003,
        month = may,
       volume = {402},
        pages = {985-1001},
          doi = {10.1051/0004-6361:20030296},
archivePrefix = {arXiv},
       eprint = {astro-ph/0303057},
 primaryClass = {astro-ph},
       adsurl = {https://ui.adsabs.harvard.edu/abs/2003A&A...402..985Y}
}

@ARTICLE{Straniero+2014,
       author = {{Straniero}, O. and {Cristallo}, S. and {Piersanti}, L.},
        title = "{Heavy Elements in Globular Clusters: The Role of Asymptotic Giant Branch Stars}",
      journal = {\apj},
         year = 2014,
        month = apr,
       volume = {785},
       number = {1},
          eid = {77},
        pages = {77},
          doi = {10.1088/0004-637X/785/1/77},
archivePrefix = {arXiv},
       eprint = {1403.0819},
 primaryClass = {astro-ph.GA},
       adsurl = {https://ui.adsabs.harvard.edu/abs/2014ApJ...785...77S}
}

@ARTICLE{Shingles+2014,
       author = {{Shingles}, Luke J. and {Karakas}, Amanda I. and {Hirschi}, Raphael and {Fishlock}, Cherie K. and {Yong}, David and {Da Costa}, Gary S. and {Marino}, Anna F.},
        title = "{The s-Process Enrichment of the Globular Clusters M4 and M22}",
      journal = {\apj},
         year = 2014,
        month = nov,
       volume = {795},
       number = {1},
          eid = {34},
        pages = {34},
          doi = {10.1088/0004-637X/795/1/34},
archivePrefix = {arXiv},
       eprint = {1409.1227},
 primaryClass = {astro-ph.SR},
       adsurl = {https://ui.adsabs.harvard.edu/abs/2014ApJ...795...34S}
}

@ARTICLE{Karakas2010,
       author = {{Karakas}, A.~I.},
        title = "{Updated stellar yields from asymptotic giant branch models}",
      journal = {\mnras},
         year = 2010,
        month = apr,
       volume = {403},
       number = {3},
        pages = {1413-1425},
          doi = {10.1111/j.1365-2966.2009.16198.x},
archivePrefix = {arXiv},
       eprint = {0912.2142},
 primaryClass = {astro-ph.SR},
       adsurl = {https://ui.adsabs.harvard.edu/abs/2010MNRAS.403.1413K}
}

@ARTICLE{Yong+2013,
       author = {{Yong}, David and {Mel{\'e}ndez}, Jorge and {Grundahl}, Frank and {Roederer}, Ian U. and {Norris}, John E. and {Milone}, A.~P. and {Marino}, A.~F. and {Coelho}, P. and {McArthur}, Barbara E. and {Lind}, K. and {Collet}, R. and {Asplund}, Martin},
        title = "{High precision differential abundance measurements in globular clusters: chemical inhomogeneities in NGC 6752}",
      journal = {\mnras},
         year = 2013,
        month = oct,
       volume = {434},
       number = {4},
        pages = {3542-3565},
          doi = {10.1093/mnras/stt1276},
archivePrefix = {arXiv},
       eprint = {1307.4486},
 primaryClass = {astro-ph.GA},
       adsurl = {https://ui.adsabs.harvard.edu/abs/2013MNRAS.434.3542Y}
}

@ARTICLE{Marino+2009,
       author = {{Marino}, A.~F. and {Milone}, A.~P. and {Piotto}, G. and {Villanova}, S. and {Bedin}, L.~R. and {Bellini}, A. and {Renzini}, A.},
        title = "{A double stellar generation in the globular cluster NGC 6656 (M 22). Two stellar groups with different iron and s-process element abundances}",
      journal = {\aap},
         year = 2009,
        month = oct,
       volume = {505},
       number = {3},
        pages = {1099-1113},
          doi = {10.1051/0004-6361/200911827},
archivePrefix = {arXiv},
       eprint = {0905.4058},
 primaryClass = {astro-ph.SR},
       adsurl = {https://ui.adsabs.harvard.edu/abs/2009A&A...505.1099M}
}

@ARTICLE{Marino+2011,
       author = {{Marino}, A.~F. and {Sneden}, C. and {Kraft}, R.~P. and {Wallerstein}, G. and {Norris}, J.~E. and {Da Costa}, G. and {Milone}, A.~P. and {Ivans}, I.~I. and {Gonzalez}, G. and {Fulbright}, J.~P. and {Hilker}, M. and {Piotto}, G. and {Zoccali}, M. and {Stetson}, P.~B.},
        title = "{The two metallicity groups of the globular cluster M 22: a chemical perspective}",
      journal = {\aap},
         year = 2011,
        month = aug,
       volume = {532},
          eid = {A8},
        pages = {A8},
          doi = {10.1051/0004-6361/201116546},
archivePrefix = {arXiv},
       eprint = {1105.1523},
 primaryClass = {astro-ph.SR},
       adsurl = {https://ui.adsabs.harvard.edu/abs/2011A&A...532A...8M}
}

@ARTICLE{Roederer+2011,
       author = {{Roederer}, I.~U. and {Marino}, A.~F. and {Sneden}, C.},
        title = "{Characterizing the Heavy Elements in Globular Cluster M22 and an Empirical s-process Abundance Distribution Derived from the Two Stellar Groups}",
      journal = {\apj},
         year = 2011,
        month = nov,
       volume = {742},
       number = {1},
          eid = {37},
        pages = {37},
          doi = {10.1088/0004-637X/742/1/37},
archivePrefix = {arXiv},
       eprint = {1108.3868},
 primaryClass = {astro-ph.SR},
       adsurl = {https://ui.adsabs.harvard.edu/abs/2011ApJ...742...37R}
}

@ARTICLE{Alves-Brito+2012,
       author = {{Alves-Brito}, A. and {Yong}, D. and {Mel{\'e}ndez}, J. and {V{\'a}squez}, S. and {Karakas}, A.~I.},
        title = "{CNO and F abundances in the globular cluster M 22 (NGC 6656)}",
      journal = {\aap},
         year = 2012,
        month = apr,
       volume = {540},
          eid = {A3},
        pages = {A3},
          doi = {10.1051/0004-6361/201118623},
archivePrefix = {arXiv},
       eprint = {1202.0797},
 primaryClass = {astro-ph.SR},
       adsurl = {https://ui.adsabs.harvard.edu/abs/2012A&A...540A...3A}
}

@ARTICLE{Doherty+2014,
       author = {{Doherty}, Carolyn L. and {Gil-Pons}, Pilar and {Lau}, Herbert H.~B. and {Lattanzio}, John C. and {Siess}, Lionel},
        title = "{Super and massive AGB stars - II. Nucleosynthesis and yields - Z = 0.02, 0.008 and 0.004}",
      journal = {\mnras},
         year = 2014,
        month = jan,
       volume = {437},
       number = {1},
        pages = {195-214},
          doi = {10.1093/mnras/stt1877},
archivePrefix = {arXiv},
       eprint = {1310.2614},
 primaryClass = {astro-ph.SR},
       adsurl = {https://ui.adsabs.harvard.edu/abs/2014MNRAS.437..195D}
}

@ARTICLE{Karakas_Lugaro2016,
       author = {{Karakas}, Amanda I. and {Lugaro}, Maria},
        title = "{Stellar Yields from Metal-rich Asymptotic Giant Branch Models}",
      journal = {\apj},
         year = 2016,
        month = jul,
       volume = {825},
       number = {1},
          eid = {26},
        pages = {26},
          doi = {10.3847/0004-637X/825/1/26},
archivePrefix = {arXiv},
       eprint = {1604.02178},
 primaryClass = {astro-ph.SR},
       adsurl = {https://ui.adsabs.harvard.edu/abs/2016ApJ...825...26K}
}

@ARTICLE{Sneden_1973,
       author = {{Sneden}, C.},
        title = "{The nitrogen abundance of the very metal-poor star HD 122563.}",
      journal = {\apj},
         year = 1973,
        month = sep,
       volume = {184},
        pages = {839},
          doi = {10.1086/152374},
       adsurl = {https://ui.adsabs.harvard.edu/abs/1973ApJ...184..839S}
}

@ARTICLE{Thygesen+2016,
       author = {{Thygesen}, A.~O. and {Sbordone}, L. and {Ludwig}, H. -G. and {Ventura}, P. and {Yong}, D. and {Collet}, R. and {Christlieb}, N. and {Melendez}, J. and {Zaggia}, S.},
        title = "{The chemical composition of red giants in 47 Tucanae. II. Magnesium isotopes and pollution scenarios}",
      journal = {\aap},
         year = 2016,
        month = apr,
       volume = {588},
          eid = {A66},
        pages = {A66},
          doi = {10.1051/0004-6361/201526643},
archivePrefix = {arXiv},
       eprint = {1602.00058},
 primaryClass = {astro-ph.SR},
       adsurl = {https://ui.adsabs.harvard.edu/abs/2016A&A...588A..66T}
}

@ARTICLE{Thygesen+2017,
       author = {{Thygesen}, Anders O. and {Kirby}, Evan N. and {Gallagher}, Andrew J. and {Ludwig}, Hans-G. and {Caffau}, Elisabetta and {Bonifacio}, Piercarlo and {Sbordone}, Luca},
        title = "{An Investigation of the Formation and Line Properties of MgH in 3D Hydrodynamical Model Stellar Atmospheres}",
      journal = {\apj},
         year = 2017,
        month = jul,
       volume = {843},
       number = {2},
          eid = {144},
        pages = {144},
          doi = {10.3847/1538-4357/aa79a0},
archivePrefix = {arXiv},
       eprint = {1706.04218},
 primaryClass = {astro-ph.SR},
       adsurl = {https://ui.adsabs.harvard.edu/abs/2017ApJ...843..144T}
}

@ARTICLE{DaCosta+2013,
       author = {{Da Costa}, G.~S. and {Norris}, John E. and {Yong}, David},
        title = "{Magnesium Isotope Ratios in {\ensuremath{\omega}} Centauri Red Giants}",
      journal = {\apj},
         year = 2013,
        month = may,
       volume = {769},
       number = {1},
          eid = {8},
        pages = {8},
          doi = {10.1088/0004-637X/769/1/8},
archivePrefix = {arXiv},
       eprint = {1304.0523},
 primaryClass = {astro-ph.GA},
       adsurl = {https://ui.adsabs.harvard.edu/abs/2013ApJ...769....8D}
}

@ARTICLE{Yong+2004,
       author = {{Yong}, David and {Lambert}, David L. and {Allende Prieto}, Carlos and {Paulson}, Diane B.},
        title = "{Magnesium Isotope Ratios in Hyades Stars}",
      journal = {\apj},
         year = 2004,
        month = mar,
       volume = {603},
       number = {2},
        pages = {697-707},
          doi = {10.1086/381701},
archivePrefix = {arXiv},
       eprint = {astro-ph/0312054},
 primaryClass = {astro-ph},
       adsurl = {https://ui.adsabs.harvard.edu/abs/2004ApJ...603..697Y}
}

@ARTICLE{DaCosta+2009,
       author = {{Da Costa}, G.~S. and {Held}, E.~V. and {Saviane}, I. and {Gullieuszik}, M.},
        title = "{M22: An [Fe/H] Abundance Range Revealed}",
      journal = {\apj},
         year = 2009,
        month = nov,
       volume = {705},
       number = {2},
        pages = {1481-1491},
          doi = {10.1088/0004-637X/705/2/1481},
archivePrefix = {arXiv},
       eprint = {0909.5265},
 primaryClass = {astro-ph.GA},
       adsurl = {https://ui.adsabs.harvard.edu/abs/2009ApJ...705.1481D}
}

@INPROCEEDINGS{Peterson_1980,
       author = {{Peterson}, R.~C.},
        title = "{Evidence for primordial inhomogeneities from abundance of giants in M 5, M 13, and M 22.}",
    booktitle = {Star Clusters},
         year = 1980,
       editor = {{Hesser}, J.~E.},
       volume = {85},
        month = jan,
        series = {IAU Symposium},
        pages = {461-464},
       adsurl = {https://ui.adsabs.harvard.edu/abs/1980IAUS...85..461P}
}

@ARTICLE{Norris_DaCosta1995,
       author = {{Norris}, John E. and {Da Costa}, G.~S.},
        title = "{The Giant Branch of omega Centauri. IV. Abundance Patterns Based on Echelle Spectra of 40 Red Giants}",
      journal = {\apj},
         year = 1995,
        month = jul,
       volume = {447},
        pages = {680},
          doi = {10.1086/175909},
       adsurl = {https://ui.adsabs.harvard.edu/abs/1995ApJ...447..680N}
}

@ARTICLE{Ferraro+2009,
       author = {{Ferraro}, F.~R. and {Dalessandro}, E. and {Mucciarelli}, A. and
         {Beccari}, G. and {Rich}, R.~M. and {Origlia}, L. and {Lanzoni}, B. and
         {Rood}, R.~T. and {Valenti}, E. and {Bellazzini}, M. and
         {Ransom}, S.~M. and {Cocozza}, G.},
        title = "{The cluster Terzan 5 as a remnant of a primordial building block of the Galactic bulge}",
      journal = {\nat},
         year = 2009,
        month = nov,
       volume = {462},
       number = {7272},
        pages = {483-486},
          doi = {10.1038/nature08581},
archivePrefix = {arXiv},
       eprint = {0912.0192},
 primaryClass = {astro-ph.GA},
       adsurl = {https://ui.adsabs.harvard.edu/abs/2009Natur.462..483F}
}

@ARTICLE{Carretta+2010_m54,
       author = {{Carretta}, E. and {Bragaglia}, A. and {Gratton}, R.~G. and {Lucatello}, S. and {Bellazzini}, M. and {Catanzaro}, G. and {Leone}, F. and {Momany}, Y. and {Piotto}, G. and {D'Orazi}, V.},
        title = "{Detailed abundances of a large sample of giant stars in M 54 and in the Sagittarius nucleus}",
      journal = {\aap},
         year = 2010,
        month = sep,
       volume = {520},
          eid = {A95},
        pages = {A95},
          doi = {10.1051/0004-6361/201014924},
archivePrefix = {arXiv},
       eprint = {1006.5866},
 primaryClass = {astro-ph.GA},
       adsurl = {https://ui.adsabs.harvard.edu/abs/2010A&A...520A..95C}
}

@ARTICLE{Yong+2014_m2,
       author = {{Yong}, David and {Roederer}, Ian U. and {Grundahl}, Frank and {Da Costa}, Gary S. and {Karakas}, Amanda I. and {Norris}, John E. and {Aoki}, Wako and {Fishlock}, Cherie K. and {Marino}, A.~F. and {Milone}, A.~P. and {Shingles}, Luke J.},
        title = "{Iron and neutron-capture element abundance variations in the globular cluster M2 (NGC 7089)$^{★}$}",
      journal = {\mnras},
         year = 2014,
        month = jul,
       volume = {441},
       number = {4},
        pages = {3396-3416},
          doi = {10.1093/mnras/stu806},
archivePrefix = {arXiv},
       eprint = {1404.6873},
 primaryClass = {astro-ph.SR},
       adsurl = {https://ui.adsabs.harvard.edu/abs/2014MNRAS.441.3396Y}
}

@ARTICLE{Lehnert+1991,
       author = {{Lehnert}, M.~D. and {Bell}, R.~A. and {Cohen}, J.~G.},
        title = "{Abundances in the Red Giants of M13 and M22}",
      journal = {\apj},
         year = 1991,
        month = feb,
       volume = {367},
        pages = {514},
          doi = {10.1086/169648},
       adsurl = {https://ui.adsabs.harvard.edu/abs/1991ApJ...367..514L}
}

@ARTICLE{Meszaros+2020,
       author = {{M{\'e}sz{\'a}ros}, Szabolcs and {Masseron}, Thomas and {Garc{\'\i}a-Hern{\'a}ndez}, D.~A. and {Allende Prieto}, Carlos and {Beers}, Timothy C. and {Bizyaev}, Dmitry and {Chojnowski}, Drew and {Cohen}, Roger E. and {Cunha}, Katia and {Dell'Agli}, Flavia and {Ebelke}, Garrett and {Fern{\'a}ndez-Trincado}, Jos{\'e} G. and {Frinchaboy}, Peter and {Geisler}, Doug and {Hasselquist}, Sten and {Hearty}, Fred and {Holtzman}, Jon and {Johnson}, Jennifer and {Lane}, Richard R. and {Lacerna}, Ivan and {Longa-Pe{\~n}a}, Penelop{\'e} and {Majewski}, Steven R. and {Martell}, Sarah L. and {Minniti}, Dante and {Nataf}, David and {Nidever}, David L. and {Pan}, Kaike and {Schiavon}, Ricardo P. and {Shetrone}, Matthew and {Smith}, Verne V. and {Sobeck}, Jennifer S. and {Stringfellow}, Guy S. and {Szigeti}, L{\'a}szl{\'o} and {Tang}, Baitian and {Wilson}, John C. and {Zamora}, Olga},
        title = "{Homogeneous analysis of globular clusters from the APOGEE survey with the BACCHUS code - II. The Southern clusters and overview}",
      journal = {\mnras},
         year = 2020,
        month = feb,
       volume = {492},
       number = {2},
        pages = {1641-1670},
          doi = {10.1093/mnras/stz3496},
archivePrefix = {arXiv},
       eprint = {1912.04839},
 primaryClass = {astro-ph.SR},
       adsurl = {https://ui.adsabs.harvard.edu/abs/2020MNRAS.492.1641M}
}

@INBOOK{Renzini_Buzzoni1986,
       author = {{Renzini}, Alvio and {Buzzoni}, Alberto},
        title = "{Global properties of stellar populations and the spectral evolution of galaxies.}",
    booktitle = {Spectral Evolution of Galaxies},
         year = 1986,
       publisher = {Dordrecht, D. Reidel Publishing Co.},
       volume = {122},
        pages = {195-231},
          doi = {10.1007/978-94-009-4598-2\_19},
       adsurl = {https://ui.adsabs.harvard.edu/abs/1986ASSL..122..195R}
}

@ARTICLE{McKenzie_Bekki2018,
       author = {{McKenzie}, M. and {Bekki}, K.},
        title = "{A new model for the multiple stellar populations within Terzan 5}",
      journal = {\mnras},
         year = 2018,
        month = sep,
       volume = {479},
       number = {3},
        pages = {3126-3141},
          doi = {10.1093/mnras/sty1557},
archivePrefix = {arXiv},
       eprint = {1806.04824},
 primaryClass = {astro-ph.GA},
       adsurl = {https://ui.adsabs.harvard.edu/abs/2018MNRAS.479.3126M}
}

@ARTICLE{McKenzie_Bekki2021,
       author = {{McKenzie}, Madeleine and {Bekki}, Kenji},
        title = "{Simulations of globular clusters within their parent galaxies: multiple stellar populations and internal kinematics}",
      journal = {\mnras},
         year = 2021,
        month = jan,
       volume = {500},
       number = {4},
        pages = {4578-4596},
          doi = {10.1093/mnras/staa3376},
archivePrefix = {arXiv},
       eprint = {2101.02348},
 primaryClass = {astro-ph.GA},
       adsurl = {https://ui.adsabs.harvard.edu/abs/2021MNRAS.500.4578M}
}

@ARTICLE{McKenzie2022,
       author = {{McKenzie}, M. and {Yong}, D. and {Marino}, A.~F. and {Monty}, S. and {Wang}, E. and {Karakas}, A.~I. and {Milone}, A.~P. and {Legnardi}, M.~V. and {Roederer}, I.~U. and {Martell}, S. and {Horta}, D.},
        title = "{The complex stellar system M 22: confirming abundance variations with high precision differential measurements}",
      journal = {\mnras},
         year = 2022,
        month = nov,
       volume = {516},
       number = {3},
        pages = {3515-3531},
          doi = {10.1093/mnras/stac2254},
       adsurl = {https://ui.adsabs.harvard.edu/abs/2022MNRAS.516.3515M}
}

@ARTICLE{Bastian_Lardo2018,
       author = {{Bastian}, Nate and {Lardo}, Carmela},
        title = "{Multiple Stellar Populations in Globular Clusters}",
      journal = {\araa},
         year = 2018,
        month = sep,
       volume = {56},
        pages = {83-136},
          doi = {10.1146/annurev-astro-081817-051839},
archivePrefix = {arXiv},
       eprint = {1712.01286},
 primaryClass = {astro-ph.SR},
       adsurl = {https://ui.adsabs.harvard.edu/abs/2018ARA&A..56...83B}
}

@ARTICLE{Cottrell_DaCosta_1981,
       author = {{Cottrell}, P.~L. and {Da Costa}, G.~S.},
        title = "{Correlated cyanogen and sodium anomalies in the globular clusters 47 TUC and NGC 6752.}",
      journal = {\apjl},
         year = 1981,
        month = apr,
       volume = {245},
        pages = {L79-L82},
          doi = {10.1086/183527},
       adsurl = {https://ui.adsabs.harvard.edu/abs/1981ApJ...245L..79C}
}

@ARTICLE{Milone+2017MSP,
       author = {{Milone}, A.~P. and {Piotto}, G. and {Renzini}, A. and {Marino}, A.~F. and
         {Bedin}, L.~R. and {Vesperini}, E. and {D'Antona}, F. and
         {Nardiello}, D. and {Anderson}, J. and {King}, I.~R. and {Yong}, D. and
         {Bellini}, A. and {Aparicio}, A. and {Barbuy}, B. and {Brown}, T.~M. and
         {Cassisi}, S. and {Ortolani}, S. and {Salaris}, M. and
         {Sarajedini}, A. and {van der Marel}, R.~P.},
        title = "{The Hubble Space Telescope UV Legacy Survey of Galactic globular clusters - IX. The Atlas of multiple stellar populations}",
      journal = {\mnras},
         year = 2017,
        month = jan,
       volume = {464},
       number = {3},
        pages = {3636-3656},
          doi = {10.1093/mnras/stw2531},
archivePrefix = {arXiv},
       eprint = {1610.00451},
 primaryClass = {astro-ph.SR},
       adsurl = {https://ui.adsabs.harvard.edu/abs/2017MNRAS.464.3636M}
}

@ARTICLE{Nissen_Gustafsson2018,
       author = {{Nissen}, Poul Erik and {Gustafsson}, Bengt},
        title = "{High-precision stellar abundances of the elements: methods and applications}",
      journal = {\aapr},
         year = 2018,
        month = oct,
       volume = {26},
       number = {1},
          eid = {6},
        pages = {6},
          doi = {10.1007/s00159-018-0111-3},
archivePrefix = {arXiv},
       eprint = {1810.06535},
 primaryClass = {astro-ph.SR},
       adsurl = {https://ui.adsabs.harvard.edu/abs/2018A&ARv..26....6N}
}

@INPROCEEDINGS{Castelli_Kurucz2003,
       author = {{Castelli}, F. and {Kurucz}, R.~L.},
        title = "{New Grids of ATLAS9 Model Atmospheres}",
    booktitle = {Modelling of Stellar Atmospheres},
         year = 2003,
       editor = {{Piskunov}, N. and {Weiss}, W.~W. and {Gray}, D.~F.},
       volume = {210},
        month = jan,
        pages = {A20},
        series= {},
archivePrefix = {arXiv},
       eprint = {astro-ph/0405087},
 primaryClass = {astro-ph},
       adsurl = {https://ui.adsabs.harvard.edu/abs/2003IAUS..210P.A20C}
}

@ARTICLE{Kraft_1994,
       author = {{Kraft}, Robert P.},
        title = "{Abundance Differences among Globular Cluster Giants: Primordial vs. Evolutionary Scenarios}",
      journal = {\pasp},
         year = 1994,
        month = jun,
       volume = {106},
        pages = {553},
          doi = {10.1086/133416},
       adsurl = {https://ui.adsabs.harvard.edu/abs/1994PASP..106..553K}
}

@ARTICLE{Gratton+2012,
       author = {{Gratton}, Raffaele G. and {Carretta}, Eugenio and {Bragaglia}, Angela},
        title = "{Multiple populations in globular clusters. Lessons learned from the Milky Way globular clusters}",
      journal = {\aapr},
         year = 2012,
        month = feb,
       volume = {20},
          eid = {50},
        pages = {50},
          doi = {10.1007/s00159-012-0050-3},
archivePrefix = {arXiv},
       eprint = {1201.6526},
 primaryClass = {astro-ph.SR},
       adsurl = {https://ui.adsabs.harvard.edu/abs/2012A&ARv..20...50G}
}

@ARTICLE{Gratton+2019,
       author = {{Gratton}, Raffaele and {Bragaglia}, Angela and {Carretta}, Eugenio and {D'Orazi}, Valentina and {Lucatello}, Sara and {Sollima}, Antonio},
        title = "{What is a globular cluster? An observational perspective}",
      journal = {\aapr},
         year = 2019,
        month = nov,
       volume = {27},
       number = {1},
          eid = {8},
        pages = {8},
          doi = {10.1007/s00159-019-0119-3},
archivePrefix = {arXiv},
       eprint = {1911.02835},
 primaryClass = {astro-ph.SR},
       adsurl = {https://ui.adsabs.harvard.edu/abs/2019A&ARv..27....8G}
}

@ARTICLE{Busso+1999,
       author = {{Busso}, M. and {Gallino}, R. and {Wasserburg}, G.~J.},
        title = "{Nucleosynthesis in Asymptotic Giant Branch Stars: Relevance for Galactic Enrichment and Solar System Formation}",
      journal = {\araa},
         year = 1999,
        month = jan,
       volume = {37},
        pages = {239-309},
          doi = {10.1146/annurev.astro.37.1.239},
       adsurl = {https://ui.adsabs.harvard.edu/abs/1999ARA&A..37..239B}
}

@ARTICLE{Karakas_Lattanzio_2014,
       author = {{Karakas}, Amanda I. and {Lattanzio}, John C.},
        title = "{The Dawes Review 2: Nucleosynthesis and Stellar Yields of Low- and Intermediate-Mass Single Stars}",
      journal = {\pasa},
         year = 2014,
        month = jul,
       volume = {31},
          eid = {e030},
        pages = {e030},
          doi = {10.1017/pasa.2014.21},
archivePrefix = {arXiv},
       eprint = {1405.0062},
 primaryClass = {astro-ph.SR},
       adsurl = {https://ui.adsabs.harvard.edu/abs/2014PASA...31...30K}
}

@ARTICLE{Carretta+2009,
       author = {{Carretta}, E. and {Bragaglia}, A. and {Gratton}, R. and {D'Orazi}, V. and {Lucatello}, S.},
        title = "{Intrinsic iron spread and a new metallicity scale for globular clusters}",
      journal = {\aap},
         year = 2009,
        month = dec,
       volume = {508},
       number = {2},
        pages = {695-706},
          doi = {10.1051/0004-6361/200913003},
archivePrefix = {arXiv},
       eprint = {0910.0675},
 primaryClass = {astro-ph.GA},
       adsurl = {https://ui.adsabs.harvard.edu/abs/2009A&A...508..695C}
}

@ARTICLE{Lee_2016,
       author = {{Lee}, Jae-Woo},
        title = "{On the Metallicity Distribution of the Peculiar Globular Cluster M22}",
      journal = {\apjs},
         year = 2016,
        month = oct,
       volume = {226},
       number = {2},
          eid = {16},
        pages = {16},
          doi = {10.3847/0067-0049/226/2/16},
archivePrefix = {arXiv},
       eprint = {1608.08297},
 primaryClass = {astro-ph.GA},
       adsurl = {https://ui.adsabs.harvard.edu/abs/2016ApJS..226...16L}
}

@ARTICLE{Pilachowski+1982,
       author = {{Pilachowski}, C. and {Leep}, E.~M. and {Wallerstein}, G. and {Peterson}, R.~C.},
        title = "{Abundances of the elements in six stars in the globular cluster M 22.}",
      journal = {\apj},
         year = 1982,
        month = dec,
       volume = {263},
        pages = {187-198},
          doi = {10.1086/160493},
       adsurl = {https://ui.adsabs.harvard.edu/abs/1982ApJ...263..187P}
}

@ARTICLE{Brown+1990,
       author = {{Brown}, Jeffery A. and {Wallerstein}, George and {Oke}, J.~B.},
        title = "{High-Resolution CCD Spectra of Stars in Globular Clusters V: Carbon, Nitrogen, and Oxygen Abundances in Stars in 47 Tuc, M4, and M22}",
      journal = {\aj},
         year = 1990,
        month = nov,
       volume = {100},
        pages = {1561},
          doi = {10.1086/115617},
       adsurl = {https://ui.adsabs.harvard.edu/abs/1990AJ....100.1561B}
}

@ARTICLE{Brown_Wallerstein1992,
       author = {{Brown}, Jeffery A. and {Wallerstein}, George},
        title = "{High-Resolution CCD Spectra of Stars in Globular Clusters. VII. Abundances of 16 Elements in 47 Tuc, M4, and M22}",
      journal = {\aj},
         year = 1992,
        month = nov,
       volume = {104},
        pages = {1818},
          doi = {10.1086/116360},
       adsurl = {https://ui.adsabs.harvard.edu/abs/1992AJ....104.1818B}
}

@ARTICLE{Placco+2021,
       author = {{Placco}, Vinicius M. and {Sneden}, Christopher and {Roederer}, Ian U. and {Lawler}, James E. and {Den Hartog}, Elizabeth A. and {Hejazi}, Neda and {Maas}, Zachary and {Bernath}, Peter},
        title = "{Linemake: An Atomic and Molecular Line List Generator}",
      journal = {Research Notes of the American Astronomical Society},
         year = 2021,
        month = apr,
       volume = {5},
       number = {4},
          eid = {92},
        pages = {92},
          doi = {10.3847/2515-5172/abf651},
archivePrefix = {arXiv},
       eprint = {2104.08286},
 primaryClass = {astro-ph.IM},
       adsurl = {https://ui.adsabs.harvard.edu/abs/2021RNAAS...5...92P}
}

@ARTICLE{Lee_2020,
       author = {{Lee}, Jae-Woo},
        title = "{Five Stellar Populations in M22 (NGC 6656)}",
      journal = {\apjl},
         year = 2020,
        month = jan,
       volume = {888},
       number = {1},
          eid = {L6},
        pages = {L6},
          doi = {10.3847/2041-8213/ab60b2},
archivePrefix = {arXiv},
       eprint = {2001.00679},
 primaryClass = {astro-ph.GA},
       adsurl = {https://ui.adsabs.harvard.edu/abs/2020ApJ...888L...6L}
}

@ARTICLE{Lee+2009,
       author = {{Lee}, Jae-Woo and {Kang}, Young-Woon and {Lee}, Jina and {Lee}, Young-Wook},
        title = "{Enrichment by supernovae in globular clusters with multiple populations}",
      journal = {\nat},
         year = 2009,
        month = nov,
       volume = {462},
       number = {7272},
        pages = {480-482},
          doi = {10.1038/nature08565},
archivePrefix = {arXiv},
       eprint = {0911.4798},
 primaryClass = {astro-ph.GA},
       adsurl = {https://ui.adsabs.harvard.edu/abs/2009Natur.462..480L}
}

@ARTICLE{Joo_Lee2013,
       author = {{Joo}, Seok-Joo and {Lee}, Young-Wook},
        title = "{Star Formation Histories of Globular Clusters with Multiple Populations. I. {\ensuremath{\omega}} CEN, M22, and NGC 1851}",
      journal = {\apj},
         year = 2013,
        month = jan,
       volume = {762},
       number = {1},
          eid = {36},
        pages = {36},
          doi = {10.1088/0004-637X/762/1/36},
archivePrefix = {arXiv},
       eprint = {1211.1688},
 primaryClass = {astro-ph.GA},
       adsurl = {https://ui.adsabs.harvard.edu/abs/2013ApJ...762...36J}
}

@ARTICLE{Gratton+2014,
       author = {{Gratton}, R.~G. and {Lucatello}, S. and {Sollima}, A. and {Carretta}, E. and {Bragaglia}, A. and {Momany}, Y. and {D'Orazi}, V. and {Cassisi}, S. and {Salaris}, M.},
        title = "{The Na-O anticorrelation in horizontal branch stars. IV. M 22}",
      journal = {\aap},
         year = 2014,
        month = mar,
       volume = {563},
          eid = {A13},
        pages = {A13},
          doi = {10.1051/0004-6361/201323101},
archivePrefix = {arXiv},
       eprint = {1401.7109},
 primaryClass = {astro-ph.SR},
       adsurl = {https://ui.adsabs.harvard.edu/abs/2014A&A...563A..13G}
}

@ARTICLE{Lim+2015,
       author = {{Lim}, Dongwook and {Han}, Sang-Il and {Lee}, Young-Wook and {Roh}, Dong-Goo and {Sohn}, Young-Jong and {Chun}, Sang-Hyun and {Lee}, Jae-Woo and {Johnson}, Christian I.},
        title = "{Low-resolution Spectroscopy for the Globular Clusters with Signs of Supernova Enrichment: M22, NGC 1851, and NGC 288}",
      journal = {\apjs},
         year = 2015,
        month = jan,
       volume = {216},
       number = {1},
          eid = {19},
        pages = {19},
          doi = {10.1088/0067-0049/216/1/19},
archivePrefix = {arXiv},
       eprint = {1412.1832},
 primaryClass = {astro-ph.GA},
       adsurl = {https://ui.adsabs.harvard.edu/abs/2015ApJS..216...19L}
}

@INPROCEEDINGS{Dekker+2000,
       author = {{Dekker}, Hans and {D'Odorico}, Sandro and {Kaufer}, Andreas and {Delabre}, Bernard and {Kotzlowski}, Heinz},
        title = "{Design, construction, and performance of UVES, the echelle spectrograph for the UT2 Kueyen Telescope at the ESO Paranal Observatory}",
    booktitle = {Optical and IR Telescope Instrumentation and Detectors},
         year = 2000,
       editor = {{Iye}, Masanori and {Moorwood}, Alan F.},
       series = {Society of Photo-Optical Instrumentation Engineers (SPIE) Conference Series},
       volume = {4008},
        month = aug,
        pages = {534-545},
          doi = {10.1117/12.395512},
       adsurl = {https://ui.adsabs.harvard.edu/abs/2000SPIE.4008..534D}
}

@ARTICLE{Melendez+2009,
       author = {{Mel{\'e}ndez}, J. and {Asplund}, M. and {Gustafsson}, B. and {Yong}, D.},
        title = "{The Peculiar Solar Composition and Its Possible Relation to Planet Formation}",
      journal = {\apjl},
         year = 2009,
        month = oct,
       volume = {704},
       number = {1},
        pages = {L66-L70},
          doi = {10.1088/0004-637X/704/1/L66},
archivePrefix = {arXiv},
       eprint = {0909.2299},
 primaryClass = {astro-ph.SR},
       adsurl = {https://ui.adsabs.harvard.edu/abs/2009ApJ...704L..66M}
}

@ARTICLE{Johnson_Pilachowski2010,
       author = {{Johnson}, Christian I. and {Pilachowski}, Catherine A.},
        title = "{Chemical Abundances for 855 Giants in the Globular Cluster Omega Centauri (NGC 5139)}",
      journal = {\apj},
         year = 2010,
        month = oct,
       volume = {722},
       number = {2},
        pages = {1373-1410},
          doi = {10.1088/0004-637X/722/2/1373},
archivePrefix = {arXiv},
       eprint = {1008.2232},
 primaryClass = {astro-ph.SR},
       adsurl = {https://ui.adsabs.harvard.edu/abs/2010ApJ...722.1373J}
}

@ARTICLE{DErcole+2008,
       author = {{D'Ercole}, Annibale and {Vesperini}, Enrico and {D'Antona}, Francesca and {McMillan}, Stephen L.~W. and {Recchi}, Simone},
        title = "{Formation and dynamical evolution of multiple stellar generations in globular clusters}",
      journal = {\mnras},
         year = 2008,
        month = dec,
       volume = {391},
       number = {2},
        pages = {825-843},
          doi = {10.1111/j.1365-2966.2008.13915.x},
archivePrefix = {arXiv},
       eprint = {0809.1438},
 primaryClass = {astro-ph},
       adsurl = {https://ui.adsabs.harvard.edu/abs/2008MNRAS.391..825D}
}

@ARTICLE{Renzini+2015,
       author = {{Renzini}, A. and {D'Antona}, F. and {Cassisi}, S. and {King}, I.~R. and {Milone}, A.~P. and {Ventura}, P. and {Anderson}, J. and {Bedin}, L.~R. and {Bellini}, A. and {Brown}, T.~M. and {Piotto}, G. and {van der Marel}, R.~P. and {Barbuy}, B. and {Dalessandro}, E. and {Hidalgo}, S. and {Marino}, A.~F. and {Ortolani}, S. and {Salaris}, M. and {Sarajedini}, A.},
        title = "{The Hubble Space Telescope UV Legacy Survey of Galactic Globular Clusters - V. Constraints on formation scenarios}",
      journal = {\mnras},
         year = 2015,
        month = dec,
       volume = {454},
       number = {4},
        pages = {4197-4207},
          doi = {10.1093/mnras/stv2268},
archivePrefix = {arXiv},
       eprint = {1510.01468},
 primaryClass = {astro-ph.GA},
       adsurl = {https://ui.adsabs.harvard.edu/abs/2015MNRAS.454.4197R}
}

@ARTICLE{Lacchin+2021,
       author = {{Lacchin}, E. and {Calura}, F. and {Vesperini}, E.},
        title = "{On the role of Type Ia supernovae in the second-generation star formation in globular clusters}",
      journal = {\mnras},
         year = 2021,
        month = oct,
       volume = {506},
       number = {4},
        pages = {5951-5968},
          doi = {10.1093/mnras/stab2061},
archivePrefix = {arXiv},
       eprint = {2107.07962},
 primaryClass = {astro-ph.GA},
       adsurl = {https://ui.adsabs.harvard.edu/abs/2021MNRAS.506.5951L}
}

@ARTICLE{DAntona+2016,
       author = {{D'Antona}, F. and {Vesperini}, E. and {D'Ercole}, A. and {Ventura}, P. and {Milone}, A.~P. and {Marino}, A.~F. and {Tailo}, M.},
        title = "{A single model for the variety of multiple-population formation(s) in globular clusters: a temporal sequence}",
      journal = {\mnras},
         year = 2016,
        month = may,
       volume = {458},
       number = {2},
        pages = {2122-2139},
          doi = {10.1093/mnras/stw387},
archivePrefix = {arXiv},
       eprint = {1602.05412},
 primaryClass = {astro-ph.GA},
       adsurl = {https://ui.adsabs.harvard.edu/abs/2016MNRAS.458.2122D}
}

@INPROCEEDINGS{DaCosta_2016,
       author = {{Da Costa}, G.~S.},
        title = "{Are the globular clusters with significant internal [Fe/H] spreads all former dwarf galaxy nuclei?}",
    booktitle = {The General Assembly of Galaxy Halos: Structure, Origin and Evolution},
         year = 2016,
       editor = {{Bragaglia}, A. and {Arnaboldi}, M. and {Rejkuba}, M. and {Romano}, D.},
       volume = {317},
       series = {},
        month = aug,
        pages = {110-115},
          doi = {10.1017/S174392131500678X},
archivePrefix = {arXiv},
       eprint = {1510.00873},
 primaryClass = {astro-ph.GA},
       adsurl = {https://ui.adsabs.harvard.edu/abs/2016IAUS..317..110D}
}

@ARTICLE{Sneden+1997,
       author = {{Sneden}, Christopher and {Kraft}, Robert P. and {Shetrone}, Matthew D. and {Smith}, Graeme H. and {Langer}, G.~E. and {Prosser}, Charles F.},
        title = "{Star-To-Star Abundance Variations Among Bright Giants in the Metal-Poor Globular Cluster M15}",
      journal = {\aj},
         year = 1997,
        month = nov,
       volume = {114},
        pages = {1964},
          doi = {10.1086/118618},
       adsurl = {https://ui.adsabs.harvard.edu/abs/1997AJ....114.1964S}
}

@ARTICLE{Ventura+2009,
       author = {{Ventura}, P. and {Caloi}, V. and {D'Antona}, F. and {Ferguson}, J. and {Milone}, A. and {Piotto}, G.~P.},
        title = "{The C+N+O abundances and the splitting of the subgiant branch in the globular cluster NGC 1851}",
      journal = {\mnras},
         year = 2009,
        month = oct,
       volume = {399},
       number = {2},
        pages = {934-943},
          doi = {10.1111/j.1365-2966.2009.15335.x},
archivePrefix = {arXiv},
       eprint = {0907.1765},
 primaryClass = {astro-ph.SR},
       adsurl = {https://ui.adsabs.harvard.edu/abs/2009MNRAS.399..934V}
}

@ARTICLE{Reggiani+2017,
       author = {{Reggiani}, Henrique and {Mel{\'e}ndez}, Jorge and {Kobayashi}, Chiaki and {Karakas}, Amanda and {Placco}, Vinicius},
        title = "{Constraining cosmic scatter in the Galactic halo through a differential analysis of metal-poor stars}",
      journal = {\aap},
         year = 2017,
        month = dec,
       volume = {608},
          eid = {A46},
        pages = {A46},
          doi = {10.1051/0004-6361/201730750},
archivePrefix = {arXiv},
       eprint = {1709.03750},
 primaryClass = {astro-ph.SR},
       adsurl = {https://ui.adsabs.harvard.edu/abs/2017A&A...608A..46R}
}

@ARTICLE{DeBievre_Barnes1985,
       author = {{De Bi{\`e}vre}, P. and {Barnes}, I.~L.},
        title = "{Table of the isotopic composition of the elements as determined by mass spectrometry}",
      journal = {International Journal of Mass Spectrometry and Ion Processes},
         year = 1985,
        month = may,
       volume = {65},
       number = {1-2},
        pages = {211-230},
          doi = {10.1016/0168-1176(85)85065-5},
       adsurl = {https://ui.adsabs.harvard.edu/abs/1985IJMSI..65..211D}
}

@ARTICLE{Carlos+2018,
       author = {{Carlos}, Mar{\'\i}lia and {Karakas}, Amanda I. and {Cohen}, Judith G. and {Kobayashi}, Chiaki and {Mel{\'e}ndez}, Jorge},
        title = "{A Formation Timescale of the Galactic Halo from Mg Isotopes in Dwarf Stars}",
      journal = {\apj},
         year = 2018,
        month = apr,
       volume = {856},
       number = {2},
          eid = {161},
        pages = {161},
          doi = {10.3847/1538-4357/aab36c},
archivePrefix = {arXiv},
       eprint = {1804.01280},
 primaryClass = {astro-ph.SR},
       adsurl = {https://ui.adsabs.harvard.edu/abs/2018ApJ...856..161C}
}

@ARTICLE{Karakas_Lattanzio2003,
       author = {{Karakas}, A.~I. and {Lattanzio}, J.~C.},
        title = "{Production of Aluminium and the Heavy Magnesium Isotopes in Asymptotic Giant Branch Stars}",
      journal = {\pasa},
         year = 2003,
        month = jan,
       volume = {20},
       number = {3},
        pages = {279-293},
          doi = {10.1071/AS03010},
       adsurl = {https://ui.adsabs.harvard.edu/abs/2003PASA...20..279K}
}

@ARTICLE{Fishlock+2014,
       author = {{Fishlock}, Cherie K. and {Karakas}, Amanda I. and {Lugaro}, Maria and {Yong}, David},
        title = "{Evolution and Nucleosynthesis of Asymptotic Giant Branch Stellar Models of Low Metallicity}",
      journal = {\apj},
         year = 2014,
        month = dec,
       volume = {797},
       number = {1},
          eid = {44},
        pages = {44},
          doi = {10.1088/0004-637X/797/1/44},
archivePrefix = {arXiv},
       eprint = {1410.7457},
 primaryClass = {astro-ph.SR},
       adsurl = {https://ui.adsabs.harvard.edu/abs/2014ApJ...797...44F}
}

@ARTICLE{DErcole+2016,
       author = {{D'Ercole}, A. and {D'Antona}, F. and {Vesperini}, E.},
        title = "{Accretion of pristine gas and dilution during the formation of multiple-population globular clusters}",
      journal = {\mnras},
         year = 2016,
        month = oct,
       volume = {461},
       number = {4},
        pages = {4088-4098},
          doi = {10.1093/mnras/stw1583},
archivePrefix = {arXiv},
       eprint = {1607.00951},
 primaryClass = {astro-ph.GA},
       adsurl = {https://ui.adsabs.harvard.edu/abs/2016MNRAS.461.4088D}
}

@ARTICLE{VandenBerg+2013,
       author = {{VandenBerg}, Don A. and {Brogaard}, K. and {Leaman}, R. and {Casagrande}, L.},
        title = "{The Ages of 55 Globular Clusters as Determined Using an Improved \textbackslashDelta V\^HB\_TO Method along with Color-Magnitude Diagram Constraints, and Their Implications for Broader Issues}",
      journal = {\apj},
         year = 2013,
        month = oct,
       volume = {775},
       number = {2},
          eid = {134},
        pages = {134},
          doi = {10.1088/0004-637X/775/2/134},
archivePrefix = {arXiv},
       eprint = {1308.2257},
 primaryClass = {astro-ph.GA},
       adsurl = {https://ui.adsabs.harvard.edu/abs/2013ApJ...775..134V}
}

@ARTICLE{Yong+2006,
       author = {{Yong}, David and {Aoki}, Wako and {Lambert}, David L.},
        title = "{Mg Isotope Ratios in Giant Stars of the Globular Clusters M13 and M71}",
      journal = {\apj},
         year = 2006,
        month = feb,
       volume = {638},
       number = {2},
        pages = {1018-1027},
          doi = {10.1086/498974},
archivePrefix = {arXiv},
       eprint = {astro-ph/0510591},
 primaryClass = {astro-ph},
       adsurl = {https://ui.adsabs.harvard.edu/abs/2006ApJ...638.1018Y}
}

@ARTICLE{Piotto+2015,
       author = {{Piotto}, G. and {Milone}, A.~P. and {Bedin}, L.~R. and {Anderson}, J. and {King}, I.~R. and {Marino}, A.~F. and {Nardiello}, D. and {Aparicio}, A. and {Barbuy}, B. and {Bellini}, A. and {Brown}, T.~M. and {Cassisi}, S. and {Cool}, A.~M. and {Cunial}, A. and {Dalessandro}, E. and {D'Antona}, F. and {Ferraro}, F.~R. and {Hidalgo}, S. and {Lanzoni}, B. and {Monelli}, M. and {Ortolani}, S. and {Renzini}, A. and {Salaris}, M. and {Sarajedini}, A. and {van der Marel}, R.~P. and {Vesperini}, E. and {Zoccali}, M.},
        title = "{The Hubble Space Telescope UV Legacy Survey of Galactic Globular Clusters. I. Overview of the Project and Detection of Multiple Stellar Populations}",
      journal = {\aj},
         year = 2015,
        month = mar,
       volume = {149},
       number = {3},
          eid = {91},
        pages = {91},
          doi = {10.1088/0004-6256/149/3/91},
archivePrefix = {arXiv},
       eprint = {1410.4564},
 primaryClass = {astro-ph.SR},
       adsurl = {https://ui.adsabs.harvard.edu/abs/2015AJ....149...91P}
}

@ARTICLE{Gay_Lambert2000,
       author = {{Gay}, Pamela L. and {Lambert}, David L.},
        title = "{The Isotopic Abundances of Magnesium in Stars}",
      journal = {\apj},
         year = 2000,
        month = apr,
       volume = {533},
       number = {1},
        pages = {260-270},
          doi = {10.1086/308653},
archivePrefix = {arXiv},
       eprint = {astro-ph/9911217},
 primaryClass = {astro-ph},
       adsurl = {https://ui.adsabs.harvard.edu/abs/2000ApJ...533..260G}
}

@ARTICLE{McWilliam_Lambert1988,
       author = {{McWilliam}, Andrew and {Lambert}, David L.},
        title = "{Isotopic magnesium abundances in stars.}",
      journal = {\mnras},
         year = 1988,
        month = feb,
       volume = {230},
        pages = {573-585},
          doi = {10.1093/mnras/230.4.573},
       adsurl = {https://ui.adsabs.harvard.edu/abs/1988MNRAS.230..573M}
}

@ARTICLE{Hinkle+2013,
       author = {{Hinkle}, Kenneth H. and {Wallace}, Lloyd and {Ram}, Ram S. and {Bernath}, Peter F. and {Sneden}, Christopher and {Lucatello}, Sara},
        title = "{The Magnesium Isotopologues of MgH in the A $^{2}${\ensuremath{\Pi}}-X $^{2}${\ensuremath{\Sigma}}$^{+}$ System}",
      journal = {\apjs},
         year = 2013,
        month = aug,
       volume = {207},
       number = {2},
          eid = {26},
        pages = {26},
          doi = {10.1088/0067-0049/207/2/26},
       adsurl = {https://ui.adsabs.harvard.edu/abs/2013ApJS..207...26H}
}

@ARTICLE{Sobeck+2011,
       author = {{Sobeck}, Jennifer S. and {Kraft}, Robert P. and {Sneden}, Christopher and {Preston}, George W. and {Cowan}, John J. and {Smith}, Graeme H. and {Thompson}, Ian B. and {Shectman}, Stephen A. and {Burley}, Gregory S.},
        title = "{The Abundances of Neutron-capture Species in the Very Metal-poor Globular Cluster M15: A Uniform Analysis of Red Giant Branch and Red Horizontal Branch Stars}",
      journal = {\aj},
         year = 2011,
        month = jun,
       volume = {141},
       number = {6},
          eid = {175},
        pages = {175},
          doi = {10.1088/0004-6256/141/6/175},
archivePrefix = {arXiv},
       eprint = {1103.1008},
 primaryClass = {astro-ph.SR},
       adsurl = {https://ui.adsabs.harvard.edu/abs/2011AJ....141..175S}
}

@INPROCEEDINGS{Adamow_2017,
       author = {{Adamow}, Monika M.},
        title = "{pyMOOGi - python wrapper for MOOG}",
    booktitle = {American Astronomical Society Meeting Abstracts \#230},
         year = 2017,
       series = {American Astronomical Society Meeting Abstracts},
       volume = {230},
        month = jun,
          eid = {216.07},
        pages = {216.07},
       adsurl = {https://ui.adsabs.harvard.edu/abs/2017AAS...23021607A}
}

@BOOK{Bevington_Robinson1992,
       author = {{Bevington}, Philip R. and {Robinson}, D. Keith},
        title = "{Data reduction and error analysis for the physical sciences}",
         year = 1992,
       adsurl = {https://ui.adsabs.harvard.edu/abs/1992drea.book.....B}
}

@ARTICLE{Carney+2008,
       author = {{Carney}, Bruce W. and {Gray}, David F. and {Yong}, David and {Latham}, David W. and {Manset}, Nadine and {Zelman}, Rachel and {Laird}, John B.},
        title = "{Rotation and Macroturbulence in Metal-Poor Field Red Giant and Red Horizontal Branch Stars}",
      journal = {\aj},
         year = 2008,
        month = mar,
       volume = {135},
       number = {3},
        pages = {892-906},
          doi = {10.1088/0004-6256/135/3/892},
archivePrefix = {arXiv},
       eprint = {0711.4984},
 primaryClass = {astro-ph},
       adsurl = {https://ui.adsabs.harvard.edu/abs/2008AJ....135..892C}
}

@ARTICLE{Milone_Marino_2022,
       author = {{Milone}, Antonino P. and {Marino}, Anna F.},
        title = "{Multiple Populations in Star Clusters}",
      journal = {Universe},
         year = 2022,
        month = jun,
       volume = {8},
       number = {7},
        pages = {359},
          doi = {10.3390/universe8070359},
archivePrefix = {arXiv},
       eprint = {2206.10564},
 primaryClass = {astro-ph.GA},
       adsurl = {https://ui.adsabs.harvard.edu/abs/2022Univ....8..359M}
}

@ARTICLE{Carretta+2009UVES,
       author = {{Carretta}, E. and {Bragaglia}, A. and {Gratton}, R. and {Lucatello}, S.},
        title = "{Na-O anticorrelation and HB. VIII. Proton-capture elements and metallicities in 17 globular clusters from UVES spectra}",
      journal = {\aap},
         year = 2009,
        month = oct,
       volume = {505},
       number = {1},
        pages = {139-155},
          doi = {10.1051/0004-6361/200912097},
archivePrefix = {arXiv},
       eprint = {0909.2941},
 primaryClass = {astro-ph.GA},
       adsurl = {https://ui.adsabs.harvard.edu/abs/2009A&A...505..139C}
}

@ARTICLE{Carretta+2009GIRAFFE,
       author = {{Carretta}, E. and {Bragaglia}, A. and {Gratton}, R.~G. and {Lucatello}, S. and {Catanzaro}, G. and {Leone}, F. and {Bellazzini}, M. and {Claudi}, R. and {D'Orazi}, V. and {Momany}, Y. and {Ortolani}, S. and {Pancino}, E. and {Piotto}, G. and {Recio-Blanco}, A. and {Sabbi}, E.},
        title = "{Na-O anticorrelation and HB. VII. The chemical composition of first and second-generation stars in 15 globular clusters from GIRAFFE spectra}",
      journal = {\aap},
         year = 2009,
        month = oct,
       volume = {505},
       number = {1},
        pages = {117-138},
          doi = {10.1051/0004-6361/200912096},
archivePrefix = {arXiv},
       eprint = {0909.2938},
 primaryClass = {astro-ph.GA},
       adsurl = {https://ui.adsabs.harvard.edu/abs/2009A&A...505..117C}
}

@ARTICLE{Foreman-Mackey+2013,
       author = {{Foreman-Mackey}, Daniel and {Hogg}, David W. and {Lang}, Dustin and {Goodman}, Jonathan},
        title = "{emcee: The MCMC Hammer}",
      journal = {\pasp},
         year = 2013,
        month = mar,
       volume = {125},
       number = {925},
        pages = {306},
          doi = {10.1086/670067},
archivePrefix = {arXiv},
       eprint = {1202.3665},
 primaryClass = {astro-ph.IM},
       adsurl = {https://ui.adsabs.harvard.edu/abs/2013PASP..125..306F}
}

@ARTICLE{Hinton2016,
   author = {{Hinton}, S.~R.},
    title = "{ChainConsumer}",
  journal = {The Journal of Open Source Software},
     year = 2016,
    month = aug,
   volume = 1,
      eid = {00045},
    pages = {00045},
      doi = {10.21105/joss.00045},
   adsurl = {http://adsabs.harvard.edu/abs/2016JOSS....1...45H},
}

@ARTICLE{Andrae2010,
       author = {{Andrae}, Rene},
        title = "{Error estimation in astronomy: A guide}",
      journal = {arXiv e-prints},
         year = 2010,
        month = sep,
          eid = {arXiv:1009.2755},
        pages = {arXiv:1009.2755},
archivePrefix = {arXiv},
       eprint = {1009.2755},
 primaryClass = {astro-ph.IM},
       adsurl = {https://ui.adsabs.harvard.edu/abs/2010arXiv1009.2755A}
}

@BOOK{Hinkle+2000,
       author = {{Hinkle}, Kenneth and {Wallace}, Lloyd and {Valenti}, Jeff and {Harmer}, Dianne},
        title = "{Visible and Near Infrared Atlas of the Arcturus Spectrum 3727-9300 A}",
         year = 2000,
       adsurl = {https://ui.adsabs.harvard.edu/abs/2000vnia.book.....H}
}

@ARTICLE{Mashonkina+2007,
       author = {{Mashonkina}, L. and {Korn}, A.~J. and {Przybilla}, N.},
        title = "{A non-LTE study of neutral and singly-ionized calcium in late-type stars}",
      journal = {\aap},
         year = 2007,
        month = jan,
       volume = {461},
       number = {1},
        pages = {261-275},
          doi = {10.1051/0004-6361:20065999},
archivePrefix = {arXiv},
       eprint = {astro-ph/0609527},
 primaryClass = {astro-ph},
       adsurl = {https://ui.adsabs.harvard.edu/abs/2007A&A...461..261M}
}

@ARTICLE{Bloecker+1995,
       author = {{Bloecker}, T.},
        title = "{Stellar evolution of low and intermediate-mass stars. I. Mass loss on the AGB and its consequences for stellar evolution.}",
      journal = {\aap},
         year = 1995,
        month = may,
       volume = {297},
        pages = {727},
       adsurl = {https://ui.adsabs.harvard.edu/abs/1995A&A...297..727B}
}

@ARTICLE{Vassiliadis_Wood1993,
       author = {{Vassiliadis}, E. and {Wood}, P.~R.},
        title = "{Evolution of Low- and Intermediate-Mass Stars to the End of the Asymptotic Giant Branch with Mass Loss}",
      journal = {\apj},
         year = 1993,
        month = aug,
       volume = {413},
        pages = {641},
          doi = {10.1086/173033},
       adsurl = {https://ui.adsabs.harvard.edu/abs/1993ApJ...413..641V}
}

@ARTICLE{Meyer1994,
       author = {{Meyer}, Bradley S.},
        title = "{The r-, s-, and p-Processes in Nucleosynthesis}",
      journal = {\araa},
         year = 1994,
        month = jan,
       volume = {32},
        pages = {153-190},
          doi = {10.1146/annurev.aa.32.090194.001101},
       adsurl = {https://ui.adsabs.harvard.edu/abs/1994ARA&A..32..153M}
}

@ARTICLE{Iben_Renzini1983,
       author = {{Iben}, I., Jr. and {Renzini}, A.},
        title = "{Asymptotic giant branch evolution and beyond.}",
      journal = {\araa},
         year = 1983,
        month = jan,
       volume = {21},
        pages = {271-342},
          doi = {10.1146/annurev.aa.21.090183.001415},
       adsurl = {https://ui.adsabs.harvard.edu/abs/1983ARA&A..21..271I}
}

@ARTICLE{Prantzos_Charbonnel_Iliadis2007,
       author = {{Prantzos}, N. and {Charbonnel}, C. and {Iliadis}, C.},
        title = "{Light nuclei in galactic globular clusters: constraints on the self-enrichment scenario from nucleosynthesis}",
      journal = {\aap},
         year = 2007,
        month = jul,
       volume = {470},
       number = {1},
        pages = {179-190},
          doi = {10.1051/0004-6361:20077205},
archivePrefix = {arXiv},
       eprint = {0704.3331},
 primaryClass = {astro-ph},
       adsurl = {https://ui.adsabs.harvard.edu/abs/2007A&A...470..179P}
}

@ARTICLE{Ventura+2011,
       author = {{Ventura}, P. and {Carini}, R. and {D'Antona}, F.},
        title = "{A deep insight into the Mg-Al-Si nucleosynthesis in massive asymptotic giant branch and super-asymptotic giant branch stars}",
      journal = {\mnras},
         year = 2011,
        month = aug,
       volume = {415},
       number = {4},
        pages = {3865-3871},
          doi = {10.1111/j.1365-2966.2011.18997.x},
archivePrefix = {arXiv},
       eprint = {1105.0603},
 primaryClass = {astro-ph.SR},
       adsurl = {https://ui.adsabs.harvard.edu/abs/2011MNRAS.415.3865V}
}

@ARTICLE{Ventura+2018,
       author = {{Ventura}, P. and {D'Antona}, F. and {Imbriani}, G. and {Di Criscienzo}, M. and {Dell'Agli}, F. and {Tailo}, M.},
        title = "{Magnesium isotopes: a tool to understand self-enrichment in globular clusters}",
      journal = {\mnras},
         year = 2018,
        month = jun,
       volume = {477},
       number = {1},
        pages = {438-449},
          doi = {10.1093/mnras/sty635},
archivePrefix = {arXiv},
       eprint = {1803.04127},
 primaryClass = {astro-ph.SR},
       adsurl = {https://ui.adsabs.harvard.edu/abs/2018MNRAS.477..438V}
}

@ARTICLE{Renzini_2008,
       author = {{Renzini}, Alvio},
        title = "{Origin of multiple stellar populations in globular clusters and their helium enrichment}",
      journal = {\mnras},
         year = 2008,
        month = nov,
       volume = {391},
       number = {1},
        pages = {354-362},
          doi = {10.1111/j.1365-2966.2008.13892.x},
archivePrefix = {arXiv},
       eprint = {0808.4095},
 primaryClass = {astro-ph},
       adsurl = {https://ui.adsabs.harvard.edu/abs/2008MNRAS.391..354R}
}

@ARTICLE{Marino+2012,
       author = {{Marino}, A.~F. and {Milone}, A.~P. and {Sneden}, C. and {Bergemann}, M. and {Kraft}, R.~P. and {Wallerstein}, G. and {Cassisi}, S. and {Aparicio}, A. and {Asplund}, M. and {Bedin}, R.~L. and {Hilker}, M. and {Lind}, K. and {Momany}, Y. and {Piotto}, G. and {Roederer}, I.~U. and {Stetson}, P.~B. and {Zoccali}, M.},
        title = "{The double sub-giant branch of NGC 6656 (M 22): a chemical characterization}",
      journal = {\aap},
         year = 2012,
        month = may,
       volume = {541},
          eid = {A15},
        pages = {A15},
          doi = {10.1051/0004-6361/201118381},
archivePrefix = {arXiv},
       eprint = {1202.2825},
 primaryClass = {astro-ph.SR},
       adsurl = {https://ui.adsabs.harvard.edu/abs/2012A&A...541A..15M}
}

@ARTICLE{Shetrone1996,
       author = {{Shetrone}, M.~D.},
        title = "{Magnesium and Carbon isotopes in Globular Cluster Giants. Test of Deep Mixing.II.}",
      journal = {\aj},
         year = 1996,
        month = dec,
       volume = {112},
        pages = {2639},
          doi = {10.1086/118208},
       adsurl = {https://ui.adsabs.harvard.edu/abs/1996AJ....112.2639S}
}

@ARTICLE{Yong+2003_coolstars,
       author = {{Yong}, David and {Lambert}, David L. and {Ivans}, Inese I.},
        title = "{Magnesium Isotopic Abundance Ratios in Cool Stars}",
      journal = {\apj},
         year = 2003,
        month = dec,
       volume = {599},
       number = {2},
        pages = {1357-1371},
          doi = {10.1086/379369},
archivePrefix = {arXiv},
       eprint = {astro-ph/0309079},
 primaryClass = {astro-ph},
       adsurl = {https://ui.adsabs.harvard.edu/abs/2003ApJ...599.1357Y}
}

@ARTICLE{Melendez_Cohen2009,
       author = {{Mel{\'e}ndez}, J. and {Cohen}, J.~G.},
        title = "{The Rise of the AGB in the Galactic Halo: Mg Isotopic Ratios and High Precision Elemental Abundances in M71 Giants}",
      journal = {\apj},
         year = 2009,
        month = jul,
       volume = {699},
       number = {2},
        pages = {2017-2025},
          doi = {10.1088/0004-637X/699/2/2017},
archivePrefix = {arXiv},
       eprint = {0905.1872},
 primaryClass = {astro-ph.GA},
       adsurl = {https://ui.adsabs.harvard.edu/abs/2009ApJ...699.2017M}
}

@ARTICLE{Decressin+2007,
       author = {{Decressin}, T. and {Charbonnel}, C. and {Meynet}, G.},
        title = "{Origin of the abundance patterns in Galactic globular clusters: constraints on dynamical and chemical properties of globular clusters}",
      journal = {\aap},
         year = 2007,
        month = dec,
       volume = {475},
       number = {3},
        pages = {859-873},
          doi = {10.1051/0004-6361:20078425},
archivePrefix = {arXiv},
       eprint = {0709.4160},
 primaryClass = {astro-ph},
       adsurl = {https://ui.adsabs.harvard.edu/abs/2007A&A...475..859D}
}

@ARTICLE{Pignatari+2008,
       author = {{Pignatari}, M. and {Gallino}, R. and {Meynet}, G. and {Hirschi}, R. and {Herwig}, F. and {Wiescher}, M.},
        title = "{The s-Process in Massive Stars at Low Metallicity: The Effect of Primary $^{14}$N from Fast Rotating Stars}",
      journal = {\apjl},
         year = 2008,
        month = nov,
       volume = {687},
       number = {2},
        pages = {L95},
          doi = {10.1086/593350},
archivePrefix = {arXiv},
       eprint = {0810.0182},
 primaryClass = {astro-ph},
       adsurl = {https://ui.adsabs.harvard.edu/abs/2008ApJ...687L..95P}
}

@ARTICLE{Frischknecht+2012,
       author = {{Frischknecht}, U. and {Hirschi}, R. and {Thielemann}, F. -K.},
        title = "{Non-standard s-process in low metallicity massive rotating stars}",
      journal = {\aap},
         year = 2012,
        month = feb,
       volume = {538},
          eid = {L2},
        pages = {L2},
          doi = {10.1051/0004-6361/201117794},
archivePrefix = {arXiv},
       eprint = {1112.5548},
 primaryClass = {astro-ph.SR},
       adsurl = {https://ui.adsabs.harvard.edu/abs/2012A&A...538L...2F}
}

@ARTICLE{Frischknecht+2016,
       author = {{Frischknecht}, Urs and {Hirschi}, Raphael and {Pignatari}, Marco and {Maeder}, Andr{\'e} and {Meynet}, George and {Chiappini}, Cristina and {Thielemann}, Friedrich-Karl and {Rauscher}, Thomas and {Georgy}, Cyril and {Ekstr{\"o}m}, Sylvia},
        title = "{s-process production in rotating massive stars at solar and low metallicities}",
      journal = {\mnras},
         year = 2016,
        month = feb,
       volume = {456},
       number = {2},
        pages = {1803-1825},
          doi = {10.1093/mnras/stv2723},
archivePrefix = {arXiv},
       eprint = {1511.05730},
 primaryClass = {astro-ph.SR},
       adsurl = {https://ui.adsabs.harvard.edu/abs/2016MNRAS.456.1803F}
}

@ARTICLE{Denissenkov+2015,
       author = {{Denissenkov}, P.~A. and {VandenBerg}, D.~A. and {Hartwick}, F.~D.~A. and {Herwig}, F. and {Weiss}, A. and {Paxton}, B.},
        title = "{The primordial and evolutionary abundance variations in globular-cluster stars: a problem with two unknowns}",
      journal = {\mnras},
         year = 2015,
        month = apr,
       volume = {448},
       number = {4},
        pages = {3314-3324},
          doi = {10.1093/mnras/stv211},
archivePrefix = {arXiv},
       eprint = {1409.1193},
 primaryClass = {astro-ph.SR},
       adsurl = {https://ui.adsabs.harvard.edu/abs/2015MNRAS.448.3314D}
}

@ARTICLE{Melendez_Cohen2007,
       author = {{Mel{\'e}ndez}, Jorge and {Cohen}, Judith G.},
        title = "{Magnesium Isotopes in Metal-poor Dwarfs: The Rise of AGB Stars and the Formation Timescale of the Galactic Halo}",
      journal = {\apjl},
         year = 2007,
        month = apr,
       volume = {659},
       number = {1},
        pages = {L25-L28},
          doi = {10.1086/516735},
archivePrefix = {arXiv},
       eprint = {astro-ph/0702655},
 primaryClass = {astro-ph},
       adsurl = {https://ui.adsabs.harvard.edu/abs/2007ApJ...659L..25M}
}

@ARTICLE{Karakas+2018,
       author = {{Karakas}, Amanda I. and {Lugaro}, Maria and {Carlos}, Mar{\'\i}lia and {Cseh}, Borb{\'a}la and {Kamath}, Devika and {Garc{\'\i}a-Hern{\'a}ndez}, D.~A.},
        title = "{Heavy-element yields and abundances of asymptotic giant branch models with a Small Magellanic Cloud metallicity}",
      journal = {\mnras},
         year = 2018,
        month = jun,
       volume = {477},
       number = {1},
        pages = {421-437},
          doi = {10.1093/mnras/sty625},
archivePrefix = {arXiv},
       eprint = {1803.02028},
 primaryClass = {astro-ph.SR},
       adsurl = {https://ui.adsabs.harvard.edu/abs/2018MNRAS.477..421K}
}

@ARTICLE{deMink+2009,
       author = {{de Mink}, S.~E. and {Pols}, O.~R. and {Langer}, N. and {Izzard}, R.~G.},
        title = "{Massive binaries as the source of abundance anomalies in globular clusters}",
      journal = {\aap},
         year = 2009,
        month = nov,
       volume = {507},
       number = {1},
        pages = {L1-L4},
          doi = {10.1051/0004-6361/200913205},
archivePrefix = {arXiv},
       eprint = {0910.1086},
 primaryClass = {astro-ph.SR},
       adsurl = {https://ui.adsabs.harvard.edu/abs/2009A&A...507L...1D}
}

@ARTICLE{Bekki2023,
       author = {{Bekki}, Kenji},
        title = "{Globular cluster formation with multiple stellar populations: a single-binary composite scenario}",
      journal = {\mnras},
         year = 2023,
        month = jan,
       volume = {518},
       number = {3},
        pages = {3274-3285},
          doi = {10.1093/mnras/stac3163},
archivePrefix = {arXiv},
       eprint = {2211.00344},
 primaryClass = {astro-ph.SR},
       adsurl = {https://ui.adsabs.harvard.edu/abs/2023MNRAS.518.3274B}
}

@ARTICLE{MarinFranch+2009,
       author = {{Mar{\'\i}n-Franch}, Antonio and {Aparicio}, Antonio and {Piotto}, Giampaolo and {Rosenberg}, Alfred and {Chaboyer}, Brian and {Sarajedini}, Ata and {Siegel}, Michael and {Anderson}, Jay and {Bedin}, Luigi R. and {Dotter}, Aaron and {Hempel}, Maren and {King}, Ivan and {Majewski}, Steven and {Milone}, Antonino P. and {Paust}, Nathaniel and {Reid}, I. Neill},
        title = "{The ACS Survey of Galactic Globular Clusters. VII. Relative Ages}",
      journal = {\apj},
         year = 2009,
        month = apr,
       volume = {694},
       number = {2},
        pages = {1498-1516},
          doi = {10.1088/0004-637X/694/2/1498},
archivePrefix = {arXiv},
       eprint = {0812.4541},
 primaryClass = {astro-ph},
       adsurl = {https://ui.adsabs.harvard.edu/abs/2009ApJ...694.1498M}
}

@ARTICLE{VandenBerg+2002,
       author = {{VandenBerg}, Don A. and {Richard}, O. and {Michaud}, G. and {Richer}, J.},
        title = "{Models of Metal-poor Stars with Gravitational Settling and Radiative Accelerations. II. The Age of the Oldest Stars}",
      journal = {\apj},
         year = 2002,
        month = may,
       volume = {571},
       number = {1},
        pages = {487-500},
          doi = {10.1086/339895},
       adsurl = {https://ui.adsabs.harvard.edu/abs/2002ApJ...571..487V}
}

@ARTICLE{VandenBerg+2016,
       author = {{VandenBerg}, Don A. and {Denissenkov}, P.~A. and {Catelan}, M{\'a}rcio},
        title = "{Constraints on the Distance Moduli, Helium and Metal Abundances, and Ages of Globular Clusters from their RR Lyrae and Non-variable Horizontal-branch Stars. I. M3, M15, and M92}",
      journal = {\apj},
         year = 2016,
        month = aug,
       volume = {827},
       number = {1},
          eid = {2},
        pages = {2},
          doi = {10.3847/0004-637X/827/1/2},
archivePrefix = {arXiv},
       eprint = {1607.02088},
 primaryClass = {astro-ph.SR},
       adsurl = {https://ui.adsabs.harvard.edu/abs/2016ApJ...827....2V}
}

@ARTICLE{Salaris_Weiss2002,
       author = {{Salaris}, M. and {Weiss}, A.},
        title = "{Homogeneous age dating of 55 Galactic globular clusters. Clues to the Galaxy formation mechanisms}",
      journal = {\aap},
         year = 2002,
        month = jun,
       volume = {388},
        pages = {492-503},
          doi = {10.1051/0004-6361:20020554},
archivePrefix = {arXiv},
       eprint = {astro-ph/0204410},
 primaryClass = {astro-ph},
       adsurl = {https://ui.adsabs.harvard.edu/abs/2002A&A...388..492S}
}

@ARTICLE{Gieles+2018,
       author = {{Gieles}, Mark and {Charbonnel}, Corinne and {Krause}, Martin G.~H. and {H{\'e}nault-Brunet}, Vincent and {Agertz}, Oscar and {Lamers}, Henny J.~G.~L.~M. and {Bastian}, Nathan and {Gualandris}, Alessia and {Zocchi}, Alice and {Petts}, James A.},
        title = "{Concurrent formation of supermassive stars and globular clusters: implications for early self-enrichment}",
      journal = {\mnras},
         year = 2018,
        month = aug,
       volume = {478},
       number = {2},
        pages = {2461-2479},
          doi = {10.1093/mnras/sty1059},
archivePrefix = {arXiv},
       eprint = {1804.04682},
 primaryClass = {astro-ph.GA},
       adsurl = {https://ui.adsabs.harvard.edu/abs/2018MNRAS.478.2461G}
}

@article{corner,
      doi = {10.21105/joss.00024},
      url = {https://doi.org/10.21105/joss.00024},
      year  = {2016},
      month = {jun},
      publisher = {The Open Journal},
      volume = {1},
      number = {2},
      pages = {24},
      author = {Daniel Foreman-Mackey},
      title = {corner.py: Scatterplot matrices in Python},
      journal = {The Journal of Open Source Software}
    }

@ARTICLE{Gallino+1998,
       author = {{Gallino}, Roberto and {Arlandini}, Claudio and {Busso}, Maurizio and {Lugaro}, Maria and {Travaglio}, Claudia and {Straniero}, Oscar and {Chieffi}, Alessandro and {Limongi}, Marco},
        title = "{Evolution and Nucleosynthesis in Low-Mass Asymptotic Giant Branch Stars. II. Neutron Capture and the S-Process}",
      journal = {\apj},
         year = 1998,
        month = apr,
       volume = {497},
       number = {1},
        pages = {388-403},
          doi = {10.1086/305437},
       adsurl = {https://ui.adsabs.harvard.edu/abs/1998ApJ...497..388G}
}

@ARTICLE{Marigo2022,
       author = {{Marigo}, Paola},
        title = "{The Initial-Final Mass Relation of White Dwarfs: A Tool to Calibrate the Third Dredge-Up}",
      journal = {Universe},
         year = 2022,
        month = apr,
       volume = {8},
       number = {4},
        pages = {243},
          doi = {10.3390/universe8040243},
archivePrefix = {arXiv},
       eprint = {2204.06470},
 primaryClass = {astro-ph.SR},
       adsurl = {https://ui.adsabs.harvard.edu/abs/2022Univ....8..243M}
}

@ARTICLE{Hirschi_2007,
       author = {{Hirschi}, R.},
        title = "{Very low-metallicity massive stars:. Pre-SN evolution models and primary nitrogen production}",
      journal = {\aap},
         year = 2007,
        month = jan,
       volume = {461},
       number = {2},
        pages = {571-583},
          doi = {10.1051/0004-6361:20065356},
archivePrefix = {arXiv},
       eprint = {astro-ph/0608170},
 primaryClass = {astro-ph},
       adsurl = {https://ui.adsabs.harvard.edu/abs/2007A&A...461..571H}
}

@ARTICLE{Marino+2015_5286,
       author = {{Marino}, A.~F. and {Milone}, A.~P. and {Karakas}, A.~I. and {Casagrande}, L. and {Yong}, D. and {Shingles}, L. and {Da Costa}, G. and {Norris}, J.~E. and {Stetson}, P.~B. and {Lind}, K. and {Asplund}, M. and {Collet}, R. and {Jerjen}, H. and {Sbordone}, L. and {Aparicio}, A. and {Cassisi}, S.},
        title = "{Iron and s-elements abundance variations in NGC 5286: comparison with `anomalous' globular clusters and Milky Way satellites}",
      journal = {\mnras},
         year = 2015,
        month = jun,
       volume = {450},
       number = {1},
        pages = {815-845},
          doi = {10.1093/mnras/stv420},
archivePrefix = {arXiv},
       eprint = {1502.07438},
 primaryClass = {astro-ph.SR},
       adsurl = {https://ui.adsabs.harvard.edu/abs/2015MNRAS.450..815M}
}

@ARTICLE{Monty+2023,
       author = {{Monty}, Stephanie and {Yong}, David and {Marino}, Anna F. and {Karakas}, Amanda I. and {McKenzie}, Madeleine and {Grundahl}, Frank and {Mura-Guzm{\'a}n}, Aldo},
        title = "{Peeking beneath the precision floor - I. Metallicity spreads and multiple elemental dispersions in the globular clusters NGC 288 and NGC 362}",
      journal = {\mnras},
         year = 2023,
        month = jan,
       volume = {518},
       number = {1},
        pages = {965-986},
          doi = {10.1093/mnras/stac3040},
archivePrefix = {arXiv},
       eprint = {2210.15061},
 primaryClass = {astro-ph.GA},
       adsurl = {https://ui.adsabs.harvard.edu/abs/2023MNRAS.518..965M}
}

@ARTICLE{Johnson+2020,
       author = {{Johnson}, Christian I. and {Dupree}, Andrea K. and {Mateo}, Mario and {Bailey}, John I., III and {Olszewski}, Edward W. and {Walker}, Matthew G.},
        title = "{The Most Metal-poor Stars in Omega Centauri (NGC 5139)}",
      journal = {\aj},
         year = 2020,
        month = jun,
       volume = {159},
       number = {6},
          eid = {254},
        pages = {254},
          doi = {10.3847/1538-3881/ab8819},
archivePrefix = {arXiv},
       eprint = {2004.09023},
 primaryClass = {astro-ph.SR},
       adsurl = {https://ui.adsabs.harvard.edu/abs/2020AJ....159..254J}
}

@ARTICLE{Nitschai+2023,
       author = {{Nitschai}, M.~S. and {Neumayer}, N. and {Clontz}, C. and {H{\"a}berle}, M. and {Seth}, A.~C. and {Husser}, T. -O. and {Kamann}, S. and {Alfaro-Cuello}, M. and {Kacharov}, N. and {Bellini}, A. and {Dotter}, A. and {Dreizler}, S. and {Feldmeier-Krause}, A. and {Latour}, M. and {Libralato}, M. and {Milone}, A.~P. and {Pechetti}, R. and {van de Ven}, G. and {Voggel}, K. and {Weisz}, Daniel R.},
        title = "{oMEGACat I: MUSE spectroscopy of 300,000 stars within the half-light radius of $\omega$ Centauri}",
      journal = {arXiv e-prints},
         year = 2023,
        month = sep,
          eid = {arXiv:2309.02503},
        pages = {arXiv:2309.02503},
          doi = {10.48550/arXiv.2309.02503},
archivePrefix = {arXiv},
       eprint = {2309.02503},
 primaryClass = {astro-ph.GA},
       adsurl = {https://ui.adsabs.harvard.edu/abs/2023arXiv230902503N}
}

@ARTICLE{Marino+2021,
       author = {{Marino}, A.~F. and {Milone}, A.~P. and {Renzini}, A. and {Yong}, D. and {Asplund}, M. and {Da Costa}, G.~S. and {Jerjen}, H. and {Cordoni}, G. and {Carlos}, M. and {Dondoglio}, E. and {Lagioia}, E.~P. and {Jang}, S. and {Tailo}, M.},
        title = "{Spectroscopy and Photometry of the Least Massive Type II Globular Clusters: NGC 1261 and NGC 6934}",
      journal = {\apj},
         year = 2021,
        month = dec,
       volume = {923},
       number = {1},
          eid = {22},
        pages = {22},
          doi = {10.3847/1538-4357/ac282c},
archivePrefix = {arXiv},
       eprint = {2106.15978},
 primaryClass = {astro-ph.SR},
       adsurl = {https://ui.adsabs.harvard.edu/abs/2021ApJ...923...22M}
}

@ARTICLE{Ram+2014,
       author = {{Ram}, Ram S. and {Brooke}, James S.~A. and {Bernath}, Peter F. and {Sneden}, Christopher and {Lucatello}, Sara},
        title = "{Improved Line Data for the Swan System $^{12}$C$^{13}$C Isotopologue}",
      journal = {\apjs},
         year = 2014,
        month = mar,
       volume = {211},
       number = {1},
          eid = {5},
        pages = {5},
          doi = {10.1088/0067-0049/211/1/5},
       adsurl = {https://ui.adsabs.harvard.edu/abs/2014ApJS..211....5R}
}

@Article{         harris2020array,
 title         = {Array programming with {NumPy}},
 author        = {Charles R. Harris and K. Jarrod Millman and St{\'{e}}fan J.
                 van der Walt and Ralf Gommers and Pauli Virtanen and David
                 Cournapeau and Eric Wieser and Julian Taylor and Sebastian
                 Berg and Nathaniel J. Smith and Robert Kern and Matti Picus
                 and Stephan Hoyer and Marten H. van Kerkwijk and Matthew
                 Brett and Allan Haldane and Jaime Fern{\'{a}}ndez del
                 R{\'{i}}o and Mark Wiebe and Pearu Peterson and Pierre
                 G{\'{e}}rard-Marchant and Kevin Sheppard and Tyler Reddy and
                 Warren Weckesser and Hameer Abbasi and Christoph Gohlke and
                 Travis E. Oliphant},
 year          = {2020},
 month         = sep,
 journal       = {Nature},
 volume        = {585},
 number        = {7825},
 pages         = {357--362},
 doi           = {10.1038/s41586-020-2649-2},
 publisher     = {Springer Science and Business Media {LLC}},
 url           = {https://doi.org/10.1038/s41586-020-2649-2}
}

@Article{Hunter2007,
  Author    = {Hunter, J. D.},
  Title     = {Matplotlib: A 2D graphics environment},
  Journal   = {Computing in Science \& Engineering},
  Volume    = {9},
  Number    = {3},
  Pages     = {90--95},
  publisher = {IEEE COMPUTER SOC},
  doi       = {10.1109/MCSE.2007.55},
  year      = 2007
}

@software{reback2020pandas,
    author       = {The pandas development team},
    title        = {pandas-dev/pandas: Pandas},
    month        = feb,
    year         = 2020,
    publisher    = {Zenodo},
    version      = {latest},
    doi          = {10.5281/zenodo.3509134},
    url          = {https://doi.org/10.5281/zenodo.3509134}
}

@conference{Kluyver2016jupyter,
Title = {Jupyter Notebooks -- a publishing format for reproducible computational workflows},
Author = {Thomas Kluyver and Benjamin Ragan-Kelley and Fernando P{\'e}rez and Brian Granger and Matthias Bussonnier and Jonathan Frederic and Kyle Kelley and Jessica Hamrick and Jason Grout and Sylvain Corlay and Paul Ivanov and Dami{\'a}n Avila and Safia Abdalla and Carol Willing},
Booktitle = {Positioning and Power in Academic Publishing: Players, Agents and Agendas},
Editor = {F. Loizides and B. Schmidt},
Organization = {IOS Press},
Pages = {87 - 90},
Year = {2016}
}

%%%%%%%%%%%%%%%%%%%%%%%%%%%%%%%%%%%%%%%%%%%%%%%%%%

%%%%%%%%%%%%%%%%% APPENDICES %%%%%%%%%%%%%%%%%%%%%

\appendix

\section{Methods for calculating the isotopic ratio}
\label{app:alternative_methods}

For this work, we test a variety of algorithms to determine the most accurate method of analysis which also produces realistic errors. Initial versions of \textsc{ratio} used a simple hill-climbing algorithm which minimised the reduced $\chi^2$ value based on some initial value determined using pyMOOGi. Initial guesses were based on the closest fit to the blue wing (i.e., the side not influenced by isotopic splitting) of each line individually. Variations of the code tested the impact of holding one or multiple of these variables constant (e.g. fixing the radial velocity and/or continuum). However, we found that the final isotopic ratio was dependent on the initial value provided to the program and did not yield consistent results with different initial guesses. This hinted at a degeneracy between the total Mg value and the broadening from macroturbulence.

Subsequent versions utilised the Python module \texttt{scipy.optimize.minimize}, testing several built-in minimisation techniques including `Nelder-Mead', `BFGS' and `Newton-CG'. When optimising all six parameters, the code either did not return a satisfactory fit to the spectra (as judged by eye) or failed to converge entirely. To reduce the dimensionality of the problem, we tried fixing the broadening value based on the \ion{Ti}{I} 5145.5 \AA{} and \ion{Ni}{I} 5115.4 \AA{} lines as first suggested in \cite{McWilliam_Lambert1988}. For our sample stars, the Ti I is blended with a neighbouring Fe I line, making it unsuitable to use independently to determine the blending. Running \texttt{scipy.optimize.minimize} on the \ion{Ni}{I} line while varying the total Ni and the broadening after fixing the continuum and radial velocity by eye produced well-fitting synthetic spectra. However, when calculating the isotopic ratio with the \ion{Ni}{I} broadening value, the MgH lines were too narrow to produce an accurate fit of the line (i.e., the \ion{Ni}{I} overestimated the broadening). 

Additional attempts to fit these lines were to fit regions R1, R2, and R3 simultaneously. Our intention for this was twofold: (1) the increased number of data points may help the code to converge and (2) to find one isotopic ratio that satisfied all regions. As noted in previous studies, known and unknown blends within the line list result in no isotopic ratio being able to describe all three lines. Therefore we conclude this is not a feasible method for determining the global isotopic ratio within our stars.

%therefore we conclude that it is not realistic to fit all regions simultaneously. 

Given the degeneracies between the parameters, a brute-force approach to fitting the isotopes would result in the most accurate and unbiased fits to each line. To test this we synthesised a grid of models for each of the three MgH lines, varying the total Mg abundance, broadening, $^{24}$Mg, $^{25}$Mg and $^{26}$Mg. Each line is then compared against upwards of 200,000 MOOG models with the stellar parameters of \teff = 4070 K, \logg = 0.4 cm s$^{-2}$ and [Fe/H] = -1.8. As in previous methods, the accuracy of the fit is based on a reduced $\chi^2$ estimation. The top 10\% of these models are then re-run, but now shifting the observed spectra over an additional grid with different radial velocity and continuum values. The synthetic spectra produced by this method were by far the best match to the observed spectra compared to the previous methods. Unfortunately, this grid was created based on one model atmosphere and line list. Although \cite{Yong+2003} illustrated that changing the model atmosphere does not result in a significant change in the final isotopic ratios, testing illustrated that when using a different model atmosphere to the stellar parameters as calculated in Paper I, the synthetic spectra from this grid method no longer resulted in a good fit. Therefore based on the computational time and memory storage requirements of this method, it is not realistic to create custom grids of this size for each of our stars.

\section{Additional light element abundances}
\label{app:light_elements}
\begin{figure}
    \includegraphics[width=\columnwidth]{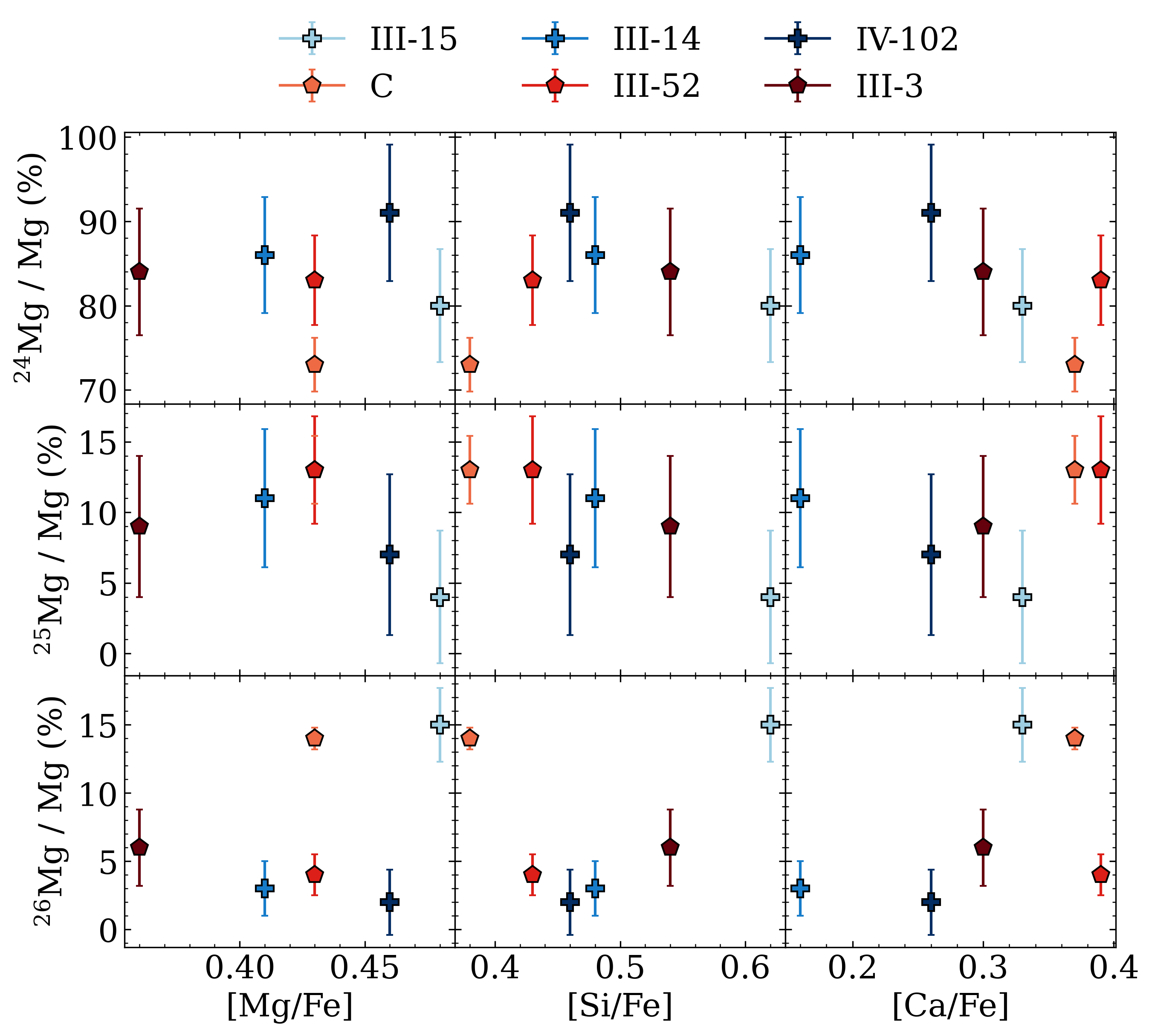}
    \caption{Mg isotope ratios as a function of the light elements Mg, Si and Ca using abundances from \protect\cite{Marino+2011}. We do not find any significant trends in any of these elements.}
    \label{fig:iso_anti_light_no_corr}
\end{figure}

\cite{Marino+2011} provides Mg, Si and Ca abundances for our target stars. However, we do not see any significant trends in the isotopic ratios for these elements. We include Fig. \ref{fig:iso_anti_light_no_corr} here for completeness as the (lack of) trends in these elements may prove useful for constraining nucleosynthetic models. 

\begin{figure}
    \includegraphics[width=\columnwidth]{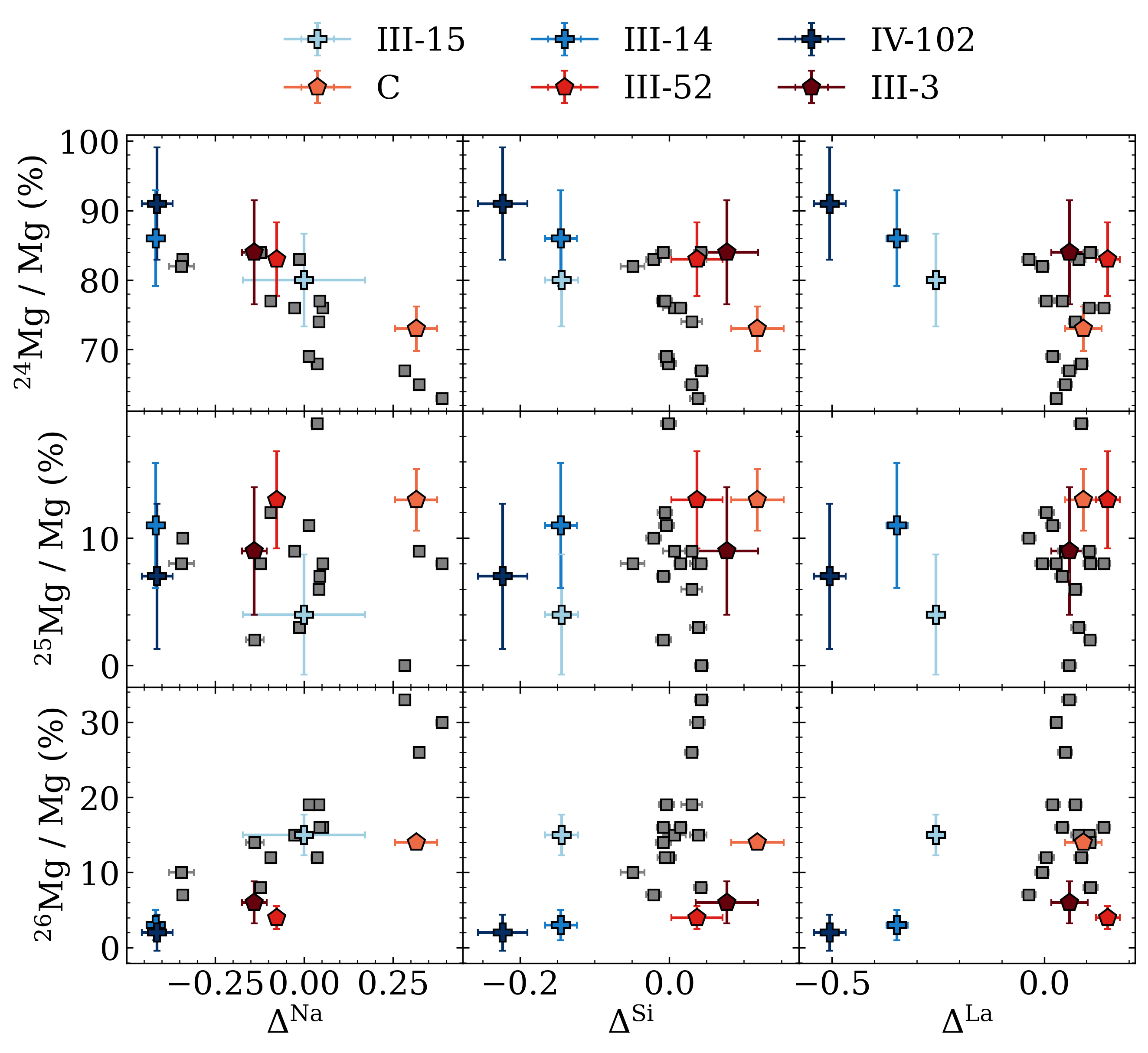}
    \caption{Mg isotope ratios as a function of the light elements Na, Si and La using differential abundances from Paper I. In grey squares, we plot the isotopic ratios from \protect\cite{Yong+2003} and differential abundances from \protect\cite{Yong+2013} for the Galactic GC NGC6752. For Na, we see similar correlations in $^{26}$Mg and anticorrelations in $^{24}$Mg. For Si and La there are no discernible trends in either cluster.}
    \label{fig:iso_anti_NGC6752}
\end{figure}

To compare with the literature results from \cite{Yong+2003} and \cite{Yong+2013}, we plot the isotopic ratios and differential abundances found for NGC6752 alongside M 22 in Fig. \ref{fig:iso_anti_NGC6752}. 

% Don't change these lines
\bsp	% typesetting comment
\label{lastpage}
\end{document}